\documentclass[fleqn,10pt]{wlscirep}
\usepackage[utf8]{inputenc}
\usepackage[T1]{fontenc}
\usepackage{authblk}
\usepackage{amssymb}
\usepackage{amsmath} 
\usepackage{amsfonts}
\usepackage{latexsym}
\usepackage{amsthm} 
\usepackage{eucal}

\usepackage{rotating}
\usepackage[ruled,vlined,resetcount,algosection,boxed,algo2e]{algorithm2e}
\DontPrintSemicolon  
\usepackage{subfig}
\usepackage{graphicx}
\usepackage[english]{babel}
\usepackage[english]{varioref}
\usepackage{relsize}%
\usepackage{url}
\usepackage{multirow}
\usepackage{hhline}
\usepackage{colortbl}
\usepackage{wrapfig}
\usepackage{microtype}
\usepackage{tikz}
\usepackage{pgfplots}
\usepackage{caption}
\usepackage[subfigure]{tocloft}
\pgfplotsset{compat=1.18}

\addto\captionsenglish{
  \renewcommand{\contentsname}%
    {Supplementary Notes}%
}

\newcommand{\listequationsname}{Supplementary Equations}
\newlistof{myequations}{equ}{\listequationsname}
\newcommand{\myequations}[1]{%
\addcontentsline{equ}{myequations}{\protect\numberline{\theequation}#1}\par}
\addto\captionsenglish{%
  }
\def\l@section{\@tocline{1}{0,2pt}{2pc}{12mm}{\ \ }} 

\newcommand{\startsupplement}{
    \setcounter{figure}{0}
    \setcounter{table}{0}
    \setcounter{equation}{0}
    \renewcommand{\thetable}{S\arabic{table}}
    \renewcommand{\theequation}{S\arabic{equation}}
    \captionsetup[figure]{name=Figure S}
        \makeatother
    \renewcommand{\cftfigpresnum}{Fig. S}
    \setlength{\cftfignumwidth}{5.5em}
    \setlength{\cftfigindent}{0em}
}

\title{Time-space dynamics of income segregation: a case study of Milan's neighbourhoods}

\makeatletter

\author[1,2]{{Lavinia} {Rossi Mori}}
\author[3,4,1,5]{{Vittorio} {Loreto}}
\author[6,7,8*]{{Riccardo} {Di Clemente}}

\affil[1]{Centro Ricerche Enrico Fermi, Via Panisperna 89/A, 00184, Rome, Italy}

\affil[2]{Physics Department, Universit\`a di Roma Tor Vergata, 00133, Rome, Italy}

\affil[3]{Sony Computer Science Laboratories Rome, Joint Initiative CREF-Sony, Centro Ricerche Enrico Fermi, Via Panisperna 89/A, 00184, Rome, Italy}

\affil[4]{Physics Department, Sapienza University of Rome, Piazzale Aldo Moro 2, 00185, Rome, Italy
}

\affil[5]{Complexity Science Hub, Josefst\"{a}dter Strasse 39, A 1080 Vienna, Austria}

\affil[6]{Complex Connections Lab, Network Science Institute, Northeastern University London, London, E1W 1LP, United Kingdom.}

\affil[7]{The Alan Turing Institute, London, NW12DB, United Kingdom}

\affil[8]{ISI Foundation, 10126, Torino, Italy}

\affil[*]{corresponding author: riccardo.diclemente@nulondon.ac.uk}


\begin{abstract}
Traditional approaches to urban income segregation focus on static residential patterns, often failing to capture the dynamic nature of social mixing at the neighborhood level.
Leveraging high-resolution location-based data from mobile phones, we capture the interplay of three different income groups (high, medium, low) based on their daily routines. 
We propose a three-dimensional space to analyze social mixing, which is embedded in the temporal dynamics of urban activities. This framework offers a more detailed perspective on social interactions, closely linked to the geographical features of each neighborhood.
While residential areas fail to encourage social mixing in the nighttime, the working hours foster inclusion, with the city center showing a heightened level of interaction.
As evening sets in, leisure areas emerge as potential facilitators for social interactions, depending on urban features such as public transport and a variety of Points Of Interest.
These characteristics significantly modulate the magnitude and type of social stratification involved in social mixing, also underscoring the significance of urban design in either bridging or widening socio-economic divides.

\end{abstract}

\usepackage{hyperref}
\begin{document}

\flushbottom
\maketitle
\thispagestyle{empty}

\section*{Introduction}\label{sec1}

Cities are vibrant organisms in constant evolution, continuously reshaped by economic, technological, and social forces~\cite{Sharifi2019}. These dynamics shape the form and function of urban neighborhoods, influencing where individuals live, work, and
interact ~\cite{He2018}. While technological advances and economic developments have redefined these environments, the benefits have not been uniformly distributed across cities~\cite{Ravallion2007} or neighbourhoods~\cite{Galster2007}. This inequitable allocation of opportunities~\cite{Chetty2016} and services~\cite{Tammaru2021} leads to socio-economic disparity~\cite{Thomas2014}. A recent study ~\cite{Xu2022} has shown that residents in diverse residential contexts often experience limited exposure to varied social contacts due to the spatial and social confines of their neighbourhoods. This highlights a disparity in opportunities for social interactions, despite the mobility afforded by urban living, particularly in areas of high residential segregation. Wealthier neighbourhoods often enjoy better access to essential services, such as healthcare~\cite{Bor2017}, education~\cite{Owens2019}, and job opportunities~\cite{Bosquet2019}, resulting in income-based urban segregation and limited interaction between different income groups~\cite{Echenique2007}.
The growing income disparity~\cite{Musterd2017} further compounds these inequalities, leading to spatial divisions within urban areas. These divisions go beyond mere physical boundaries~\cite{ric1}, influencing how people interact with their environment, affecting mobility patterns~\cite{Wissink2016}, and social encounters~\cite{Xu2019}.

The study of income segregation has expanded from an initial focus on residential patterns~\cite{Massey1988} to a more dynamic perspective that includes workplace dynamics. This shift acknowledges the diverse mobility behaviours of different income groups. Wealthier individuals, endowed with a broader range of transportation options, are more selective about their destinations and generally commute less~\cite{Macedo2022}.
Meanwhile, the place of work can mitigate some of the segregation dynamics by acting as a social bridge and fostering interaction among people of different city areas~\cite{dong}. 
The interplay between residential and workplace mobility significantly impacts the socio-economic fabric of urban areas~\cite{Chetty2022}.
Recent technological developments have significantly benefited the study of urban segregation and social dynamics.
A study analyzing urban mobility patterns in American cities has shown that, despite physical mobility, residents of disadvantaged neighbourhoods continue to experience segregation, often remaining less exposed to more affluent or diverse areas~\cite{Wang2018}.
This insight into persistent segregation patterns informs our study's focus on exploring urban dynamics through Location-Based Services (LBS)~\cite{lbs} trajectory data to capture the broader dynamics of income segregation. We extend the analytical scope beyond the traditionally studied 'first' and 'second' places (i.e., home and work) to also include 'third' places~\cite{Oldenburg1999} – public spaces where people spend leisure or community time. This extension allows for a more comprehensive understanding of segregation dynamics, accounting for individuals' diverse activities across various urban environments, studied at various spatial scales—from Points of Interest (POIs)~\cite{Moro2021} to streets~\cite{Moro2023} and neighbourhoods~\cite{Wong2011}.

Recent studies have underscored the significant role of unique Points of Interest (POIs) and diverse street functions in promoting social mixing~\cite{Moro2021,nilforoshan2023human,Moro2023}. However, it's essential to recognise that while these elements are crucial, they offer only a microcosmic perspective on urban segregation dynamics. A street or a single point of interest represents just a fragment of a larger picture. The intricate interplay of more expansive neighbourhoods, vital community hubs, and interconnected transport systems emerges as a central theme in understanding broader integration patterns and division within urban settings~\cite{Sampson}. Furthermore, by examining larger urban expanses, we often unearth deep-rooted socio-economic disparities and historical and cultural facets, which frequently underpin the complexities of urban segregation.

The longitudinal aspect of the spatial dimension of urban dynamics is pivotal, but the temporal aspects offer equal insights. The same urban spaces may exhibit diverse social interactions at different times, raising the following question. If individuals are socially integrated in the spatial dimension, does this integration persist or transform across the temporal dimension? The dynamics of mobility within cities can foster social inclusion, but the extent of this inclusion varies with time and the distinct characteristics of the neighbourhood~\cite{sweden}. For example, some neighbourhoods maintain a stable mix of income groups throughout the day, contributing to a consistent level of social inclusion. Conversely, others show variable degrees of social mixing depending on the time of the day: they might offer an inclusive atmosphere during the day, facilitating interactions among diverse social groups, but revert to a state of higher segregation after nightfall. The challenge is to dynamically observe these changes and understand the topological characteristics that can promote more inclusive neighbourhoods. 

We aim to shed light on the relationship between income segregation and topological features of neighbourhoods, examining the POIs that provide urban services within the same geographical space and the mixing of people in different time bands. To describe how citizens interact with the urban texture, we leverage trajectory LBS data from  94,000 users in Milan over ten months, integrated with a dataset from rent as a proxy for income~\cite{Xu2019}. Through this comprehensive data integration, we gain insights into the hourly mobility patterns of residents. 
Central to our analysis is the notion of "income triad," which represents the distribution of the three income groups in the city. This triad allows us to capture the change of social mixing, indicating whether there's a balanced representation of all income groups (perfect mixing) or dominance by a single group (complete segregation).
We group and classify neighbourhoods based on city structure-dependent features such as the efficiency of public transport,  category diversity, or median price, defined by Accessibility, Liveability, and Attractivity (ALA clustering).
When we overlay the social mixing patterns — defined through the income triad — with the summarised urban fabric characteristics from ALA clustering, we capture the neighbourhood level of income segregation. 
We find that the magnitude of segregation is influenced by the neighbourhood features represented in the ALA cluster, and the time of day. The variety of facilities a neighbourhood offers and their utilisation by different income groups at different time bands significantly impact the levels of segregation.
We find a high level of spatial segregation at night due to residential segregation but observe increased social mixing during the day when residential segregation relaxes. Neighbourhoods exhibiting interaction with middle-income groups appear to be more inclusive, as individuals from these groups commonly attend places frequented by low- and high-income groups.

Additionally, our study incorporates temporal dynamics into the segregation analysis, offering a novel concept of 'temporal mixing' at the neighbourhood level. This approach acknowledges that neighbourhoods and their socioeconomic profiles do not remain static throughout the day but rather fluctuate depending on daily rhythms and patterns of human activities. By examining these temporal shifts, we distinguish three groups - inclusive, mixed, and segregated - and study the dis-similarity between these groups and those determined by the ALA metrics.
Finally, through regression analysis, we could identify the key neighbourhood features driving inclusivity and high social mixing. We find that the most influential characteristics of a neighbourhood are its accessibility and diversity of amenities.

\section*{Results}\label{sec2}

To comprehensively characterise the dimensions of segregation, we present a novel set of space-time metrics.
The purpose is to capture the intricate spatial and temporal patterns that shape divisions within urban environments. The new metrics, rooted in a combination of topographical, socioeconomic, and human mobility data, provide a robust framework for understanding the subtle intricacies of urban interactions and disparities.

\subsection*{Data}
Leveraging individual Location-Based Service (LBS) trajectories, our study encompasses 650,000 users over ten months, utilising anonymised, high-resolution mobile location pings from Milan — Italy's second-largest city with a populace of 1.352 million (ISTAT 2011\footnote{\url{http://dati.istat.it/Index.aspx?DataSetCode=DCIS_POPRES1}}).
We detect the daily stay locations (Supplementary Note \ref{stays}) using the time duration of 20 minutes and a spatial radius of 200 m~ as the Hariharan and Toyama algorithm.
These locations are then classified into three categories: home, work, and third places. The identification of home and work locations is based on the analysis of each user's periodic daily activities~\cite{marta} (refer to Supplementary Note~\ref{homework}). We define third places as any stops that are neither home nor work. These are typically public spaces where individuals engage in leisure or community activities~\cite{Oldenburg1999}. Post-data cleansing (refer to Supplementary Note~\ref{clean}), our sample size narrows to 94,000 users, which represents approximately $10\%$ of Milan's population.
We observed no spatial bias in the dataset with a robust correlation - Pearson coefficient= 0.88 -  between the users and the official census population in each section defined by ISTAT\footnote{\url{https://www.istat.it/it/archivio/104317.}} (see Supplementary Figure S\ref{fig:correlation_istat}).
To capture the users' whereabouts linked to the social mixing by income stratification, we coupled the users' residential area, extracted from the LBS, with rental data for Milan from Caasa.it\footnote{\url{https://www.caasa.it/} (Accessed: 5 February 2024)} for the period September 2022 to June 2023 (elaboration in Supplementary Note~\ref{sec:pearson2} and Figure S\ref{fig:ocse_income} (upper)). 
As rent price near the home location has been proven a good proxy to infer users' income~\cite{Xu2019} (see Supplementary Figure S\ref{fig:income_rap}(down)), each user is associated with the average of the ten closest home rent values within a 200-meter radius of their home location extracted from LBS.
Users inferred income is consistent with census data concerning median income levels\footnote{{\url{https://www1.finanze.gov.it/finanze/analisi_stat/public/index.php?}}} with 0.89 of Pearson Coefficient at ZIP code level (see Supplementary Figure S\ref{fig:income_rap}).
To effectively capture the behavioural differences in visited locations across various income groups and measure economic segregation levels within the city, we utilise k-means to classify individuals into three income groups: low, medium, and high (Supplementary Figure S\ref{fig:kmeans}).
Our decision for a tripartite division was influenced by methodological considerations and the objective to present a clear yet encompassing representation of socio-economic strata (see Methods Section "Clusterization" for details).
To portray the interplay of social mixing dynamics within urban spaces, we leverage the three income clusters to create a 3D income vector space, the \textit{income triade} $\mathbb{I}$. We represent each income group with a basis vector (see Supplementary~\ref{basis}). This framework allows us to classify users, denoted by $\vec{u_k}$  (Eq.~\ref{utente} in the Methods Section).

Neighbourhood structure has an impact on the social dynamic of the urban encounters, the Points of Interest (POIs) locations~\cite{Yue2017}, the street topology~\cite{Dimaggio2012}, and the urban space can affect our daily routines and social activities~\cite{Mehta2019}.
These components are important in defining the neighbourhoods as the geographically basic component that interlinks community and human interactions with urban texture~\cite{Madden2014}.
For a more detailed urban representation, we employed a hexagonal tessellation approach~\cite{hex}, as ZIP codes provide a broad-brush picture (see Supplementary Figure S\ref{fig:maps_hex}). This tessellation allows for consistent spatial sampling, merging the benefits of full coverage (akin to a square grid) with uniform spacing. 
Throughout this study, the terms "hexagon" and "neighbourhood" are used interchangeably, each symbolizing distinct urban sections. 
We selected a 300-meter hexagon grid for its balanced representation of demographic and topological features of the city  (see Supplementary Note \ref{sec:hex}, Fig. S\ref{fig:maps_hex} -- S\ref{fig:profili_seg}).
To ascertain the robustness of our findings, we examined the correlation of the results across different grid resolutions, as illustrated in Fig. S\ref{fig:correlation_hex}.

City topology and neighbourhood characteristics influence our movement patterns and interactions with both individuals and the urban fabric~\cite{Zhong2014}. In this context, Points of Interest (POI) are pivotal to identifying services and opportunities available within these neighbourhoods~\cite{amenity}. We garner information across Milan from Google Places API\footnote{ \url{https://developers.google.com/maps/documentation/places/web-service/overview?hl=en.}} (see Methods section) on approximately 400,000 verified venues, including their latitude, longitude, amenity category, pricing, and reviews in Milan for the year 2022 (Fig.~\ref{fig:fig_1}.\textbf{A}).

\subsection*{ALA-clustering}
To capture the main feature that encourages social inclusivity within urban environments, we need to quantify neighbourhood attributes integral to social mixing. While traditional methods lean on residents' perceptions~\cite{Salesses2013} or census data~\cite{Nieuwenhuis2020}, our approach analyses neighbourhoods based on their topographical and geographical traits, providing a more quantifiable and consistent means of characterisation. These data can be easily captured, updated, and compared across different regions and cities~\cite{Hidalgo2020}.
We introduce three novel metrics to systematically evaluate these neighbourhood characteristics: Accessibility, Liveability, and Attractivity (ALA).
Accessibility defines the ease of reaching a location with public transportation~\cite{Biazzo2019} (details in the Methods section).
To quantify the Liveability metric, we consider both the number and diversity of schools and supermarkets and the value of a building (see supplementary Note \ref{sec:liability}).
How a neighbourhood is diverse in terms of POIs by type, price, and reviews plays a key role in determining a neighbourhood's attractivity~\cite{Saeidi2012}.
We describe Attractivity in terms of a neighbourhood's Fitness~\cite{Tacchella2012} (see Methods section) – capturing both amenity diversity and uniqueness – as well as price diversity, median pricing, and reviews.
The proposed ALA metrics exhibit mostly a low correlation (Supplementary Figure S\ref{fig:correlation}), ensuring a broad and distinct characterisation of neighbourhood qualities. 

\begin{figure}[ht]
\centering
\includegraphics[width=0.98\textwidth]{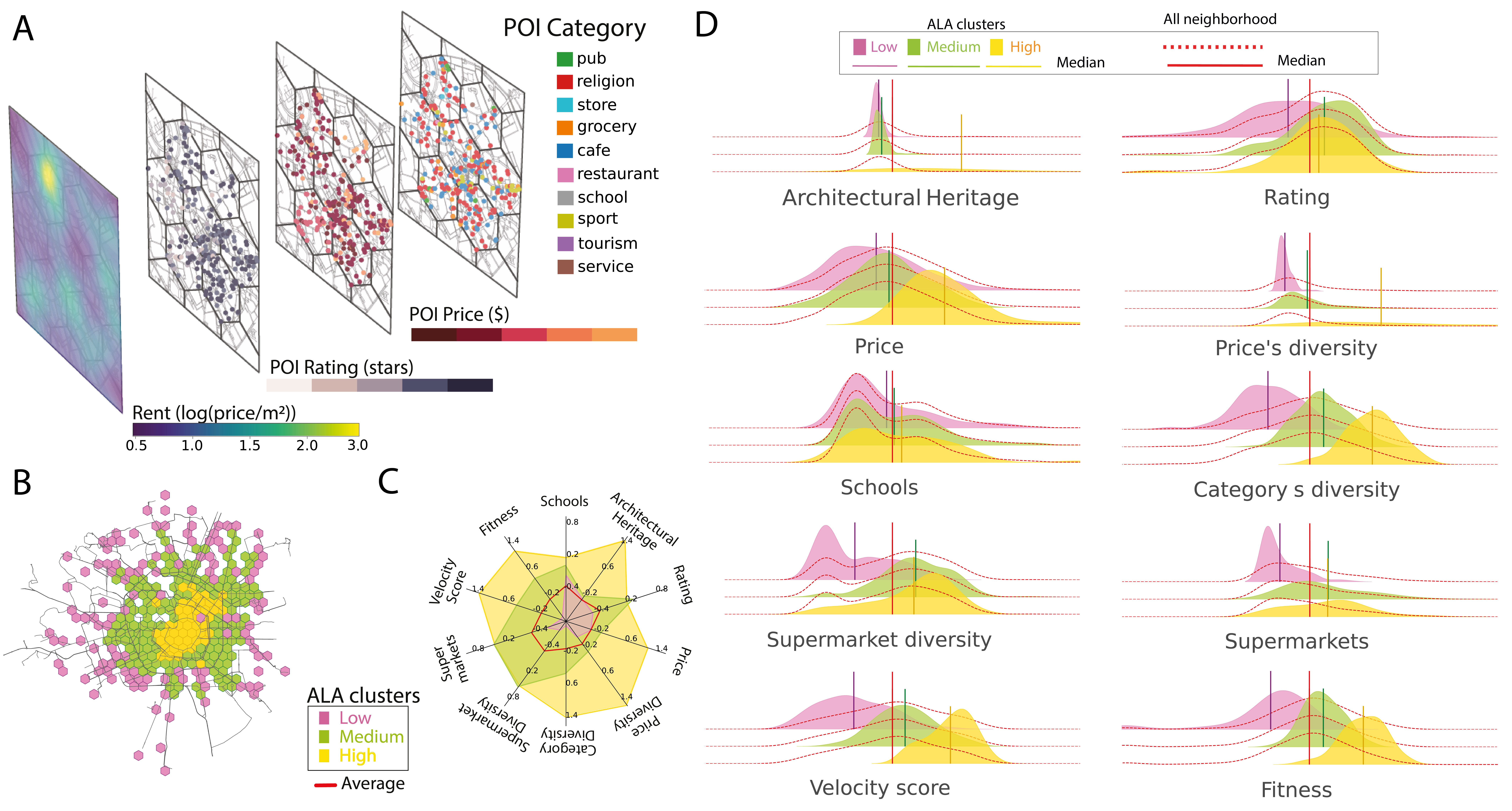}
\caption[]{\textbf{ALA clustering of neighbourhoods in Milan}.
(\textbf{A}) Panels showcase the distribution of rent per square meter, reviews, price, and categories of POIs across selected central hexagons representing individual neighbourhoods.
(\textbf{B}) Spatial map of Milan illustrating cluster distribution, created using OpenStreetMap data: central regions predominantly exhibit high ALA metric values, whereas the periphery is characterised by medium and low ALA values.
(\textbf{C}) Comparative visualisation of the median ALA metrics within each cluster against the average of all neighbourhoods, depicted in red.
(\textbf{D}) Z-score distribution of various ALA metrics, including architectural heritage, rating, price, price diversity, schools, category diversity, supermarket diversity, supermarket count, velocity score, and fitness. Each cluster is denoted by a distinct colour, with vertical lines representing its median value. The overall neighbourhood distribution, irrespective of clusters, is represented by the dotted red line, while the red vertical line signifies the median of all neighbourhoods.}
\label{fig:fig_1}
\end{figure}

We detect three distinct clusters using the k-means algorithm to the z-score normalisation of the ALA metrics (Fig.~\ref{fig:fig_1}.\textbf{B}).  The rationale behind choosing three clusters for this analysis is detailed in the Methods section. Each cluster reflects a different level of neighbourhood quality (Fig.~\ref{fig:fig_1}.\textbf{C}). 
The cluster composed of more central hexagons, labelled "high", exhibited the highest values for almost all metrics, translating to greater accessibility, a wider variety of POIs, and a higher uniqueness of categories. In contrast, the peripheral cluster recorded the lowest values, indicating limited choices of POIs, schools, and restaurants and reduced accessibility from the rest of the city (Fig.~\ref{fig:fig_1}.\textbf{D}).
Schools appear uniformly accessible across all neighbourhoods, a direct outcome of national legislative mandates that ensure equitable access to education~\cite{Belmonte2014}. This uniformity starkly contrasts the differentiation observed in other attributes, underscoring the influence of policy-driven urban planning.

The observed radial pattern in Fig.~\ref{fig:fig_1}.\textbf{B} confirms residential income segregation witnessed in many European cities~\cite{radial}, underscores the uneven distribution of amenities and opportunities among citizens, which favours the city centre to the peripheries. Spatial distributions of singular ALA metric are in Supplementary Figures (~\ref{fig:feature_all}, ~\ref{fig:feature_taglio}). 

\subsection*{Social mixing}
To evaluate how the characteristics of neighbourhoods relate to income-based social mixing over time, we define a time-sensitive metric in the income triade, the Neighbourhood Income Activity, \( \Vec{A}_{h,t} \), at a specific time \( t \) in a given neighbourhood \( h \). This vector aggregates the activities of all users from different income groups in $(h,t)$, and enables us to track the changes in the social mixing across whole cities (see Methods Section, Eq.~\ref{norm1}, Eq.~\ref{norm2}).
To ensure a fair comparison across neighbourhoods, an L1 normalisation transforms the vector components into proportions.
The three coordinates of the Neighbourhood Income Activity are in the Income Triade space \( \mathbf{e}_H \), \( \mathbf{e}_M \), and \( \mathbf{e}_L \), so $\Vec{A}_{h,t} = (0,0,1)$ represents the exclusive presence of users belonging to the "Low" income group in the neighbourhood $h$ at time $t$. Conversely, $\Vec{A}_{h,t} = (0.33,0.33,0.33)$ represents an inclusive neighbourhood that is visited equally by each income group.

\begin{figure}[ht]
\centering
\includegraphics[width=1\textwidth]{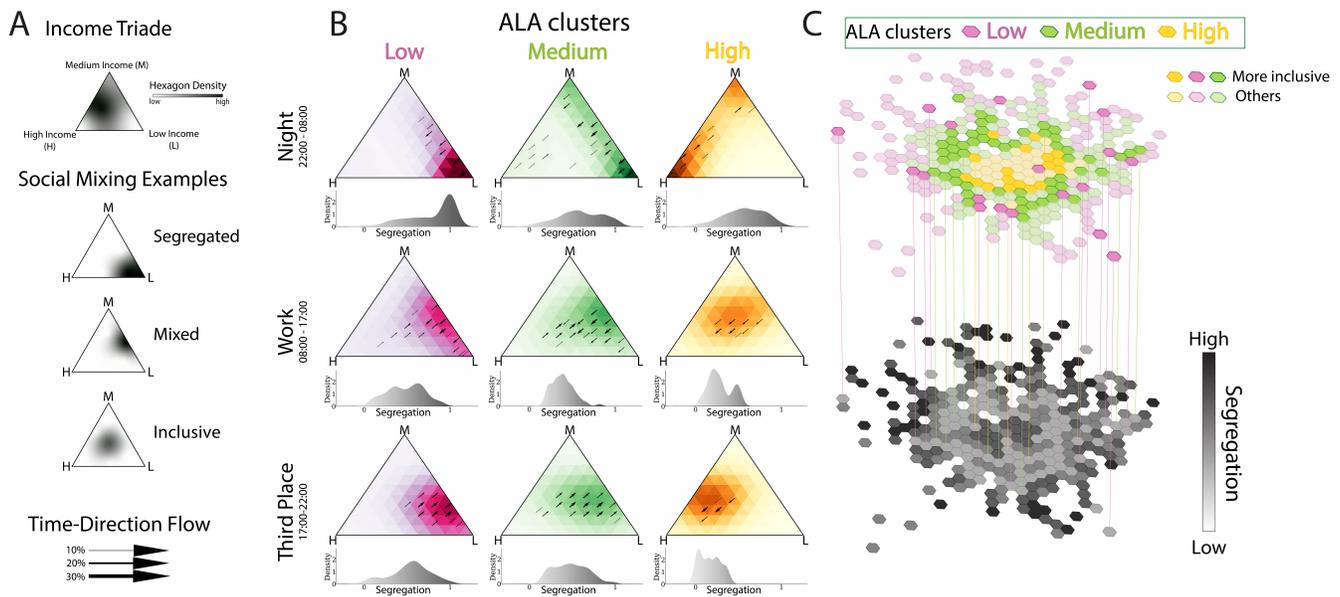}
\caption[]{\textbf{Income Triade Analysis: segregation dynamics}.
(\textbf{A}) Legend and interpretation of the income triad. The triangle's vertices represent the three income groups: high, medium, and low. The stronger the colour, the higher the density of neighbourhoods in that configuration. A neighbourhood's proximity to a vertex indicates a predominant visitation from that specific income group. When a neighbourhood's visitation is evenly distributed among the income groups, it positions near the triangle's centre. From top to bottom, the examples illustrate a segregated neighbourhood predominantly visited by the low-income group (situated at the bottom vertex), a mixed neighbourhood frequented by both low and medium incomes (located on the right edge), and an inclusive neighbourhood with equal representation from all three income groups (centred within the triangle).
To analyse segregation temporally, we employ arrows. The direction of these arrows forecasts the neighbourhood's segregation trend in the subsequent time slot, while their thickness represents the density of neighbourhoods experiencing a similar pattern.
(\textbf{B}) Each column demonstrates the spatial shift in segregation across the ALA clusters: high, medium, and low. In contrast, each row presents a temporal dimension, representing different time bands: home during night-time (10 pm - 6 am), working hours (8 am - 5 pm), and leisure activities in the evening (5 pm - 10 pm). Below each triangle, we report the histograms of the segregation values, as measured in terms of the Gini coefficient (see Methods Section), for all neighbourhoods belonging to the same ALA cluster and the same time band.
(\textbf{C}) A two-tiered map representation created with Python libraries: the upper layer emphasises the most inclusive neighbourhoods with brighter colours, while the other hexagons are opaque. On the lower tier, the colour gradient reflects the level of segregation calculated using the Gini index. All hexagons from the top tier transition into a light grey on the bottom layer, indicating low segregation.
}
\label{fig:ternary}
\end{figure}

In Fig.~\ref{fig:ternary}.\textbf{A}, we provide examples of aggregate behaviours of the neighbourhoods. When the income triade is denser at the bottom right (see Fig.~\ref{fig:ternary}.\textbf{A}, second triangle), a set of neighbourhoods at that time will be mostly visited by the low-income group. 
In the second scenario, where the density is greater on the right edge, we see a collection of mixed neighbourhoods frequented by both low and medium incomes (see Fig.~\ref{fig:ternary}.\textbf{A}, third triangle).
Finally, the fourth triangle of Fig.~\ref{fig:ternary}.\textbf{A} is denser at the centre, indicating that all three income groups equally visit the neighbourhoods.
Fig.~\ref{fig:ternary}.\textbf{B} draws the heat map of the activity of each neighbourhood within the income triad, normalised based on user density, across different ALA clusters and temporal frames.
This visualisation underscores how neighbourhoods in varying ALA clusters become the focal points of activities for citizens with distinct income stratification at different times of the day.
Here the ALA clusters are examined in distinct scenarios and timeframes: during the night (10 p.m. - 8 a.m.), where we analyse only the stops made at home, throughout the day (8 a.m. - 5 p.m.), focusing solely on stops at workplaces, and in the evening (5 p.m. - 10 p.m.), where our attention is on third-places. This methodical separation allows us to dissect the dynamics of social mixing in a targeted manner, isolating specific types of interactions that occur in these respective settings. By analyzing only home stops at night, we gain insight into patterns of static residential segregation. 
Our approach aligns with Ellis et al~\cite{Ellis2004}., who observed less segregation in workplace areas compared to residential neighborhoods, suggesting more diverse interactions during work hours. Similarly, Candipan et al~\cite{Candipan2021}. highlighted the significant variation in urban racial segregation between residential and employment settings, reinforcing the importance of considering both domains in studies of urban segregation to explore the role of employment settings as a connection for social interaction among diverse income groups. 
In the evening, our focus on third places shows how people from different economic backgrounds mingle in social settings outside their homes and workplaces. This approach underscores the urban social dynamics, revealing how the nature and location of interactions shift distinctly across different times of the day.
Within each representation of income distribution, the arrows pinpoint the gradient shifts in the subsequent time section, providing a predictive insight into temporal social dynamics.
Spatial segregation peaks at night, largely attributed to the inherent residential income disparities~\cite{Massey1988}, exacerbating the differences in the home location of different income groups. 
The pink cluster, predominantly inhabited by the low-income group, contrasts with the green cluster — home to both low and medium incomes - and the yellow one that hosts the high and medium-income groups (see Supplementary Figure S\ref{fig:ternary_diff}). 
However, a more diverse interaction appears during the day: residential segregation is relaxed across all clusters, albeit with varying intensities.
In the first row of Fig.~\ref{fig:ternary}.\textbf{B}, the arrows indicate a transition from stringent segregation to an increased social income mixing. 
During work hours, people from different income groups converge in their workplaces, highlighting the role of work in temporarily bridging economic gaps~\cite{dong}. 
The yellow cluster benefits most from the inclusive effects of work hours. 
One potential interpretation is that individuals with lower incomes might be seeking employment opportunities within high ALA cluster areas due to the prospect of better social capital and 'economic connectedness'~\cite{Chetty2022}. While these areas can offer substantial benefits, the trade-off frequently materialises in increased commuting times for these individuals and consequent worse quality of life~\cite{Blumenberg2019}.
The evening transition, characterised by third-place activities such as leisure amenities, exhibits a drift in the income triads towards the periphery, outlining a more pairwise social mixing with interactions between adjacent income groups. This trend leans towards residential segregation in the night hours.
The income triad permits us to capture the subtler interplay of income-based social mixing during transition hours.
Specific attributes render a neighbourhood attractive not just to its inhabitants but to outsiders as well. This allure is contingent on the amenities a neighbourhood offers, leading to distinct behaviours in the three ALA clusters, and it resonates differently across income groups. The high ALA cluster emerges as more inclusive, although the presence of medium income is more conspicuous than that of the low income— they exhibit limited mobility from their primary neighbourhood.
Interestingly, the data shows that the middle-income group visits areas that are common to both low and high income groups, as evidenced by the dense distribution in the middle-income corner across all three clusters (Fig.~\ref{fig:ternary}.\textbf{B} and Supplementary Figure S\ref{fig:quadrato} and S\ref{fig:pentagono}).

We can employ the Gini function~\cite{gini} to have a quantifiable metric to gauge the income social mixing in the neighbourhood. The Gini coefficient (Eq.~\ref{gini} in the Methods section) ranges between 0 (implying perfect mixing) and 1 (indicating complete segregation), serving as a quantifiable reflection of our observations in the ternary plot.
The distributions positioned below each triangle in Fig.~\ref{fig:ternary}.\textbf{B} emphasise the contrasting behaviours of the three ALA clusters. 
While the low ALA cluster invariably exhibits pronounced segregation in all temporal situations, the other two clusters display greater inclusivity.
Having analysed the behaviour of the three income groups using fixed time windows, our focus now shifts to the temporal evolution of the neighbourhood through the Gini index.
This dynamic lens offers a more granular insight into income-based social interactions, and observing how segregation ebbs and flows throughout the day enhances our comprehension of day-long shifts in income segregation.
The locations of home and work exhibit a constancy over time (as depicted in the Supplementary Figure S\ref{fig:home}, S~\ref{fig:work}). Thus, we direct our focus toward analysing the dynamics of third places, extending our analysis across the entire week, examining data over 48-hour periods, weekdays and weekends.
This analysis is illustrated in the two-level map (Fig.~\ref{fig:ternary}.\textbf{C)}, where we observe the distribution of segregation within the city (below) and identify the most inclusive hexagons (above) with the presence of all clusters. This heterogeneity suggests a more complex underlying division of hexagons, different from the one done with the ALA metrics. 

\subsection*{Social mixing Temporal Evolution}
\begin{figure}[!htp]
\centering
\includegraphics[width=\textwidth, trim = 0cm 0cm 0cm 0cm]{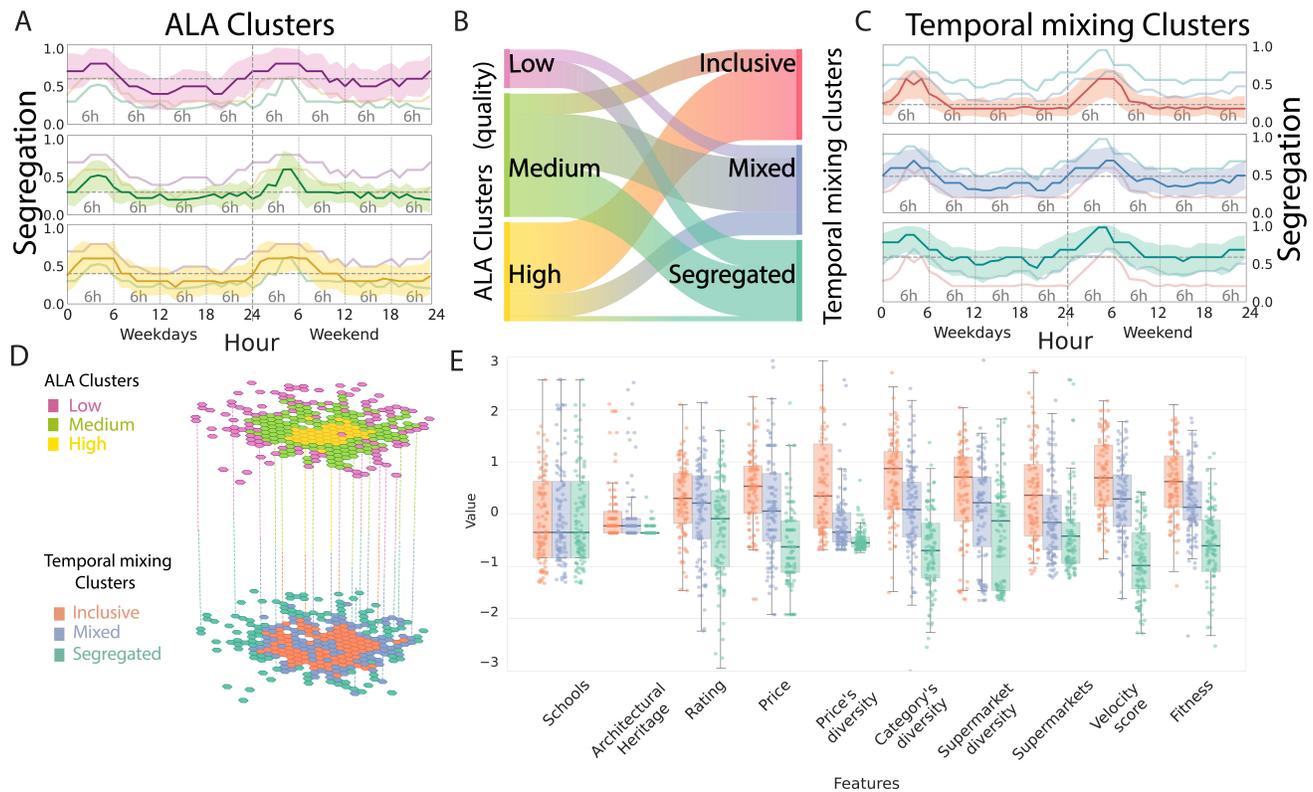}
\caption[]{\textbf{Interplay of ALA Features in Determining Income Segregation.}
(\textbf{A}) Segregation trends within the ALA clusters over weekdays and weekends. The solid line denotes the median, and the variability is captured within the shaded region representing the standard deviation. The median lines from the other two clusters are also shown for comparison in each cluster.
(\textbf{B}) Visual representation of the neighbourhood distribution across the ALA and temporal mixing clusters.
(\textbf{C}) Segregation trends for the newly defined clusters —inclusive, mixed, and segregated — are termed 'temporal mixing clusters,' across weekdays and weekends.
Again, the solid line represents the median, with the shaded region indicating the standard deviation. The median lines from the other two clusters are also shown for comparison in each cluster
(\textbf{D}) A two-tiered map representation: the upper layer highlights the ALA clusters' spatial distribution, and the segregation profiles are lower.
(\textbf{E}) Boxplot and jitter plot illustrate the distribution of the zscore for the ten ALA metrics across the three temporal mixing clusters.
}
\label{fig:segregation}
\end{figure}

We now characterise each neighbourhood by a unique time series of Gini coefficients spanning 48 hours: 24 on weekdays and 24 on weekends. 
Considering the topological classification of the city neighbourhoods, we aggregate the daily evolution of the Gini coefficient through the lens of ALA clusters, observing a consistent pattern (see Fig.~\ref{fig:segregation}.\textbf{A}).
Neighbourhoods with a low value in the ALA metrics show a stronger inclination towards segregation than those classified as medium and high.
This initial approach focuses solely on the city's topology via the ALA clusters, thereby overlooking the potential similarity in the daily evolution of the Gini coefficients. To address this, we set aside the pre-defined ALA cluster classifications. Instead, we sought to create new clusters based solely on the temporal trends of the Gini coefficient for each neighbourhood.
Through this different approach, using k-means, we discern three distinct patterns: inclusive (red), mixed (blue), and segregated (green) (as showcased in Fig.~\ref{fig:segregation}.\textbf{C}). The decision to use three clusters is elaborated upon in the Methods "Clusterization" section.
The inclusive neighbourhoods consistently register a lower Gini index throughout the day, while the segregated cluster deviates significantly from the other two.
The spatial distribution of these clusters is not trivial. Their spatial dispersal breaks from the radial patterns of the ALA Clusters, implying a multifaceted urban interplay beneath the surface (as showcased in Fig.~\ref{fig:segregation}.\textbf{D}).
Remarkably, not all neighbourhoods characterised by ALA clusters are merged into the same temporal mixing cluster.
For a deeper understanding, we examined the overlap between the two lenses, showcased in Fig.~\ref{fig:segregation}.\textbf{B}. A notable majority of the high ALA clusters are temporal-inclusive, whereas only a fraction of the low ALA clusters exhibit this trait. This evidence raises the following question. Might there be specific features within the metrics that correlate with inclusivity? 

To provide a comprehensive understanding of the urban landscape and its influence on the temporal dynamics of social mixing, we studied the distribution of the ALA metrics in relation to the temporal clusterisation (Fig.~\ref{fig:segregation}.\textbf{E}). 
Each temporal mixing cluster manifests a unique fingerprint when mapped onto the urban topology.
Interestingly, there is a wide distribution difference in 'price diversity', 'velocity score', and 'fitness' by temporal clusters.
These metrics play a pivotal role in shaping its temporal inclusivity and are strictly related to the fabric of a neighbourhood — such as type of amenities, accessibility, and availability of services. 

We utilised a regression model to identify the key factors linked to social mixing and inclusivity to evaluate how different aspects of 'accessibility', 'liveability', 'attractivity', and population density influence urban segregation. 
Observing spatial autocorrelation in our data, we incorporated a Spatial Lag model. 
For a detailed exposition of the regression analyses, including the application of Ordinary Least Squares (OLS) (Supplementary Table \ref{tab:regression_osl}) and error models (Supplementary Table \ref{tab:regression_error}), readers are referred to the supplementary materials. These extended analyses offer comprehensive insights into the statistical relationships governing urban segregation in Milan.
Adopting a Lindeman, Merenda, and Gold (LMG) approach, we discerned that 'accessibility' and 'attractivity' heavily sway the segregation trends, accounting for approximately 55\% of its variance, while 'liveability' and population density play a lesser role (Table \ref{tab:regression}).
A deeper probe using LASSO revealed pivotal features: Velocity Score, Fitness, Median Price, and Price Diversity. Subsequent application of an Ordinary Least Squares (OLS) model confirmed the statistically significant negative impact of these features on segregation (details in Table \ref{tab:regression}).
Conversely, architectural landmarks, also of statistical significance, contribute to increased segregation. 
This result highlights the city's inequality, revealing that not all citizens can enjoy its architectural marvels.
These results underscore the importance of considering multiple factors when attempting to understand and address urban segregation. Providing accessible, diverse, and attractive neighbourhoods can foster social interactions and reduce segregation within cities.

\begin{table}[h]
\caption{Regression results of the Spatial Lag model for predicting segregation using combinations of ALA metrics.}
\begin{tabular}{llllllll}
\hline
Group & Variable & Only & Only & Only & Only & All & LASSO \\
 & & Accessibility & Liveability & Attractivity & Density & Together & \\
\hline 
Accessibility & Velocity Score & -0.76*** & & & & -0.42*** & -0.42***\\
\hline
 & Schools & & -0.08 &  & & -0.02\\
Liveability & Supermarkets  & & 0.01 & & & 0.05 & \\
 & Supermarket diversity & & -0.32** & & & -0.02 & \\
& Architectural Heritage  & & 0.31** & & & 0.19*** & 0.21***\\
\hline
 & Fitness & & & -0.38** & & -0.22** & -0.30***\\
 & Category's diversity & & & -0.35* & & -0.15 & \\
Attractivity & Price's diversity & & &-0.20* &  &  -0.14*&-0.18*\\
 & Price  & & & -0.50*** & & -0.21*** & -0.23***\\
 & Rating & & & -0.23 & & -0.07 & \\
 \hline
Population & Density & & & & -0.03* & -0.03 & \\
 \hline
Autocorrelation & Segregation & 0.24*** & 0.33*** & 0.27*** & 0.26*** & 0.24*** & \\
\hline
\multicolumn{1}{r}{$R^2$} & & 0.56 & 0.31 & 0.57 & 0.34 & 0.65 & 0.62  \\
\hline
\end{tabular}
\label{tab:regression}
\end{table}

\section*{Discussion}
In recent years, urban researchers and policymakers have turned their attention to the pressing issue of income segregation in cities~\cite{florida}, recognising its implications for access to essential services and opportunities. Income segregation is not just about the surroundings of citizens' residences but critically about \textit{how much} individuals from varied economic backgrounds interact.
Our method overcomes the conventional two-dimensional representation of segregation via the Gini coefficient. By introducing the income triad, we transition into a three-dimensional space that offers a more detailed perspective on social mixing in Milan. This innovative approach does not just indicate the extent of segregation but also its composition, revealing the proportion of each income group present during specific time frames, from nightly spans to working hours and leisure periods.

Our analysis in Milan consistently highlights the deep ties between the city's structure and segregation patterns. The regression results further emphasise the decisive roles of public transportation, the architectural heritage, or the diversity of local amenities in shaping these interactions.
This last aspect of our findings resonates with the analysis presented in~\cite{Moro2023}, where the number and type of POIs were found to impact neighbourhoods' social mixing.

While our study encompasses a comprehensive set of metrics to analyse urban segregation, we acknowledge that certain factors, such as environmental pollution, were not included. Although not considered in our current analysis, the impact of pollution on urban livability and segregation presents an intriguing avenue for future research. Integrating environmental data could significantly enrich our understanding of how urban dynamics are influenced by ecological factors, adding depth to the study of urban segregation.

Regarding the geographical scope of our study, we recognise the limitations inherent in focusing solely on Milan due to data availability. This boundary issue is a common challenge in urban research, and our study's confinement to Milan highlights the need for broader data access and analysis in future studies. Expanding research to encompass areas beyond city limits could provide a more complete picture of urban dynamics and segregation patterns, addressing a critical gap in the field of urban studies.
The methodology developed, however, holds potential for broader application and can be adapted to other urban settings. Although the core methodology is expected to remain consistent, different urban environments may reveal unique segregation patterns based on local conditions. 
Beyond its current application, our methodology is versatile. It can be recalibrated to investigate other forms of segregation, such as those based on ethnicity or other sociodemographic parameters, offering a more holistic view of urban inclusivity.
Although our research paints a detailed portrait of Milan's urban dynamics over ten months, further research could extend this framework to longer timeframes, potentially revealing gentrification patterns.

Our research aligns with the growing academic interest in urban segregation. It not only offers an in-depth analysis of Milan's urban dynamics but also presents a valuable framework for both researchers and policymakers. Our tools are poised to offer profound urban insights, guiding both academic discussions to further analyse the intricate urban interactions and policymaking towards more inclusive urban futures.

\section*{Methods}\label{sec11}
\subsection*{Mobility data}
The LBS (Location-Based Services) data utilised in this study have been shared by Sony Computer Science Laboratories - Rome, under a non-disclosure agreement (NDA). This type of data has been widely used and validated in numerous studies~\cite{ric1,Moro2021,Moro2023}. The data were collected from 650,000 anonymous mobile phone users over a span of 10 months from March 2017 to December 2017, who have opted-in to provide access to their location data anonymously through a GDPR-compliant framework. In adherence to privacy compliance, all the stops belonging to the Education, Health, and Religion categories were removed from the dataset. The data were processed and analysed in a secure and privacy-compliant environment to ensure adherence to the terms of the NDA and relevant data protection laws. 

Further information about the stop extraction, the home and work identification, and the data sanitisation can be found in the Supplementary Note~\ref{sec:lbs}.

\subsection*{Clusterization}
Clusterization in our study was consistently executed using the elbow method. This technique indicated that k=3 is an optimal choice for balancing within-cluster variance against the number of clusters in all the clusterization.

\subsubsection*{Income Group}

For clustering income groups, we initially considered k=3 based on the elbow method (refer to Supplementary Figure S\ref{fig:elbow}). A silhouette score analysis was also conducted to reinforce this choice (refer to Supplementary Figure S\ref{fig:siluette}). Exploratory analysis with k values of 4 and 5 provided coherent results showing that the clusters meaning is invariant by cluster size (see Supplementary Figures S\ref{fig:quadrato}, S\ref{fig:pentagono}), but k=3 emerged as the most effective in capturing distinct patterns of economic segregation, maintaining simplicity and clarity in the model.

\subsubsection*{ALA Clustering}
For the ALA clustering we adopted k=3 based on insights from the elbow method (refer to the Supplementary Figure S\ref{fig:elbow_city}). When we experimented with k values of 4 and 5 (see Supplementary Figures \ref{fig:all_4}.\textbf{C} and \ref{fig:all_5}.\textbf{C}, the results remained coherent but resulted in a loss of cluster uniqueness, reaffirming our initial choice of k=3.

We employed different clustering methods to ensure the robustness of our findings. Specifically, we used k-means and agglomerative clustering
techniques for clusterization. After clusterization with each method, we assessed the similarity of the resulting clusters using the Fowlkes-Mallows score.
This score evaluates the similarity between two clusterings, providing a measure of how closely the clusters identified by different methods align with each other. Our analysis revealed Fowlkes-Mallows scores of 0.73 with agglomerative clustering. These high scores indicate a significant degree of similarity among the clusters generated by the different methods, validating the consistency of our clustering approach.

\subsubsection*{Temporal Mixing Clustering}
For temporal mixing clustering, k=3 was selected as the optimal number of clusters, as suggested by the elbow method (refer to the Supplementary Figure). In this analysis, Dynamic Time Warping (DTW) was employed as the distance metric for comparing time series. Evaluations with k values of 4 and 5 (see Supplementary Figures S\ref{fig:all_4}.\textbf{E} and S\ref{fig:all_5}.\textbf{E} provided oSherent results; however, these configurations led to clusters that were very similar to each other (see Supplementary Figure S\ref{fig:similitudine}. This observation further underscores the suitability of k=3 for capturing distinct and meaningful temporal patterns without overfitting the model.

We employed multiple clustering methods to ensure the robustness of our findings. Specifically, we used k-means, agglomerative and spectral clustering techniques for clusterization. After clusterization with each method, we assessed the similarity of the resulting clusters using the Fowlkes-Mallows score.
This score evaluates the similarity between two clusterings, providing a measure of how closely the clusters identified by different methods align with each other. Our analysis revealed Fowlkes-Mallows scores of 0.84 with agglomerative clustering and 0.81 with spectral clustering when compared to k-means clustering. These high scores indicate a significant degree of similarity among the clusters generated by the different methods, validating the consistency of our clustering approach.

\subsection*{Users vector}

Every user \( k \), with \( k = 1,...,K \), where \( K \) is the total number of users, can be mapped into the income triade. This space is defined over three dimensions, corresponding to the three income categories: High (H), Medium (M), and Low (L). The user vector, \( \vec{u_k} \), is described as:

\begin{equation}\label{utente}
  \vec{u_k} = \delta(u_k, \vec{u})  
\end{equation}

The function \( \delta(u_k, \vec{u}) \) maps the user \( k \) to a three-dimensional vector, where the component corresponding to the income category of the user $k$ is $1$, and all other components are $0$.

\subsection*{Neighbourhood Income Activity and Normalisation}

For the initial distribution of incomes in the general population, we normalise each component of \( \vec{A}_{h,t} \) using the total number of users in each income category. Denote these totals as \( N_H \), \( N_M \), and \( N_L \) for the High, Medium, and Low categories, respectively:

\begin{equation}
\label{norm1}
  \vec{\tilde{A}}_{h,t} = \left( \frac{\sum_{k \in U_{h,t}} \delta(u_k, \vec{u})_H}{N_H}, \frac{\sum_{k \in U_{h,t}} \delta(u_k, \vec{u})_M}{N_M}, \frac{\sum_{k \in U_{h,t}} \delta(u_k, \vec{u})_L}{N_L} \right)  
\end{equation}

Here, $U_{h,t}$ represents the set of users located in 'h' at a time 't', \( \delta(u_k, \vec{u})_i \) represents the \emph{i}-th component of the user vector, indicating whether user \( k \) belongs to income category \( i \).

We then apply the L1 normalisation to obtain a proportionate distribution:

\begin{equation}
\label{norm2}
  \vec{A}_{h,t} = \frac{\vec{\tilde{A}}_{h,t}}{\sum_i \vec{\tilde{A}}_{h,t,i}}  
\end{equation}
\phantomsection
\subsection*{POI Datasets}\label{POI}
We use the Google API to collect data on the Points of Interest (POIs).
We start by selecting a point k=(lat, long) on the maps. We fetch the first 50 POIs nearby, generating a correspondence hexagonal area of interaction from the point $A_k$.
We collect the attributes - 'latitude', 'longitude', 'category', 'price', and 'rating' within the $A_k$ for all the POIs, and we iterate the process to patch all the city areas. We collected more than 400,000 POIs, and we divided the POIs into 13 amenity categories (in Supplementary Note \ref{venue}, Fig. S\ref{fig:cat_hist}).
\subsection*{Metrics}
In this study, we introduce a novel framework for neighbourhoods analysis, in three key aspects: Accessibility, Livability, and Attractivity (ALA). The synergistic application of these dimensions allows for a comprehensive characterization of neighbourhoods. This methodology advances beyond traditional approaches that typically focus on residential perceptions or census data, offering a dynamic and data-driven perspective. By integrating Accessibility, Livability, and Attractivity, our framework is designed to assess a neighbourhood's potential to foster social inclusivity and its ability to attract diverse groups.

\subsubsection*{Velocity Score}

The velocity score evaluates the efficacy of a city's transportation systems, strongly relying on the concept of isochronic maps. An isochronic map consists of isochrones centred at a given location \( \lambda \). Each isochrone \( I(\tau, (\lambda,t_0)) \) demarcates areas that can be reached from \( \lambda \) within a travel time \( \tau \), starting at an absolute time \( t_0 \). The collective isochrones for varying \( \tau \) values represent the isochronic map for \( \lambda \) at time \( t_0 \).
To understand the transportation spread from a starting point, we analyse the rate of isochronic expansion over time. Specifically, following~\cite{Biazzo2019}, we consider the area \( A(\tau, (\lambda, t_0)) \) covered by the isochrone for a given \( \tau \). This area includes hexagons whose centres are attainable within time \( \tau \), and its growth is characterised by discrete jumps corresponding to hexagon centre acquisitions.
Considering a circle with an area equivalent to the isochrone, its radius \( \bar{r} \) represents the mean distance travelled in a random direction from the starting point:
\[ \bar{r}(\tau,(\lambda,t_0))=\sqrt{\frac{A(\tau,(\lambda,t_0))}{\pi}} \]
Consequently, the circular velocity \( \bar{v} \) is:
\[ \bar{v}(\tau,(\lambda,t_0))=\frac{\bar{r}(\tau,(\lambda,t_0))}{\tau} \]
representing the expansion speed of a circular isochrone identical in area to the actual one.

The Velocity Score (\( v \)), as introduced in~\cite{Biazzo2019}, is the average over times \( \tau \) and \( t_0 \):
\begin{equation}\label{vel}
v(\lambda)=\frac{\sum_{t_0=6\text{am}}^{10\text{pm}}\int_{0}^{\infty}\bar{v}(\tau,(\lambda,t_0))f(\tau)d\tau}{\sum_{t_0=6\text{am}}^{10\text{pm}} f(\tau)d\tau}
\end{equation}
\noindent where \( t_0 \) spans from 6 a.m. to 10 p.m. in 2-hour intervals. The average over \( \tau \) uses the empirical distribution \( f(\tau) \) as a weight. It's worth noting, as pointed out in~\cite{Biazzo2019} that the choice of distribution for average travel times holds a degree of arbitrariness.

\subsubsection*{Fitness and complexity}
\begin{figure}[]
\centering
\includegraphics[scale=0.6]{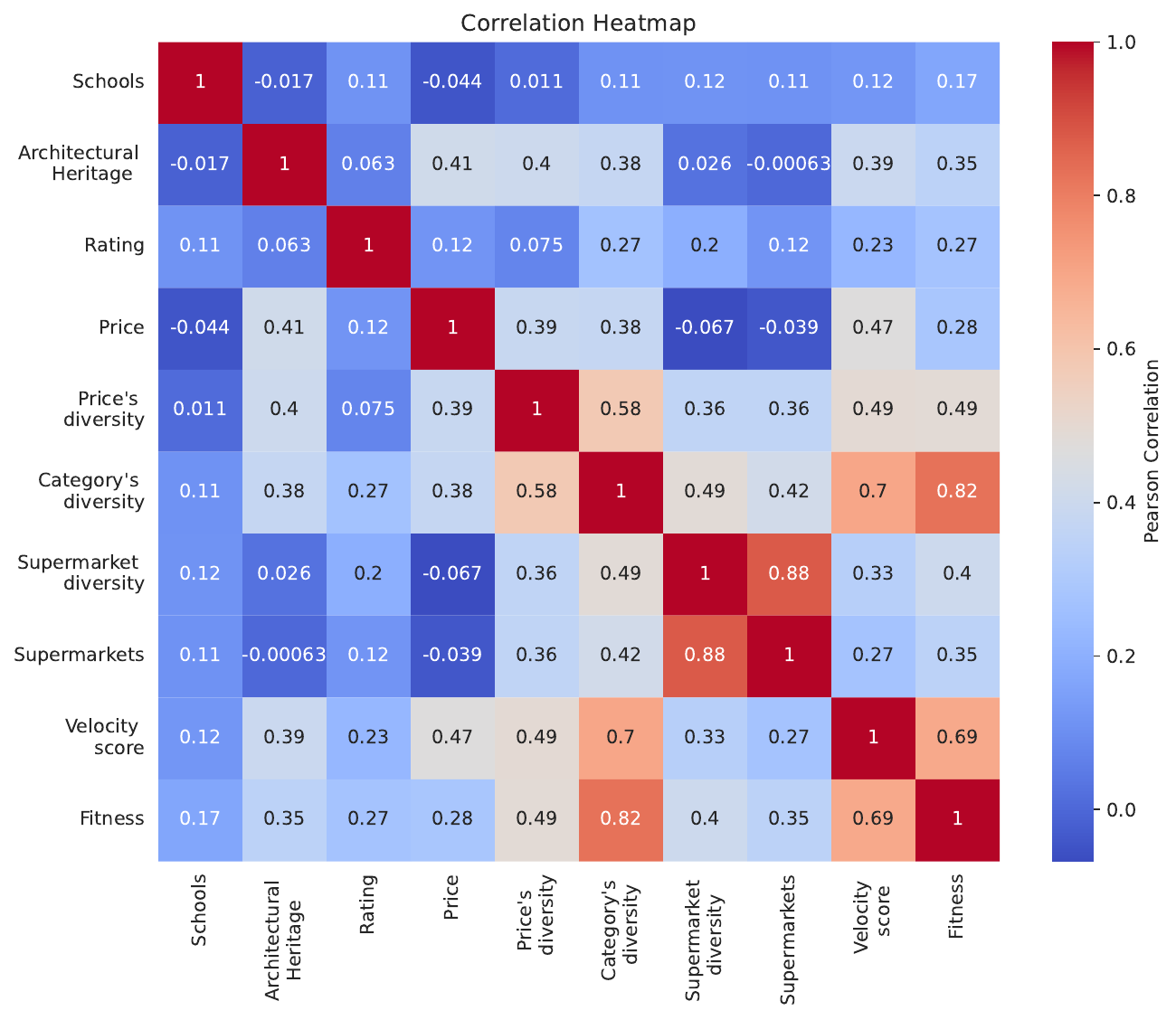}
\caption{Correlation between the ALA metrics.}
\label{fig:correlation}
\end{figure}

\begin{figure}[]
\centering

\includegraphics[scale=0.6]{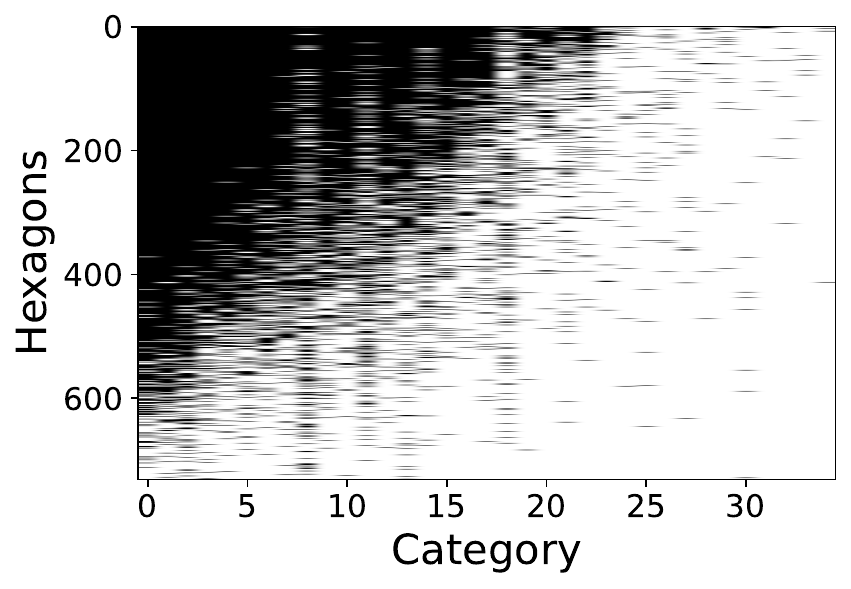}
\caption[The (bi-)adjacency matrix]{The (bi-)adjacency matrix of the bipartite hexagon-categories network where the rows (hexagons) and columns (categories) have been sorted according to the fitness-complexity algorithm }
\label{fig:fitness}
\end{figure}

In~\cite{amenity}, they use economic complexity~\cite{Hidalgo2009} to quantify the ability of locations – on the level of neighbourhoods and amenities – to attract diverse visitors from various socio-economic backgrounds across the city. 
We utilise the Fitness and Complexity framework developed by Tacchella et al.~\cite{Tacchella2012} to assess the attractivity of neighbourhoods. To do so, we implement a binary approach (Supplementary Figure S\ref{fig:fitness}) using the revealed comparative advantage (RCA) in the Fitness and Complexity algorithm. The RCA indicator enables the identification of sectors in which a country exhibits competitiveness.
In our scenario, we consider that individuals may choose to visit a particular location due to the availability of multiple options or a specific feature that is lacking in other areas, such as a stadium or a theme park. This analogy between neighbourhoods and POIs category and Fitness and Complexity of countries and products is evident. Many neighbourhoods have more common Points of Interest (POIs) that are less "complex", while unique and exceptional POIs are found in only a few areas. Therefore, a neighbourhood with high fitness, based on the binary approach with the Revealed Comparative Advantage (RCA) indicator (in the Supplementary \ref{rca}), will be more attractive to individuals seeking such unique or diverse features. 
We can use the iteration of the Supplementary \ref{rca} to calculate the fitness of a neighbourhood, as shown in the Supplementary~\ref{fitness}.
Fitness and Category Diversity have a high degree of correlation (Supplementary Figure S\ref{fig:correlation}), but they are retained in our analysis to encapsulate both the concepts of uniqueness and diversity, which are vital for a nuanced understanding of neighbourhood attractivity.

\subsubsection*{Gini Coefficent}
We utilised the following metric, which is a function of the Gini coefficient\footnote{\url{https://www.census.gov/topics/income-poverty/income-inequality/about/metrics/gini-index.html} }
, to summarise the percentage of visiting a neighbourhood from the three different income groups throughout the hours of the day.
0 implies perfect mixing, and 1 indicates complete segregation.

\begin{equation}
    G = \frac{\sum_{i=1}^{n-1} \sum_{j=i+1}^{n} |x_i - x_j|}{n},
\end{equation}
\noindent where the terms \( x_i \) and \( x_j \) denote the proportion of users in these groups for a specific neighbourhood and the denominator serves as a normalisation factor. In the specific case of $n=3$, i.e., the three income categories corresponding to the high, medium, and low-income groups, one has:
\begin{equation}\label{gini}
    G = \frac{|x_1 - x_2|+|x_1 - x_3|+|x_2 - x_3|}{3}
\end{equation}

\section*{Author contributions statement}
L.R.M., and R.D.C. conceived the research and designed the analyses,
V.L. contributed ideas for the research and participated in the development of the results.
L.R.M. gathered the rent dataset and the POIs dataset.
V.L. processed the LBS data.
L.R.M. performed the data sanitisation and the analyses.
L.R.M. created the maps. 
R.D.C. supervised the research.
All the authors discussed the results, wrote the paper, and approved the final manuscript.
\section*{Additional information}
\textbf{Competing interests} The authors declare no competing interests. 
\section*{Data availability}
The paper contains all the necessary information to assess its conclusions, including details found in both the core paper and the Supplementary Notes. The mobile phone data utilised in this study were acquired by Sony from Cuebiq. Due to contractual agreements and privacy considerations, we are unable to directly share the raw mobile phone data. However, interested researchers can negotiate access to this specific dataset through Sony, subject to the agreement and signature of a Non-Disclosure Agreement (NDA). The data supporting the findings of this study are also available from Cuebiq through their Data for Good program. Detailed information on how to request access to the data, along with its conditions and limitations, can be found at \href{https://www.cuebiq.com/about/data-for-good/}{Cuebiq - Data for Good program}.
Venues' location and category were obtained via Google using their Search API.
The rent dataset was obtained using Python libraries from \href{www.caasa.it}{www.caasa.it}.

\section*{Acknowledgments}
R.D.C acknowledges Sony Computer Science Laboratories - Paris for hosting him during part of the research. The authors thank Bernardo Monechi for the early conversations on the research.

\pagebreak[2]
\clearpage
\newpage
\startsupplement
\renewcommand{\theequation}{Eq. S\arabic{equation}}
\let\oldsection\section
\renewcommand{\section}[1]{\oldsection{Supplementary Note~\thesection \ -- \ #1}} 
\setcounter{page}{1}
\tableofcontents
\listoffigures
\listofmyequations
\newpage

\section{Location Data}
From the GPS trajectory in the LBS data, we extract 24 million stops (\textit{pings}), each with latitude and longitude with an accuracy of 20m and the associated timestamp.
\label{sec:lbs}
\subsection{Extracting stays}\label{stays}
The open-source Python software toolkit \cite{mobilkit} extracts location in three steps:
\begin{itemize}
    \item Stop detection: all the locations where users spent at least 20 minutes within a distance of 200 meters from a given point.
    \item The stops can be aggregated in location, which means that all the stop clusters refer to the same point. Adopting the method
defined in \cite{Hariharan2004}, the DBSCAN algorithm is used to group together points
given their local spatial density, with two parameters: the maximum linked distance
$\epsilon$, and s, the minimum number of
neighbours within distance $\epsilon$ for a point to be regarded as a core point.
\item At the end of the analysis, all
stops belonging to the same location: each stop is assigned to
the medoid of the sequence. 
\end{itemize}

\subsection{Identifying important places}\label{homework}
Every site has a significance, namely home, and work. 
\subsubsection[]{Identifying home and work}
\begin{algorithm2e}[H]
\SetAlgoNlRelativeSize{0}
\SetAlgoNlRelativeSize{-1}
\SetAlgoNlRelativeSize{-2}
\SetAlgoNlRelativeSize{-3}
\SetAlgoNlRelativeSize{-4}
\SetNlSty{textbf}{}{:}
\SetAlgoNlRelativeSize{-1} 
\SetAlgoNlRelativeSize{0}
\caption{3.2.1: home and work detection (user features)}
\label{alg:algorithm}

\KwIn{Stops = (UID, lat, long, time)}
\KwOut{Assigned home and work for each user}

\SetKwFunction{FMain}{home detection}
  \SetKwProg{Fn}{Function}{:}{}
  \Fn{\FMain}{
    \For{users in Users}{
        stops between 9 p.m. and 6 a.m.\;
        stops close together\;
        cluster centroids and assign home\;
        fraction of home visits $n_{home}$\;
        \If{$0.2<n_{home}<0.8$}{
            take user\;
        }
        \Else{
            drop user\;
        }
    }
  }

\SetKwFunction{FMain}{work detection}
  \SetKwProg{Fn}{Function}{:}{}
  \Fn{\FMain}{
    \For{users in Users}{
        stops between 8 a.m. and 5 p.m., no weekends\;
        most visited stops are work\;
    }
  }

\end{algorithm2e}

To derive meaningful insights from each location, it's imperative to associate them with significant places relevant to users, especially their homes and workplaces. We utilise the algorithm presented in~\ref{alg:algorithm} for this purpose. 

For our geographical scope, we adopt the Organisation for Economic Co-operation and Development (OECD) definition of urban territories~\cite{oecd}. The OECD introduces two distinct boundaries: a core region and an expansive metropolitan vicinity named the Functional Urban Area (FUA). This FUA signifies the commuting zone surrounding a city. We take all the stops in the OECD.

\subsection{Data Sanitisation}\label{clean}
A rigorous data sanitisation process was employed to guarantee statistical reliability. To isolate and exclude tourists and sporadically connected users~\cite{Phithakkitnukoon2012,Alexander2015}, we selected only those who had at least 15 unique active days with at least one recorded position. To further refine our dataset and ensure a confident understanding of users' home locations, we focused on users who return home at least one day to four. This filtering resulted in a reduction of 60\% of our initial user dataset~\cite{Xu2019,Moro2021} .

\begin{figure}[]
\centering
\includegraphics[scale=0.15]{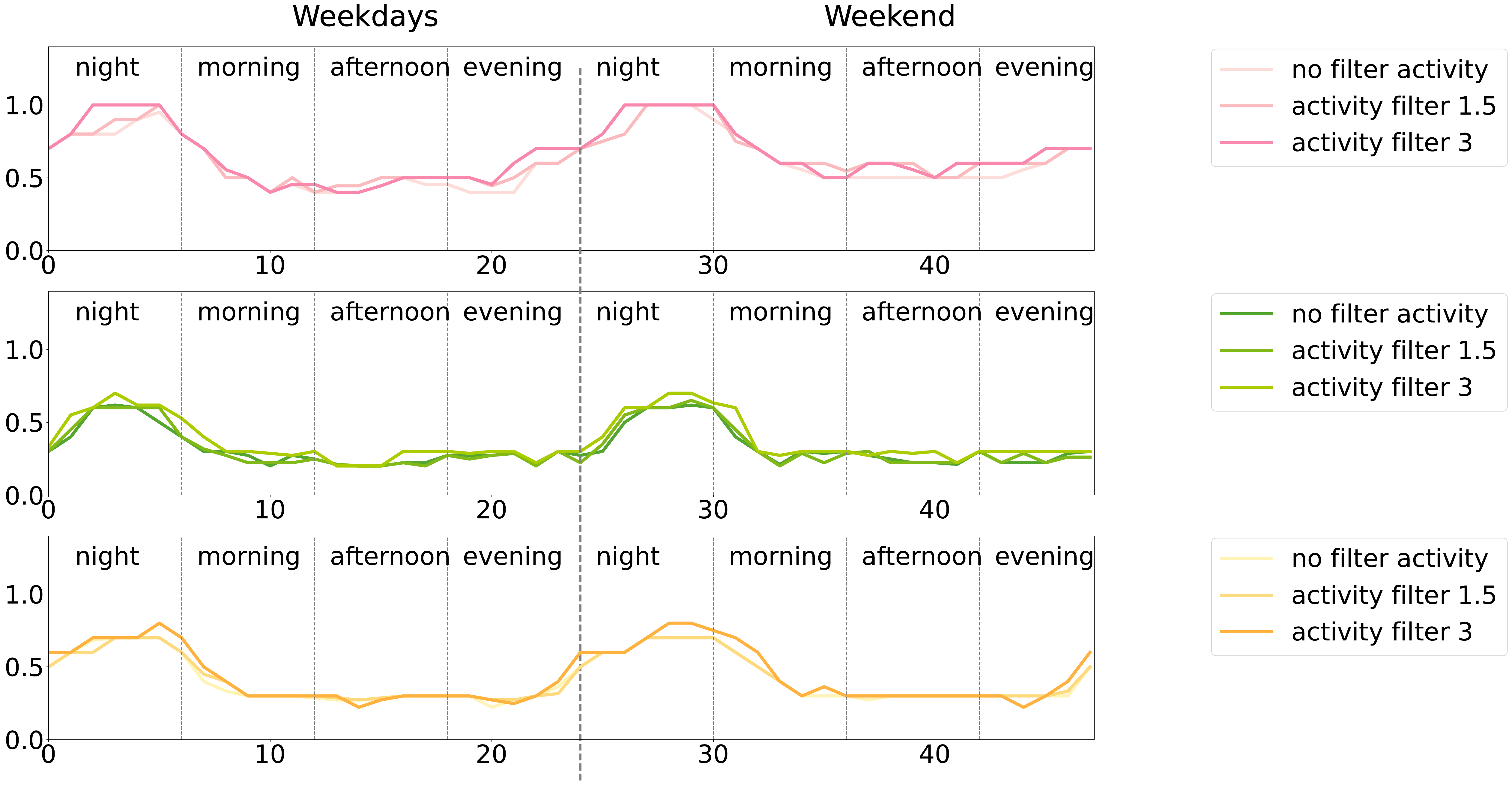}
\caption[Segregation profile with different filter]{Segregation profile depicted for each ALA cluster across a 48-hour period (24 hours on a weekday and 24 hours on a weekend), with distinct filters applied to the dataset to reflect varying levels of individual activity through the colour gradients.  The variation in segregation measure is computed using the Gini index.}

\label{fig:segregation_filter}
\end{figure}

Subsequently, we applied additional user filters to verify consistency across the various criteria ~\cite{Yabe2023}. These comparative analyses revealed no significant discrepancies in the results (Figure S\ref{fig:segregation_filter}).

This sanitisation leads us to consider 165,545 individuals in the OCSE region. By assigning work, from 8 a.m. to 5 p.m. without weekends, a significant portion of restaurant workers are excluded. To address this, the work classification was assigned to all stops made very frequently and for more than two and a half hours at a restaurant.

\section{Population Representativeness}\label{subsubsec:pearson1}
\begin{figure}[]
\centering

\includegraphics[scale=0.6]{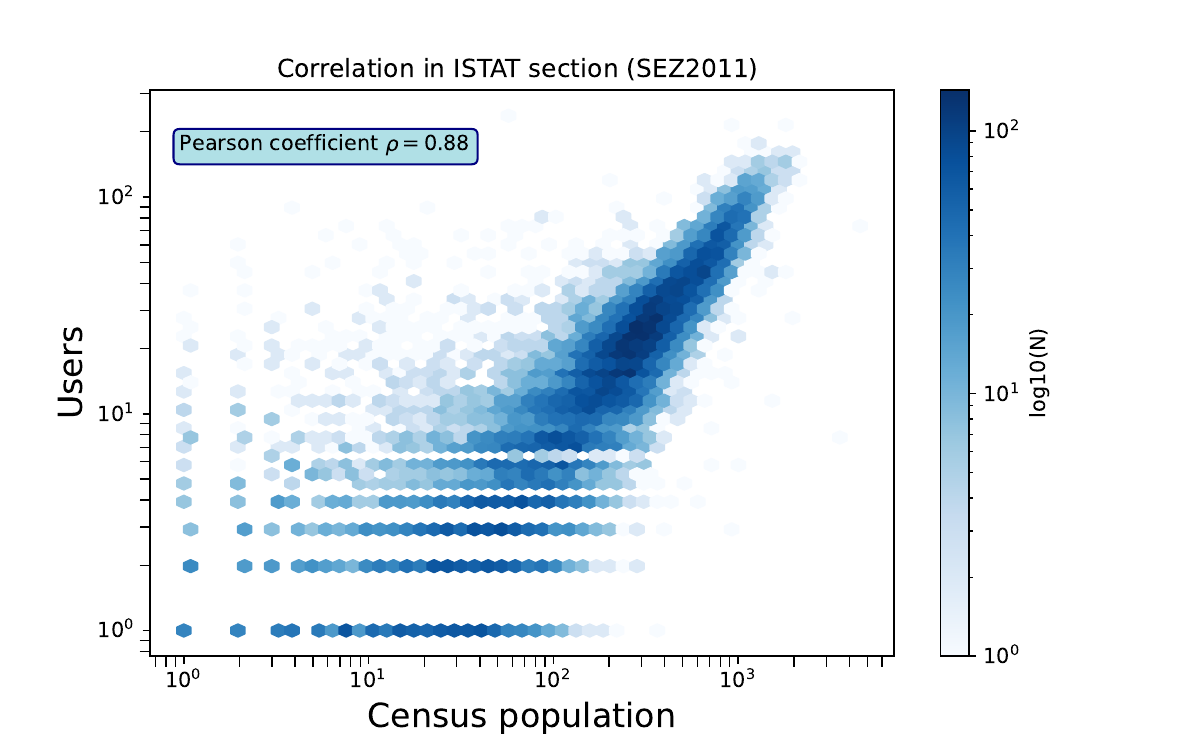}
\caption[Representativeness of data in each ISTAT section.]{Correlation between the users and the census population in each ISTAT section.}
\label{fig:correlation_istat}
\end{figure}

To assess the goodness of our dataset to capture the real population distributions across the city representatively, we measured the correlation between the number of users and the number of citizens in the ISTAT sections (Fig. S\ref{fig:correlation_istat}).

\section{Income and Rent Dataset}\label{sec:pearson2}
\begin{figure}[]
\centering
\includegraphics[scale=0.2]{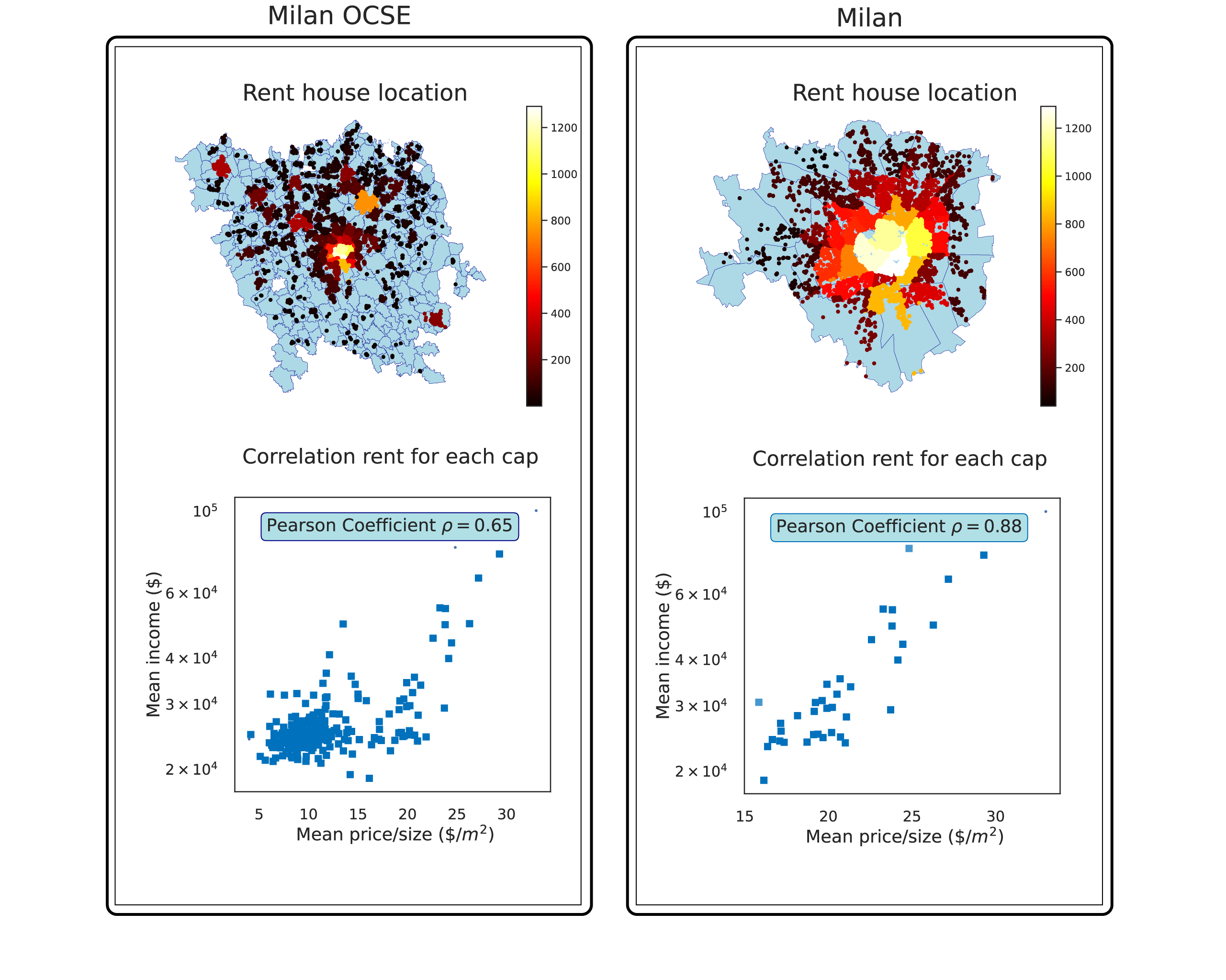}
\caption[Rent data as a proxy of income]{Rent data as a proxy of income. Maps of rented houses (upper).
Correlation in each zip code of Milan OCSE and in Milan city (down). Each point in the scatter plot represents a zip code. The x-axis represents the real mean income from census, and the y-axis represents the mean of the price on square meters. }
\label{fig:ocse_income}
\end{figure}

Income data in Italy is typically available at the broader ZIP code level, which necessitates a more granular approach for detailed analysis. In Milan, the area covered by ZIP codes varies significantly, ranging from about 1.5 km\(^2\) for central districts to as large as 18 km\(^2\) for outer areas. To achieve a finer detail in approximating individual economic statuses, we employed house rent prices per square meter as a proxy~\cite{Xu2019}. This method allows for a more precise estimation of economic conditions at a smaller spatial scale, aligning our analysis with the urban dynamics of Milan.
Data from the website Caasa~\cite{caasa}, derived from various Italian real estate platforms, provided one year of rent prices per square meter within Milan's OECD-defined region \cite{oecd}. Fig. S\ref{fig:ocse_income} indicates areas outside Milan exhibiting sparse data and lower Pearson correlation compared to inner-city regions. Consequently, we concentrated our analysis within Milan, where rent serves as a reliable income indicator.

\subsection{Income Representativeness}\label{subsubsec:pearson3}
To evaluate the representativeness of our dataset in accurately reflecting the income distributions across different ZIP codes in the city, we conducted a correlation analysis. Specifically, we compared the median income values assigned to each user in our dataset with the median actual income reported for each ZIP code. This comparison aims to validate the accuracy of our income assignment methodology and is visually depicted in Fig. S\ref{fig:income_rap}.

\begin{figure}[]
\centering
\includegraphics[scale=0.25]{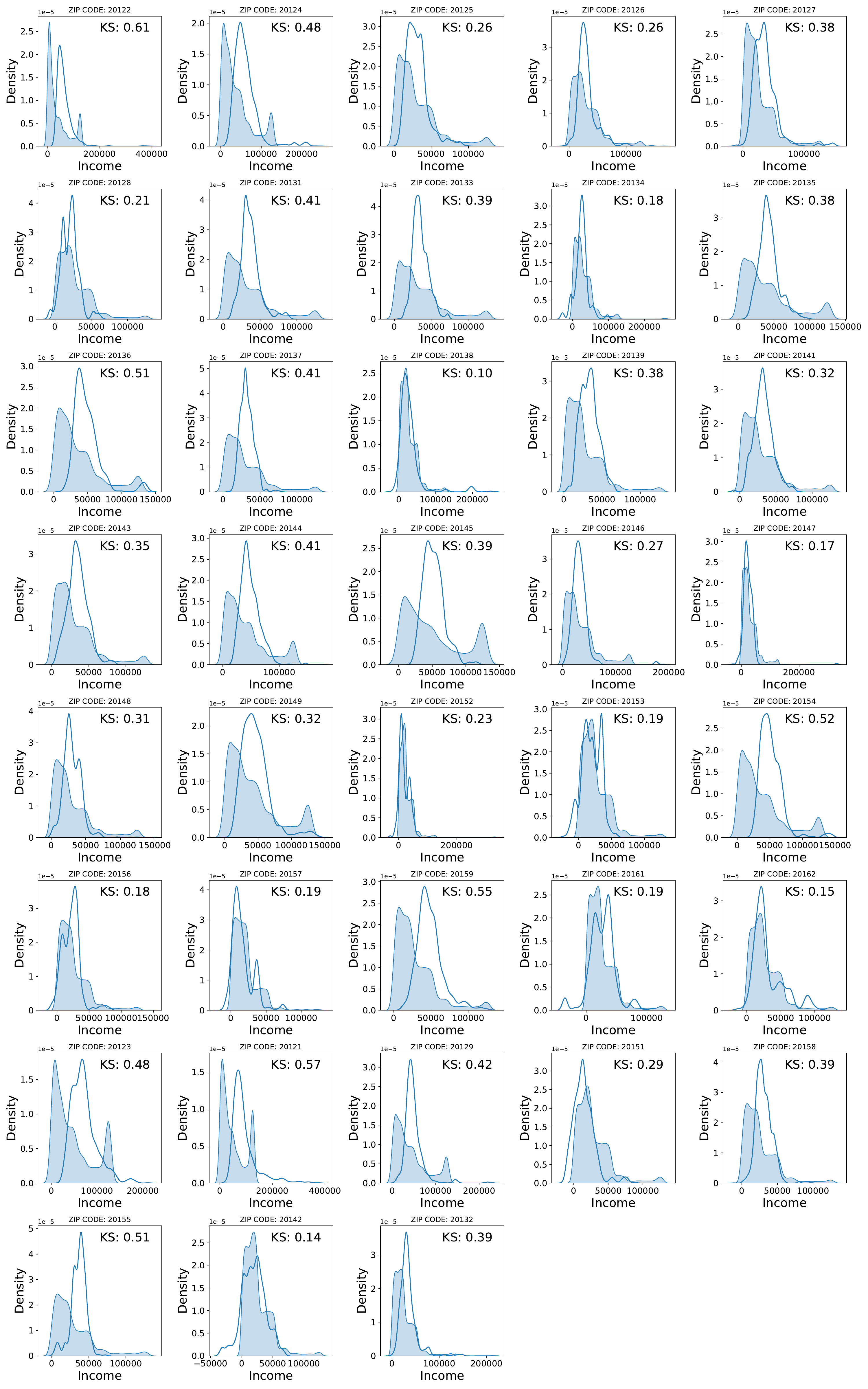}
\caption[Income Distribution]{Comparison of the actual income distribution (solid colour) and the distribution of users based on rent (dashed line) in each ZIP code. The Kolmogorov-Smirnov statistic is also provided for each ZIP code, quantifying the similarity between the two distributions.}
\label{fig:distribution_income}
\end{figure}

Furthermore, to gauge the similarity between the two income distributions, we computed the Kolmogorov-Smirnov statistic for each ZIP code. This approach provides a quantitative measure of the extent to which the income distribution in our dataset aligns with the actual income distribution in each area. The results of this comparison, illustrating the degree of similarity between the distributions, are shown in Fig. S\ref{fig:distribution_income}.

\subsection{Rent dataset}
\begin{figure}[]
\centering
\includegraphics[scale=0.3]{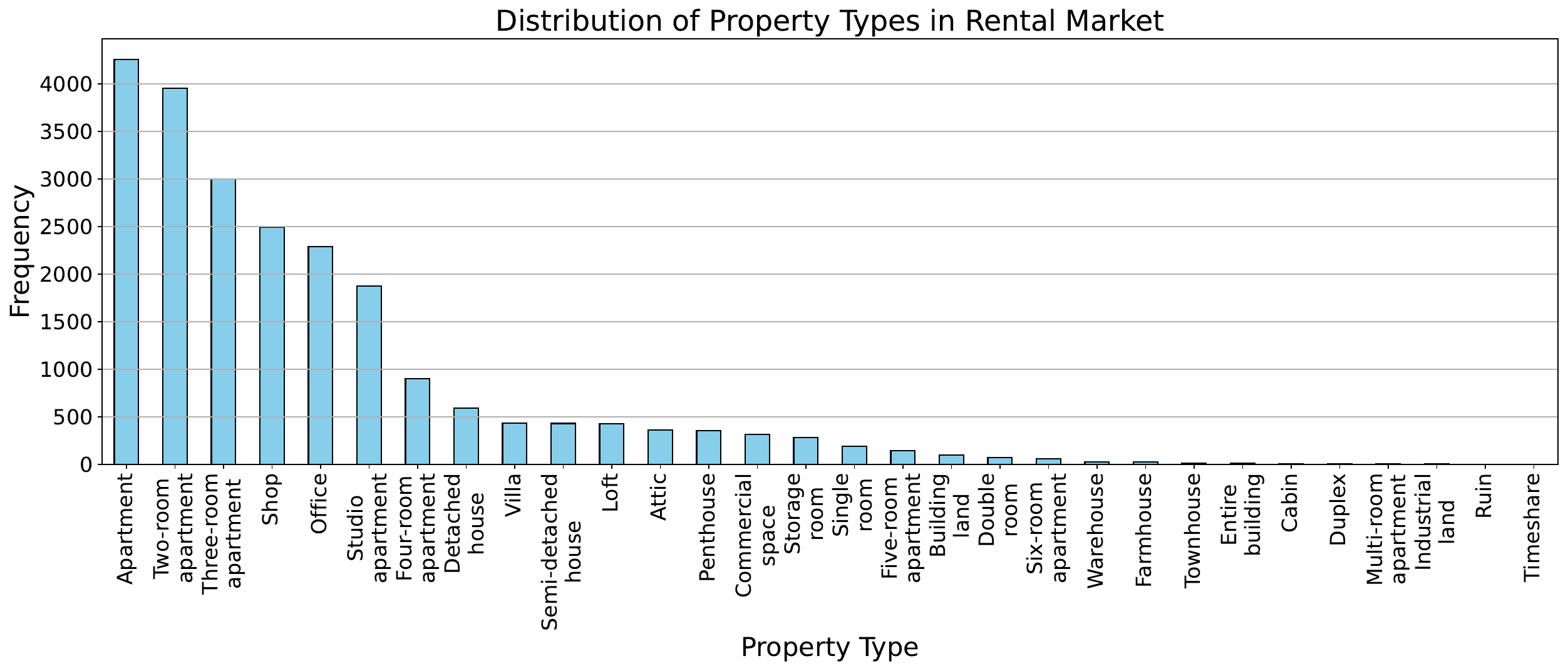}
\caption[Categories of Rental Properties]{Distribution of different categories of rental properties available in Milan. The chart illustrates the variety of property types.}
\label{fig:cat_immobili}
\end{figure}

\begin{figure}[]
\centering
\includegraphics[scale=0.5]{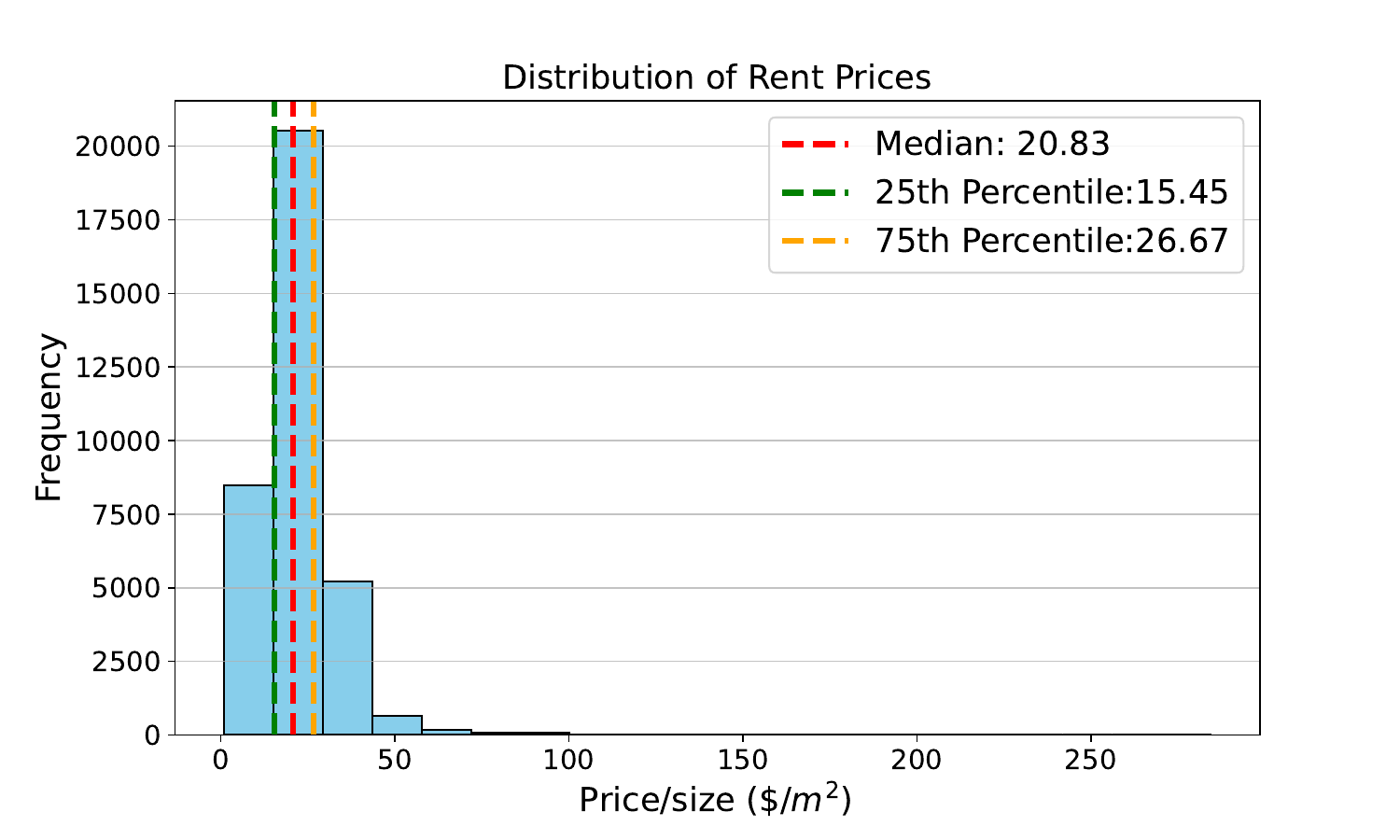}
\caption[Rent Price Distribution]{Distribution of rental prices per square meter for habitable residences in Milan. This figure presents a comprehensive view of the price range and variability, highlighting the economic diversity within the city's real estate market.}
\label{fig:price}
\end{figure}

From Caasa, we obtained data on various types of properties. Fig. S\ref{fig:cat_immobili} shows the different categories of rental properties, and we focused on habitable residences with the following distribution of prices, as depicted in Fig. S\ref{fig:price}.

\begin{figure}[]
\centering
\includegraphics[scale=0.2]{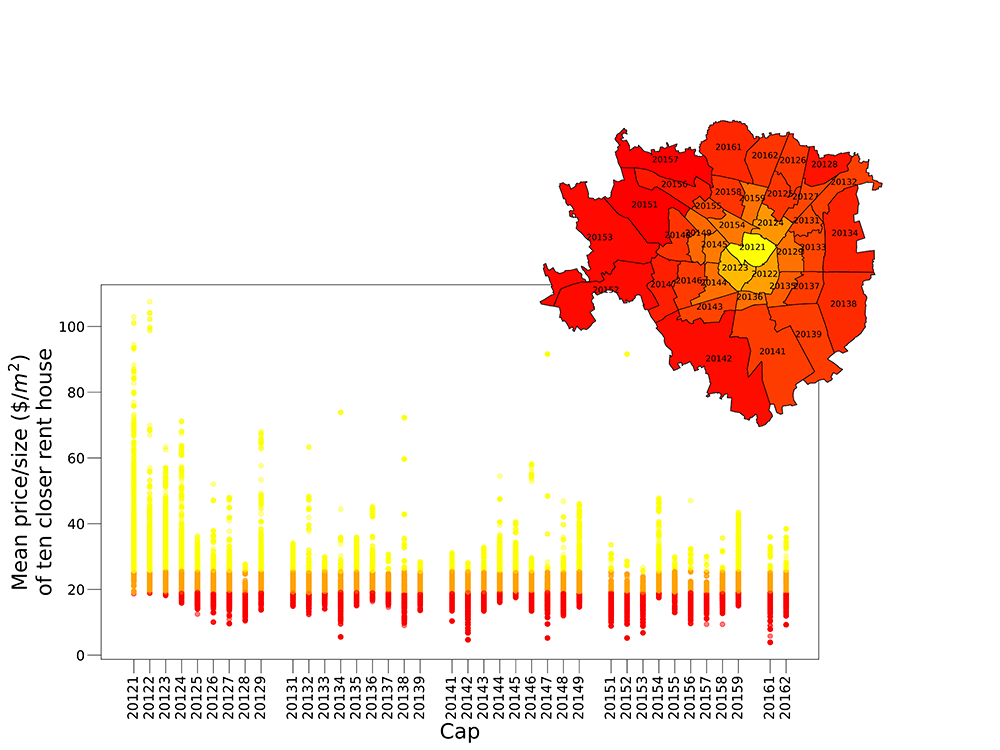}
\caption[Kmeans income groups]{Income categorisation using K-means clustering based on the average price per square meter of the nearest 10 houses within a 200-meter radius from each user's residence, plotted against ZIP codes. The plot exhibits a segregation into high (yellow), medium (orange), and low (red) income groups. Additionally, a spatial map computed using GeoPandas demonstrates the geographical distribution of ZIP codes predominantly occupied by a particular income group, thereby illustrating the spatial arrangement of economic strata within the urban fabric.}

\label{fig:kmeans}
\end{figure}

\subsection{Assigning Economic Status}
To emphasise the disparities between high and low-income groups, we employed k-means clustering. We specifically chose k-means clustering over quantile-based division of income groups to align with the actual distribution of income ~\ref{fig:distribution_income}.
Neither dividing the income into equal quantiles nor splitting the users into equal parts would accurately represent the diversity in income distribution. Dividing income into equal quantiles resulted in disproportionate representation, with values showing an uneven distribution of users across low, medium, and high-income groups (e.g., low: 92991, medium: 1451, high: 66). Kmeans instead resulting in three distinct income clusters depicted in Fig. S\ref{fig:kmeans}.

We leverage the three income clusters to create a 3D income vector space, the \textit{income triade} $\mathbb{I}$. We represent each income group with a basis vector denoted as:
\begin{equation}\label{basis}
    \mathbf{e}_H = \begin{bmatrix} 1 \\ 0 \\ 0 \end{bmatrix}, \quad
\mathbf{e}_M = \begin{bmatrix} 0 \\ 1 \\ 0 \end{bmatrix}, \quad
\mathbf{e}_L = \begin{bmatrix} 0 \\ 0 \\ 1 \end{bmatrix}
\end{equation}
    
    \myequations{Income triade basis}
where \( \mathbf{e}_H \), \( \mathbf{e}_M \), and \( \mathbf{e}_L \) represent the High, Medium, and Low-income groups, respectively.

\begin{figure}[]
\centering
\includegraphics[scale=0.2]{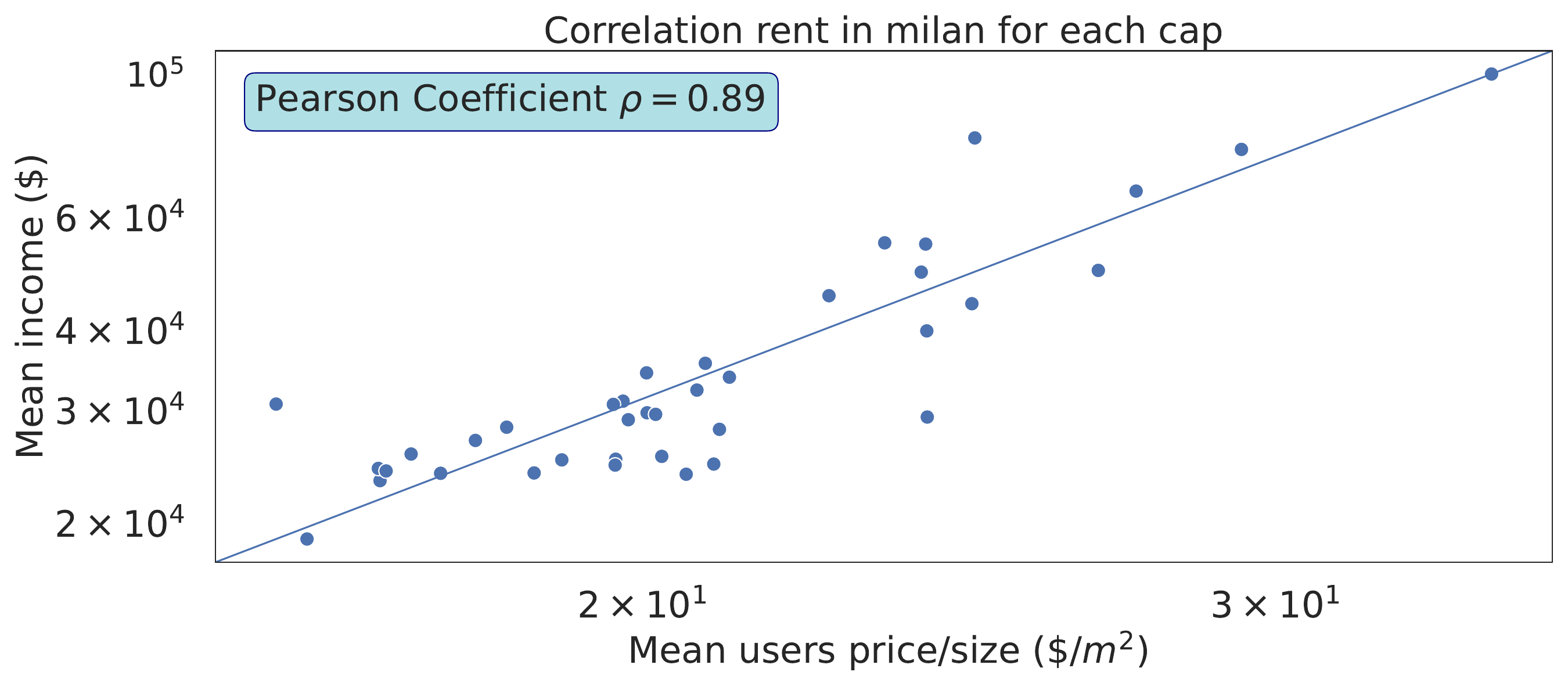}
\caption[Rappresentativness of income]{Representativeness of Income. Correlation in each ZIP code of Milan city. Each point in the scatter plot represents a ZIP code. The x-axis denotes the real median income from census data, while the y-axis represents the median value assigned to users (based on the mean rent prices of the ten closest properties to each user's home location).}
\label{fig:income_rap}
\end{figure}

\section{Hexagonal Grid}\label{sec:hex}

\begin{figure}[]
\centering
\includegraphics[scale=0.2]{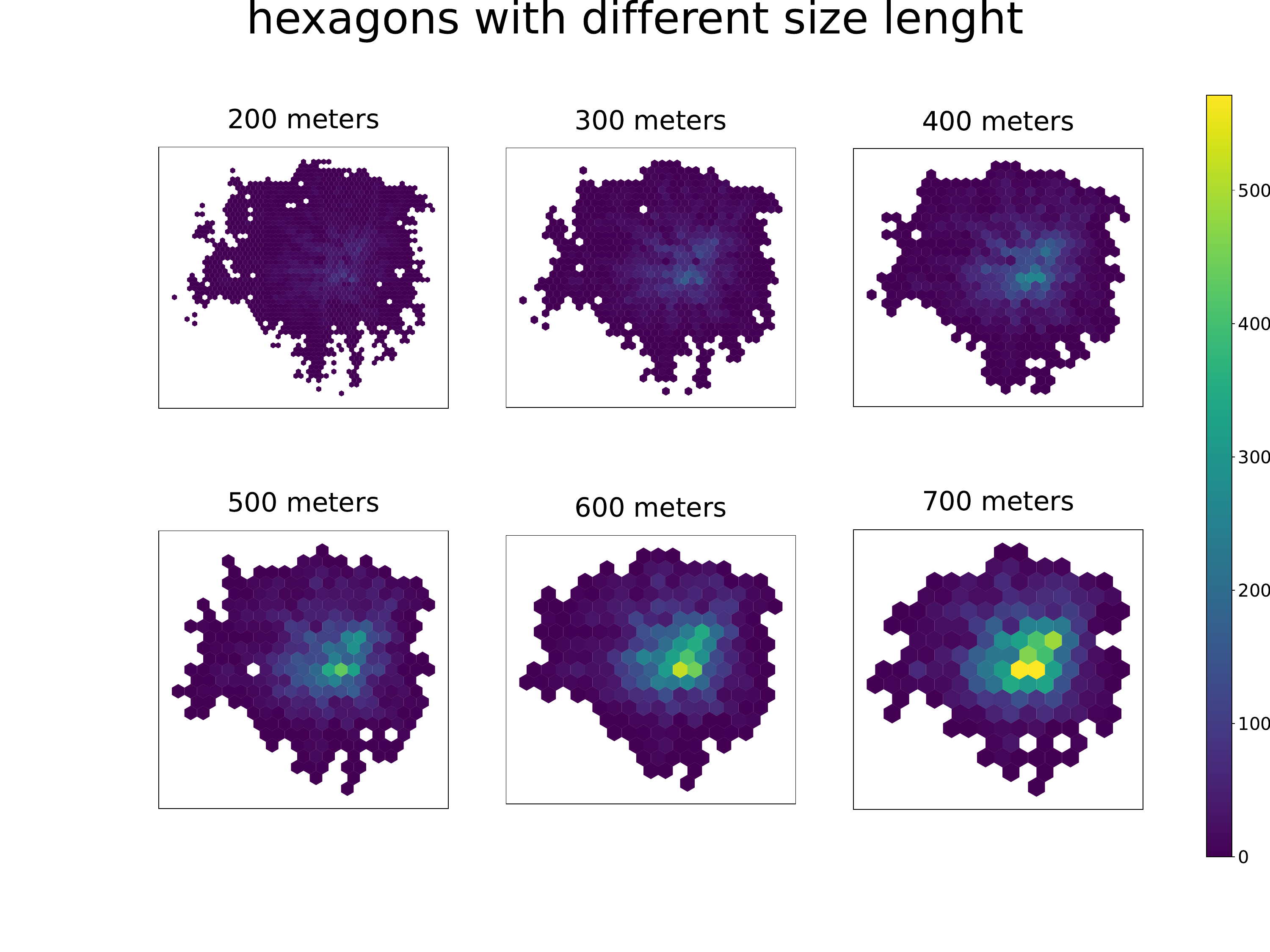}
\caption[POIs distribution maps]{Distribution Maps of Points of Interest (POIs) for Grids of Different Sizes. Each subplot represents a hexagonal grid of a particular side length overlaid on a map, with cells coloured based on the count of POIs contained within. This visual representation illustrates the variance in POI density and distribution across different grid scales.
}
\label{fig:maps_hex}
\end{figure}

To achieve a macroscopic representation of the city, we employed a hexagonal tessellation strategy, as described in \cite{hex}. The choice of grid size influenced the number and diversity of Points of Interest (POIs) encompassed within it. Larger grids,  with sides measuring 600 and 700 meters, yielded a less representative portrayal of the city, as depicted in (Fig. S\ref{fig:maps_hex}). The results show that analysing cities at broader scales (e.g., districts or ZIP codes) can miss the change in human interactions and their diverse urban settings.

\begin{figure}[]
\centering
\includegraphics[scale=0.21]{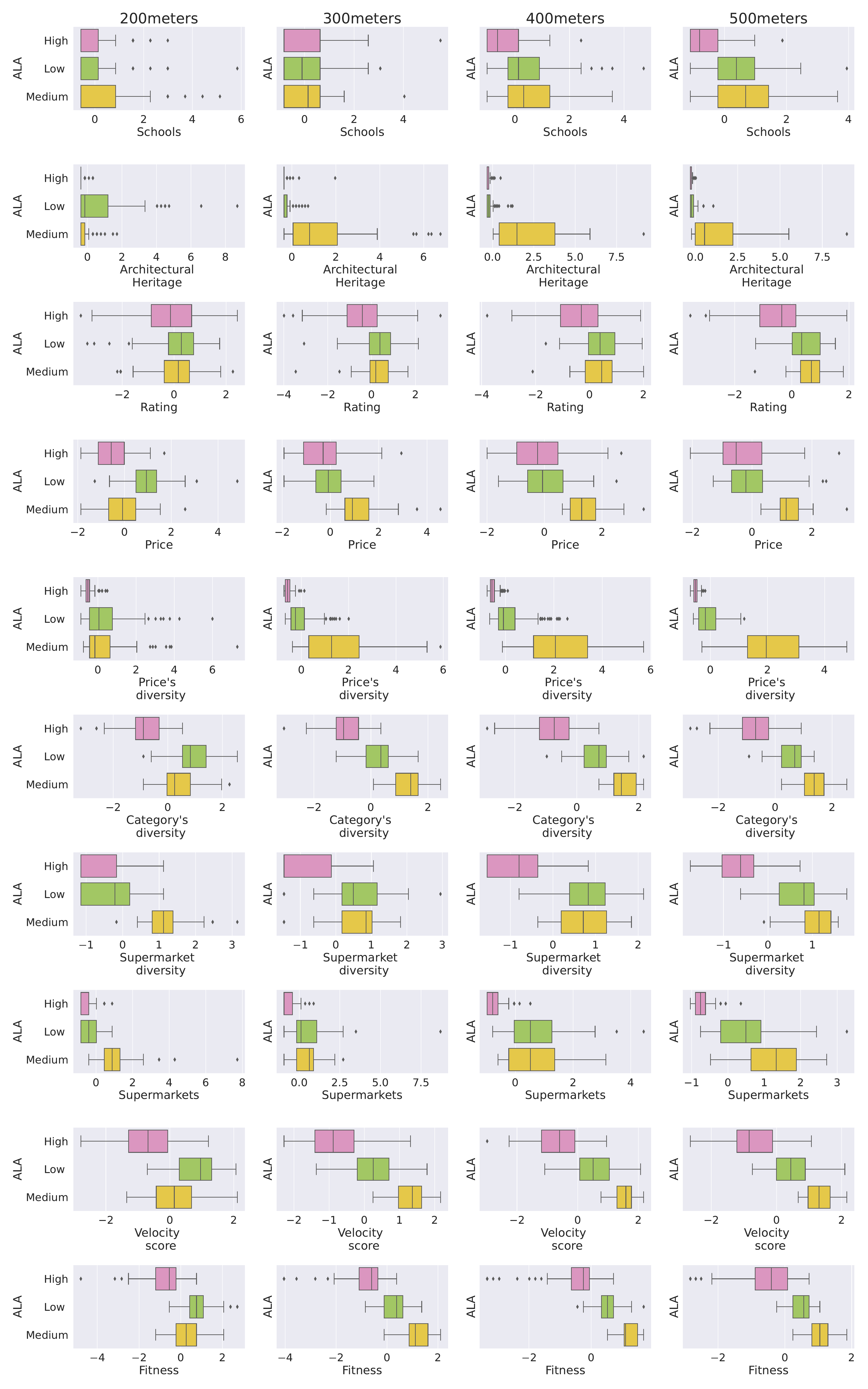}
\caption[Boxplot in each grid]{Boxplot illustrating the Z-score distribution of ALA metrics across different hexagonal grid sizes. The metrics encompass architectural heritage, ratings, price, price diversity, schools, category diversity, supermarket diversity, supermarket count, velocity score, and fitness. Each ALA cluster is represented by a distinct colour, and each column corresponds to a unique grid size, showcasing the variability in metric values across neighbourhoods and spatial resolutions.}
\label{fig:boxplot_hex}
\end{figure}

\begin{figure}[]
\centering
\includegraphics[scale=0.21]{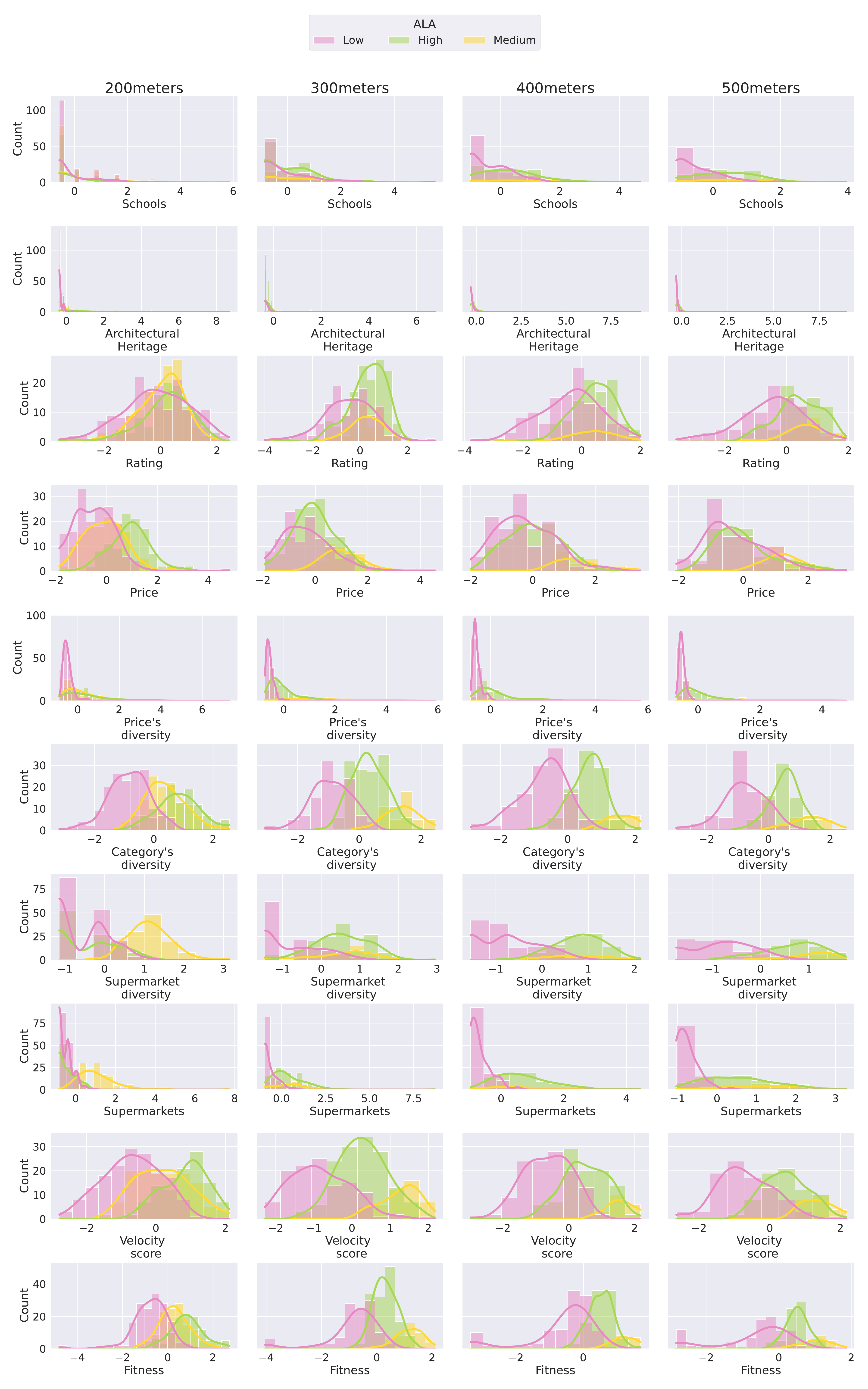}
\caption[Distribution in each grid]{Z-score distribution of ALA metrics across different hexagonal grid sizes. The metrics encompass architectural heritage, ratings, price, price diversity, schools, category diversity, supermarket diversity, supermarket count, velocity score, and fitness. Each ALA cluster is represented by a distinct colour, and each column corresponds to a unique grid size, showcasing the variability in metric values across neighbourhoods and spatial resolutions.}
\label{fig:distirbution_hex}
\end{figure}

\begin{figure}[]
\centering
\includegraphics[scale=0.17]{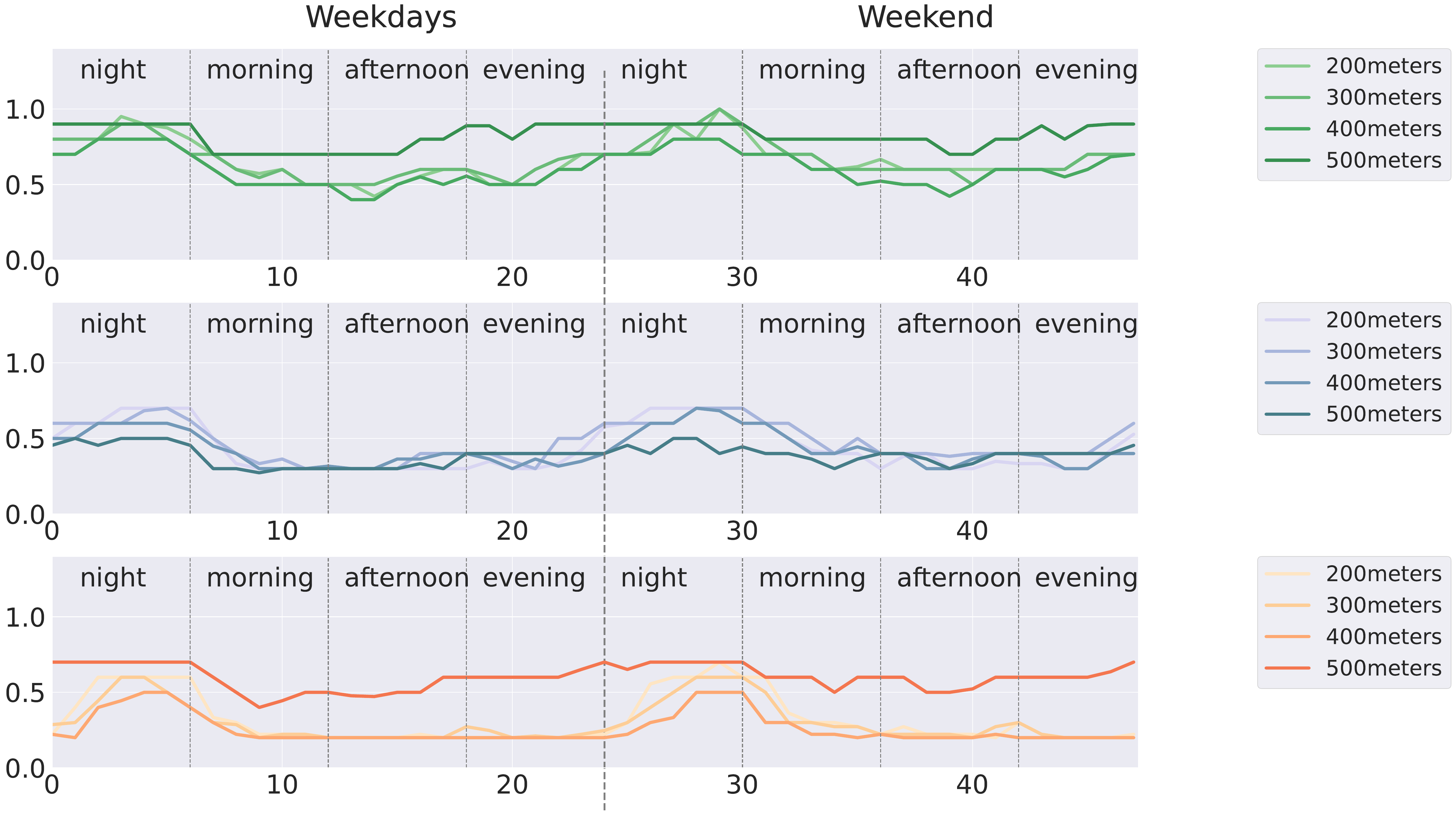}
\caption[Segregation profile for each grid]{Segregation profile depicted for each hexagonal grid across a 48-hour period (24 hours on a weekday and 24 hours on a weekend). The analysis is segmented into three social mixing clusters: segregated (red), mixed (blue), and inclusive (green). The grid size is expressed through the colour gradients. The segregation measure is computed using the Gini index.}

\label{fig:profili_seg}
\end{figure}

\begin{figure}[]
\centering
\includegraphics[scale=0.4]{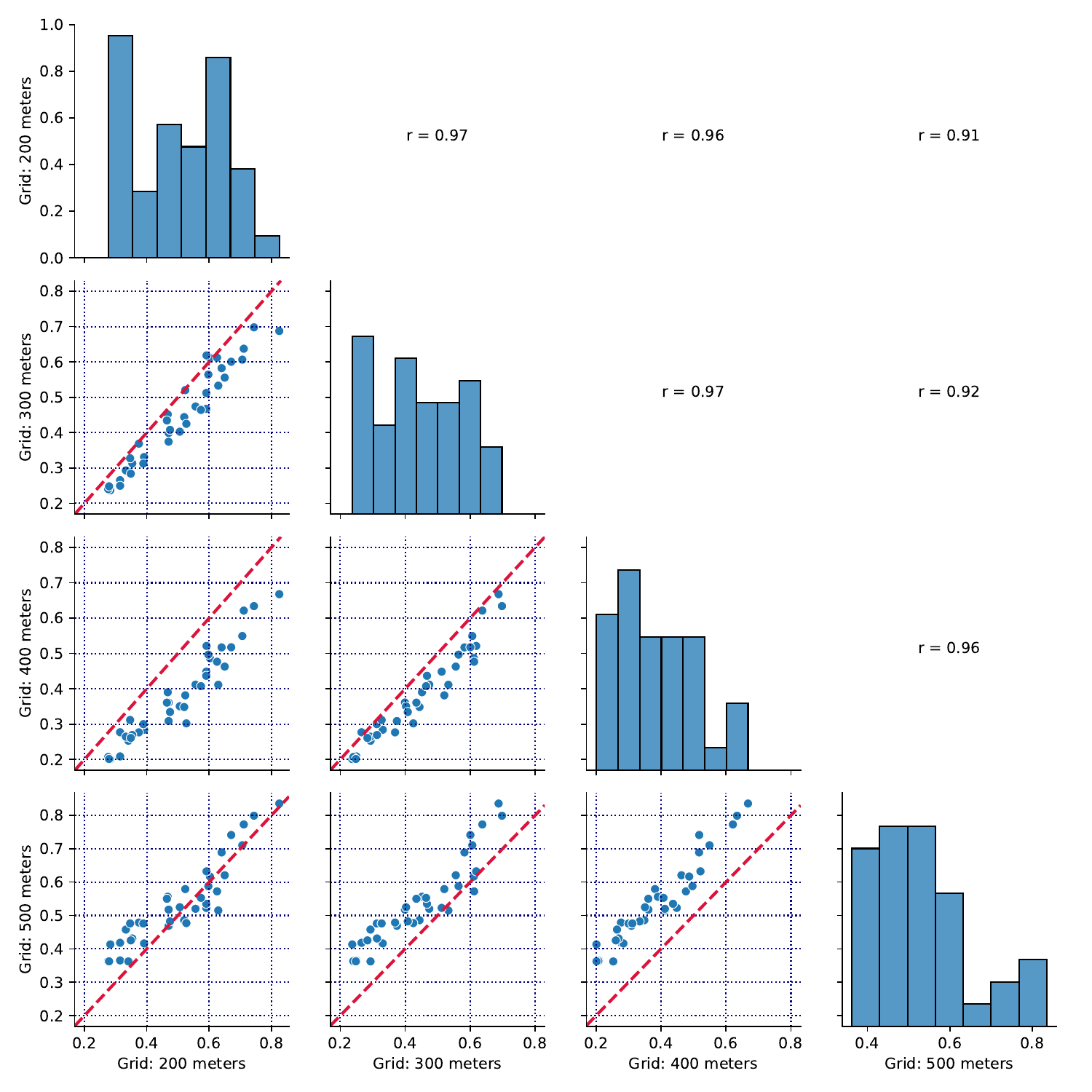}
\caption[Correlation for the hexagonal grids]{Correlation and Distribution of Segregation Across Hexagonal Grids: correlation matrix illustrating the relation between different grid configurations. The matrix shows correlation coefficients, scatter plots, and distribution along the diagonal. Each point within the scatter plots represents the median segregation value of the grids contained within a specific zipcode.}
\label{fig:correlation_hex}
\end{figure}

We cluster with the ALA metrics the remaining grid dimensions. The data presented in figures S\ref{fig:boxplot_hex} and S\ref{fig:distirbution_hex} indicate that a grid with 500-meter sides distinguishes clusters most distinctly. Nonetheless, the evaluation of segregation profiles, as presented in Fig. S\ref{fig:profili_seg}, showed a flat pattern for larger grids. Balancing between classifier quality and the segregation profile trend, we opted for a grid with sides measuring 300 meters.  We check the robustness of the results at the zipcode level. In figure S\ref{fig:correlation_hex}, each point represents the median of segregation of the grids present in a zipcode, which is a highly aggregated measure but with a significant correlation. 

\begin{figure}[]
\centering
\includegraphics[scale=0.4]{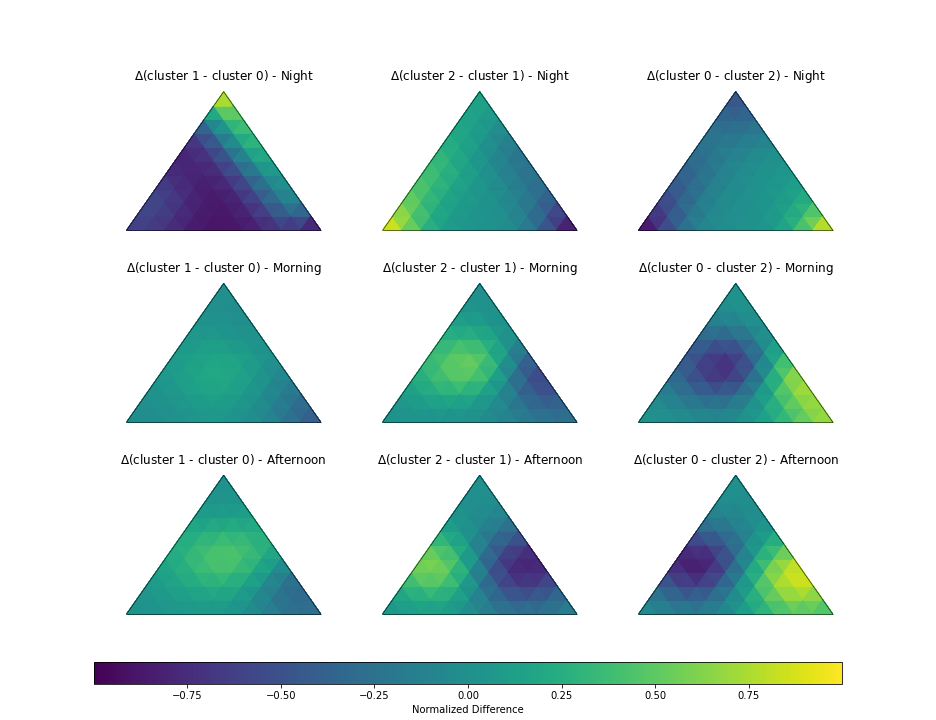}
\caption[Ternary Difference]{Ternary plot showing the differences between each income group. Lighter colours indicate positive values, while darker shades represent negative values. This visualisation helps in understanding the difference in income distribution across different clusters.}
\label{fig:ternary_diff}
\end{figure}

\section{Clusterisation}

\subsection{Income Group Clusterisation}

\begin{figure}[]
\centering
\includegraphics[scale=0.2]{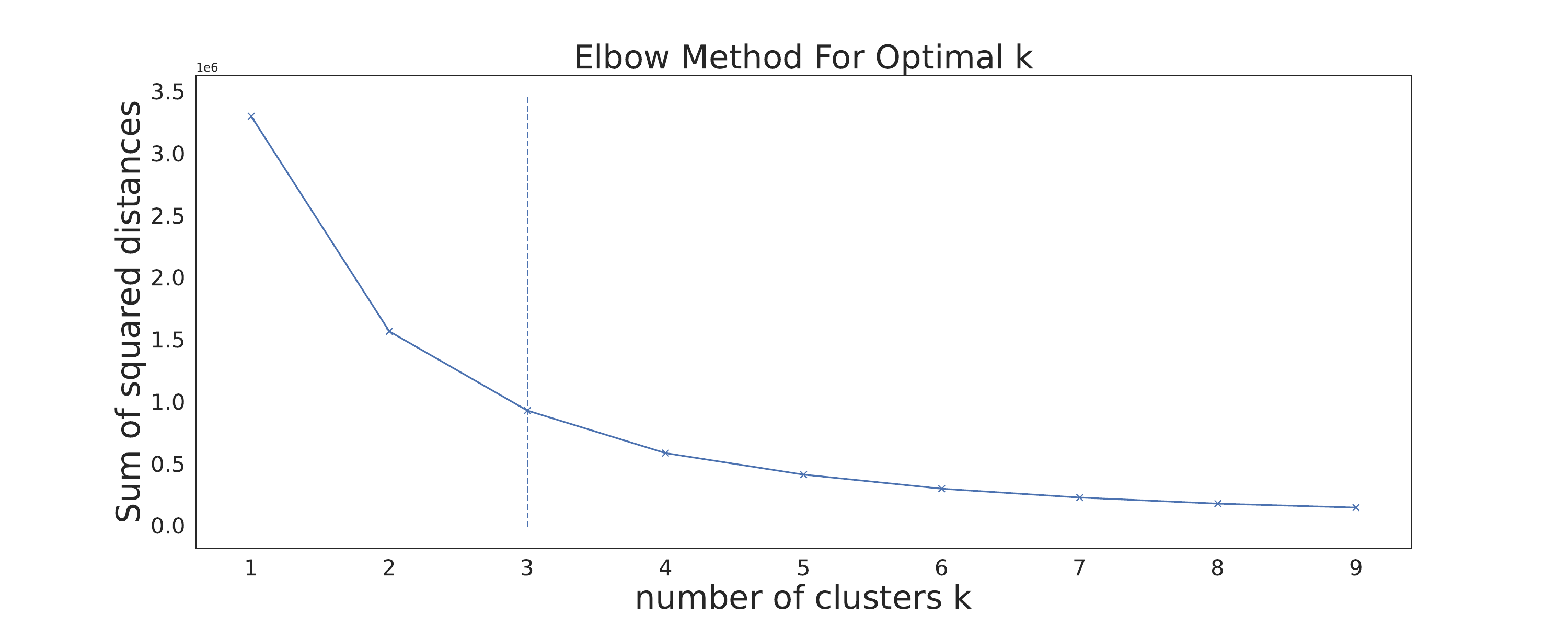}
\caption[Elbow Method Analysis]{Elbow method analysis for income group clusterisation. The elbow point suggests the optimal number of clusters for categorizing income groups.}
\label{fig:elbow}
\end{figure}

\begin{figure}[]
\centering
\includegraphics[scale=0.2]{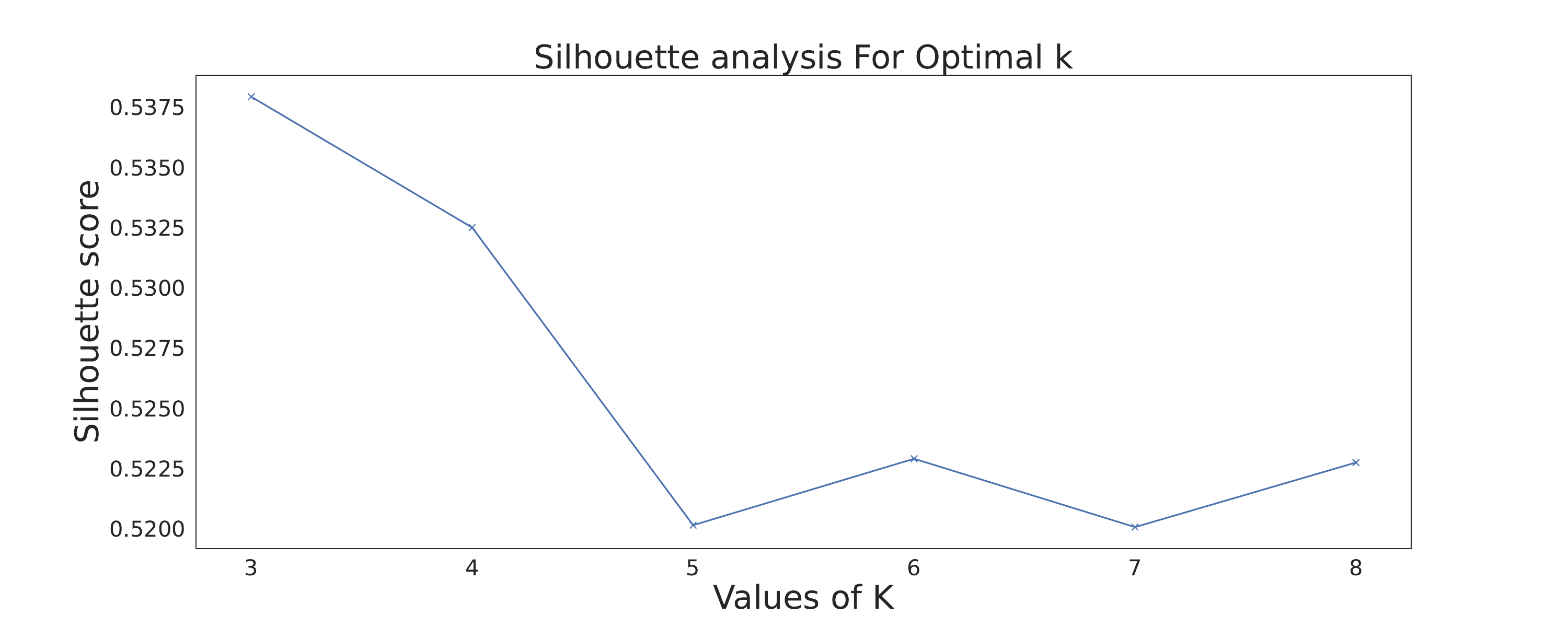}
\caption[Silhouette Score Analysis]{Silhouette score analysis for income group clusterisation. This analysis aids in validating the coherence and separation of the chosen clusters.}
\label{fig:siluette}
\end{figure}

\begin{figure}[]
\centering
\includegraphics[scale=0.2]{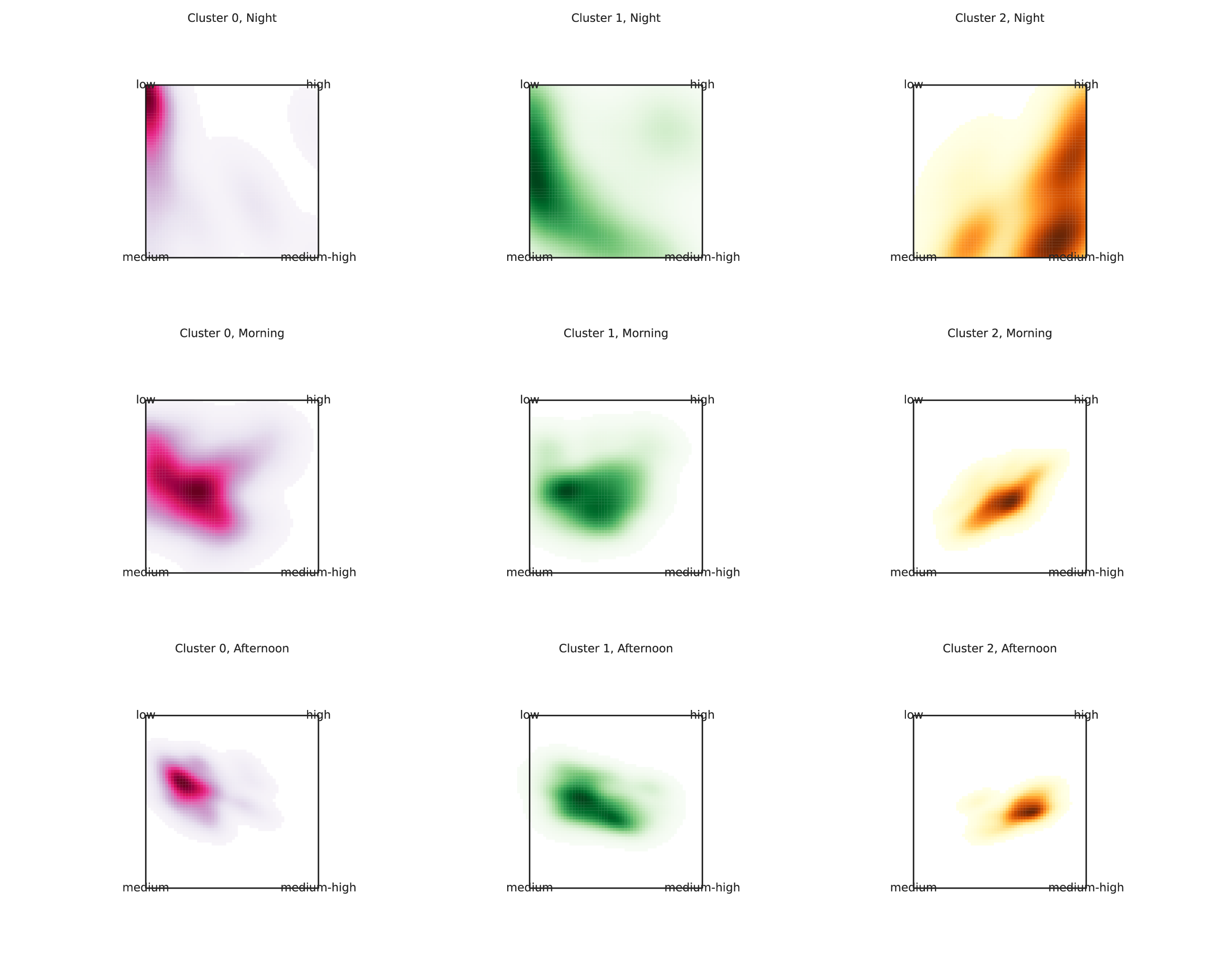}
\caption[Square-form Ternary Plot for 4 Income Groups]{Square-form ternary plot adapted from the main paper to represent 4 income groups. This visualisation format is used to accommodate an additional income group, resulting in a square layout. Each corner of the square represents one of the four income groups, with columns indicating temporal clusters and rows representing temporal windows.}
\label{fig:quadrato}
\end{figure}

\begin{figure}[]
\centering
\includegraphics[scale=0.2]{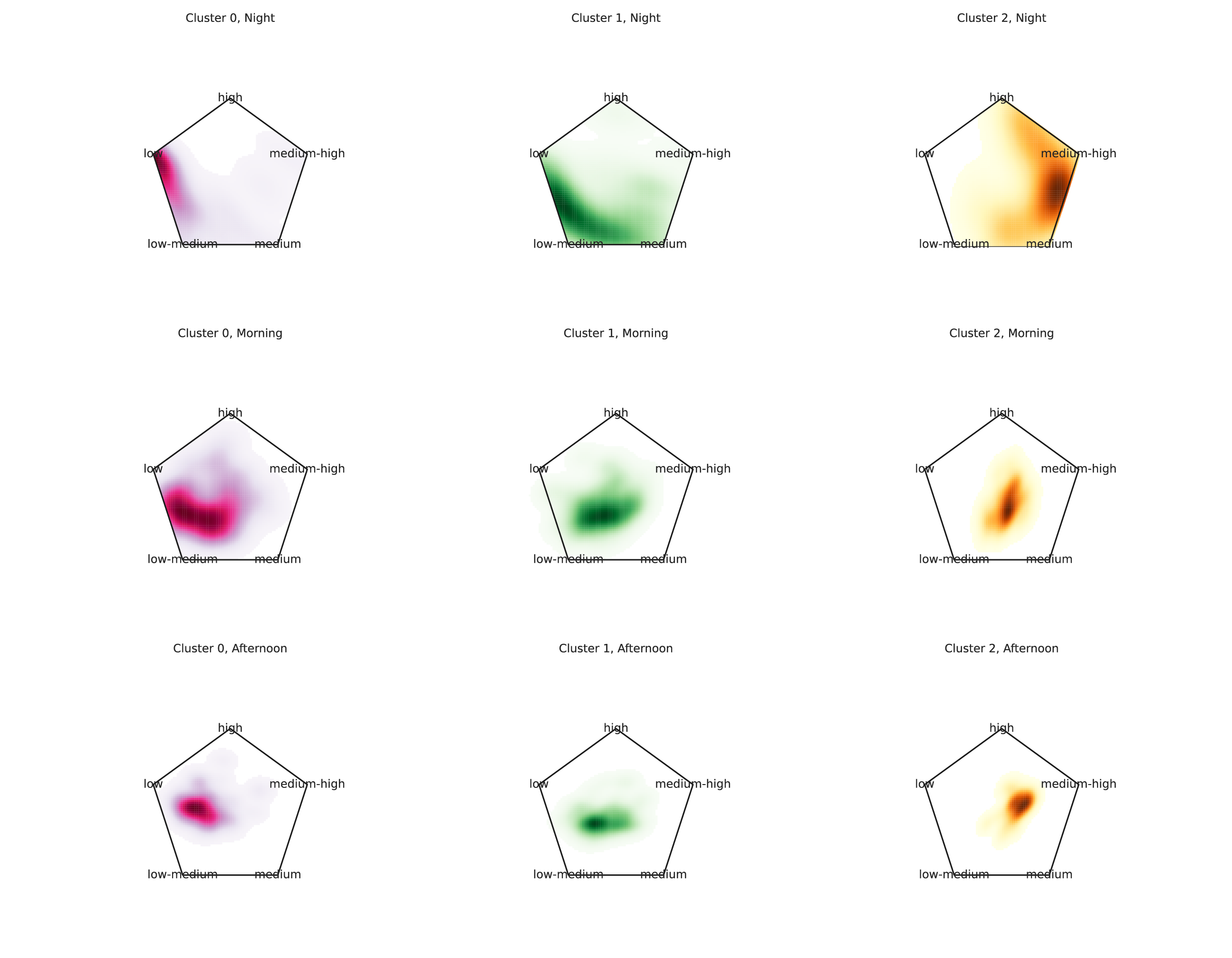}
\caption[Pentagon-form Ternary Plot for 5 Income Groups]{Pentagon-form ternary plot adapted from the main paper to represent 5 income groups. This visualisation format is used to accommodate an additional income group, resulting in a pentagon layout. Each vertex of the pentagon represents one of the five income groups, with columns indicating temporal clusters and rows depicting temporal windows.}
\label{fig:pentagono}
\end{figure}

In the process of defining income groups, our approach was grounded in both statistical methods and empirical consistency. Initially, we applied the elbow method to determine the optimal number of clusters, as illustrated in Fig. S\ref{fig:elbow}. To further validate our choice, we conducted a silhouette score analysis, shown in Fig. S\ref{fig:siluette}. Despite relying on these methodological approaches, we also aimed to ensure that our results were consistent with the actual movements of individuals across different ALA score zones. To this end, we analysed movement patterns with both 4 and 5 income groups, using squares (Fig. S\ref{fig:quadrato}) and pentagons (Fig. S\ref{fig:pentagono}) to represent different zones, observing coherent results in both cases.

However, we ultimately chose to divide the data into three income groups. This decision was based on a balance between statistical robustness and the interpretability of results. Three clusters (low, medium, and high income) offered a clear and comprehensive representation of the socio-economic diversity within the city, aligning with our research objectives.

\begin{figure}[]
\centering

\includegraphics[scale=0.2]{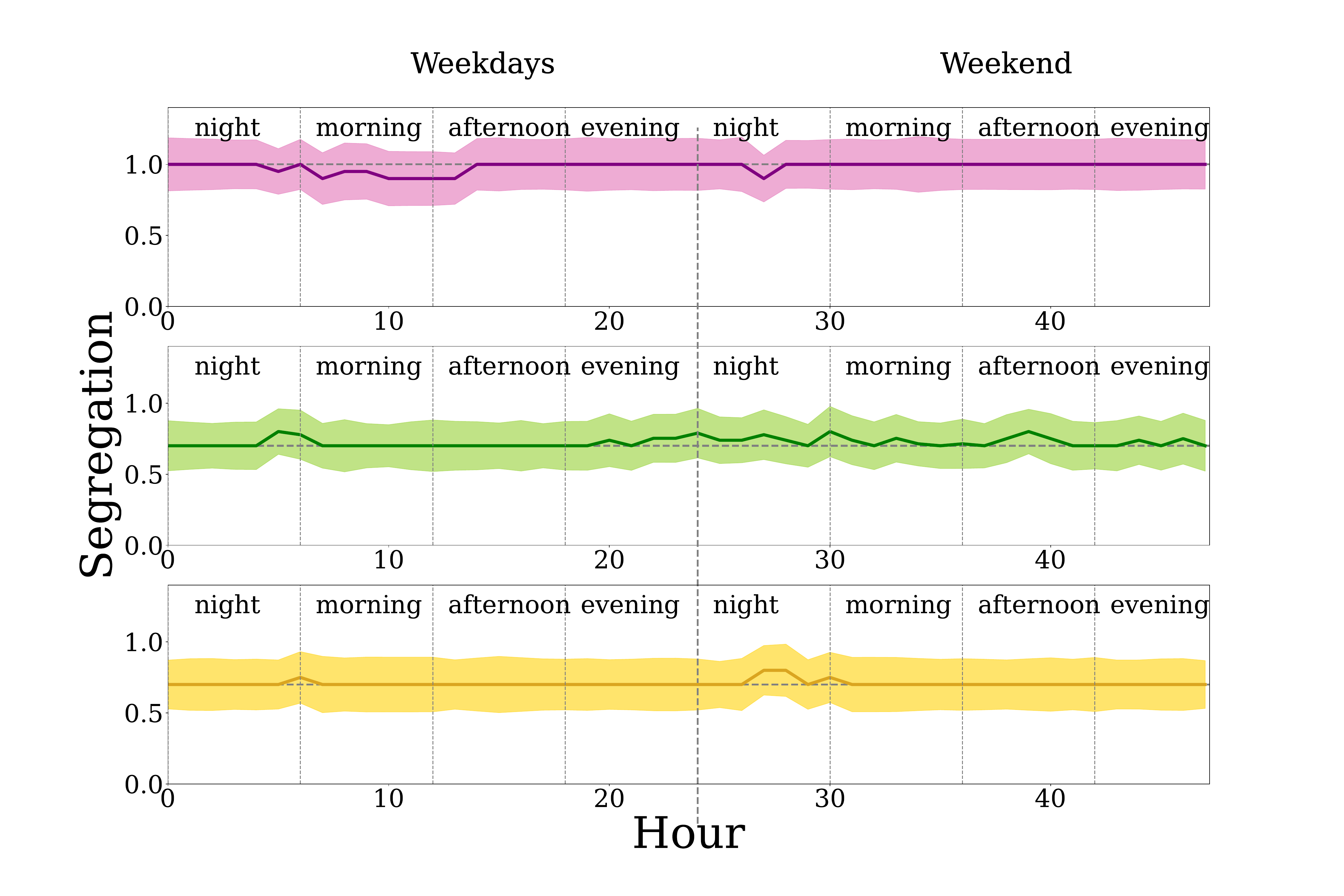}
\caption[Home Segregation profile]{Segregation profile depicted within the ALA clusters based solely on stops made at users' residences. The analysis spans a 48-hour period (24 hours on a weekday and 24 hours on a weekend), with segregation being measured using the Gini index. The segregation is segmented into three clusters: high (yellow), medium (green), and low (pink). }
\label{fig:home}
\end{figure}

\begin{figure}[]
\centering

\includegraphics[scale=0.2]{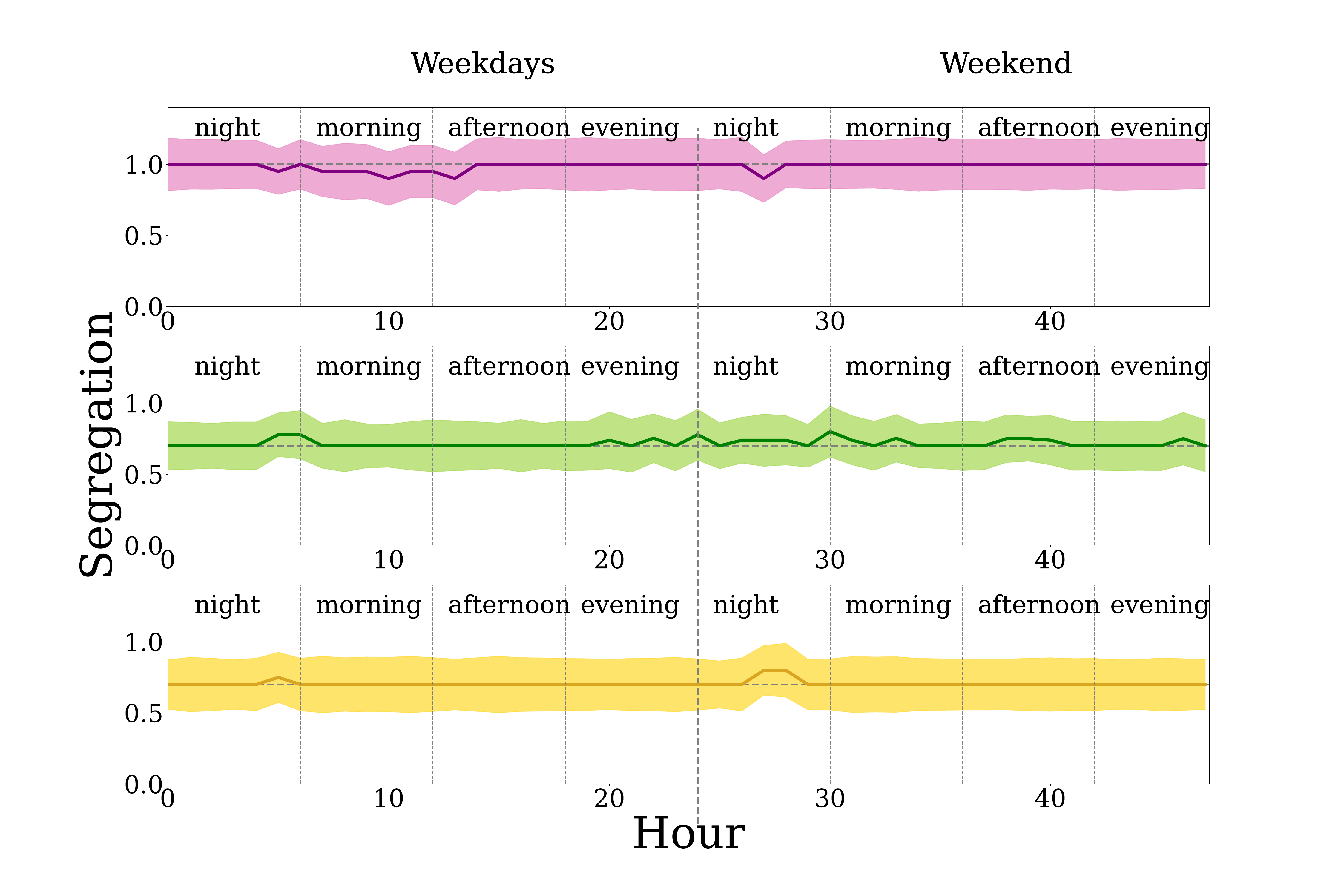}
\caption[Work Segregation profile]{Segregation profile depicted within the ALA clusters based solely on stops made at users' workplaces. The analysis spans a 48-hour period (24 hours on a weekday and 24 hours on a weekend), with segregation being measured using the Gini index. The segregation is segmented into three clusters: high (yellow), medium (green), and low (pink). }
\label{fig:work}
\end{figure}
\subsection{City Clusterisation}

\begin{figure}[]
\centering
\includegraphics[scale=1.2]{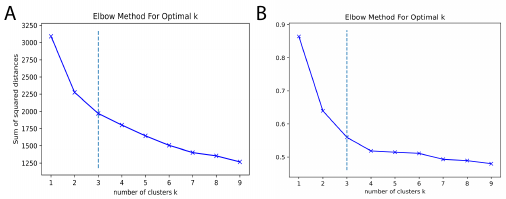}
\caption[Elbow Method Analysis]{Elbow method analysis for hexagon clusterisation. The elbow point suggests the optimal number of clusters for categorizing hexagons in ALA metrics (\textbf{A}) and in temporal mixing profile (\textbf{B}).}
\label{fig:elbow_city}
\end{figure}

\begin{figure}[]
\centering
\includegraphics[scale=0.33]{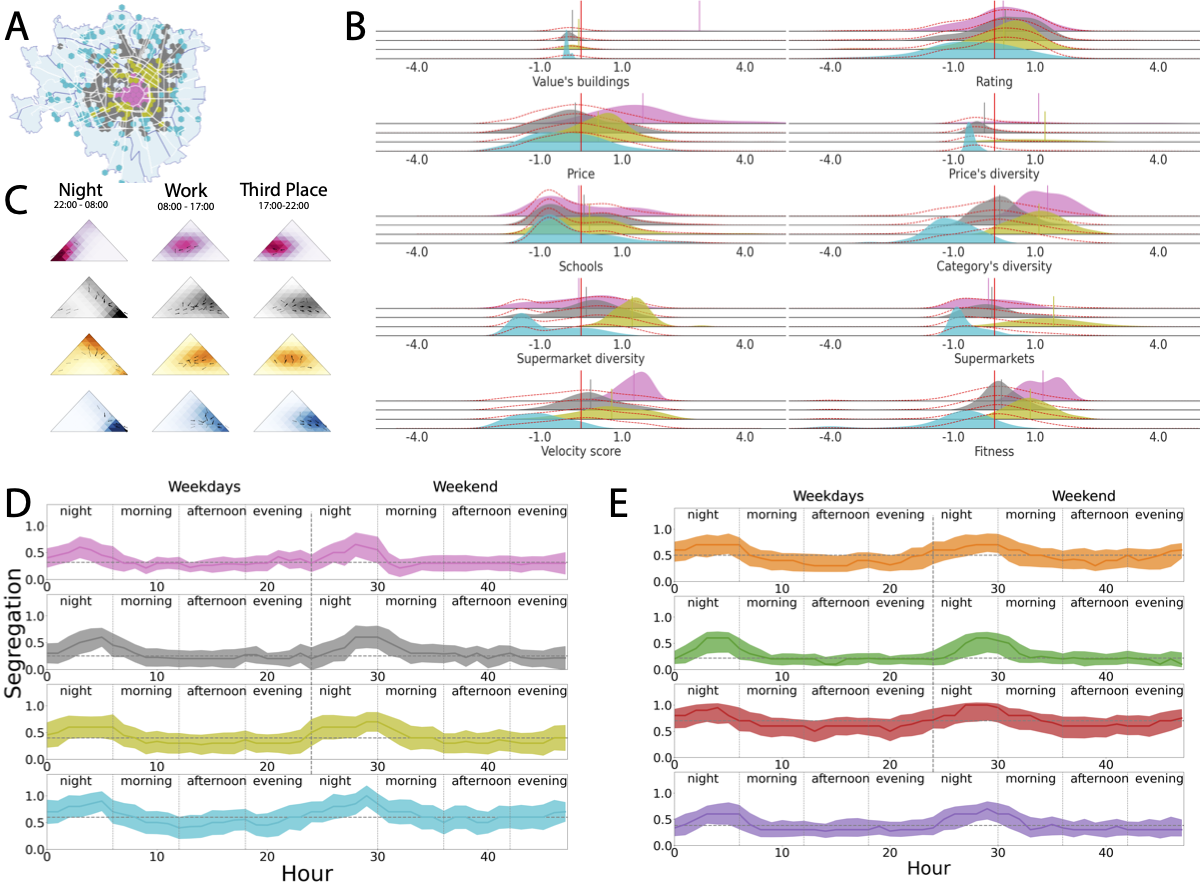}
\caption[Analysis Results for 4 Clusters]{Analysis of the ALA clusters for 4 clusters. \textbf{A}: Map of ALA clusters. \textbf{B}: Distribution of ALA metrics per cluster. \textbf{C}: Ternary plot for the 4 ALA clusters. \textbf{D}: Segregation trends within the ALA clusters over weekdays and weekends. \textbf{E}: Segregation trends for the temporal mixing clusters.}
\label{fig:all_4}
\end{figure}

\begin{figure}[]
\centering
\includegraphics[scale=0.33]{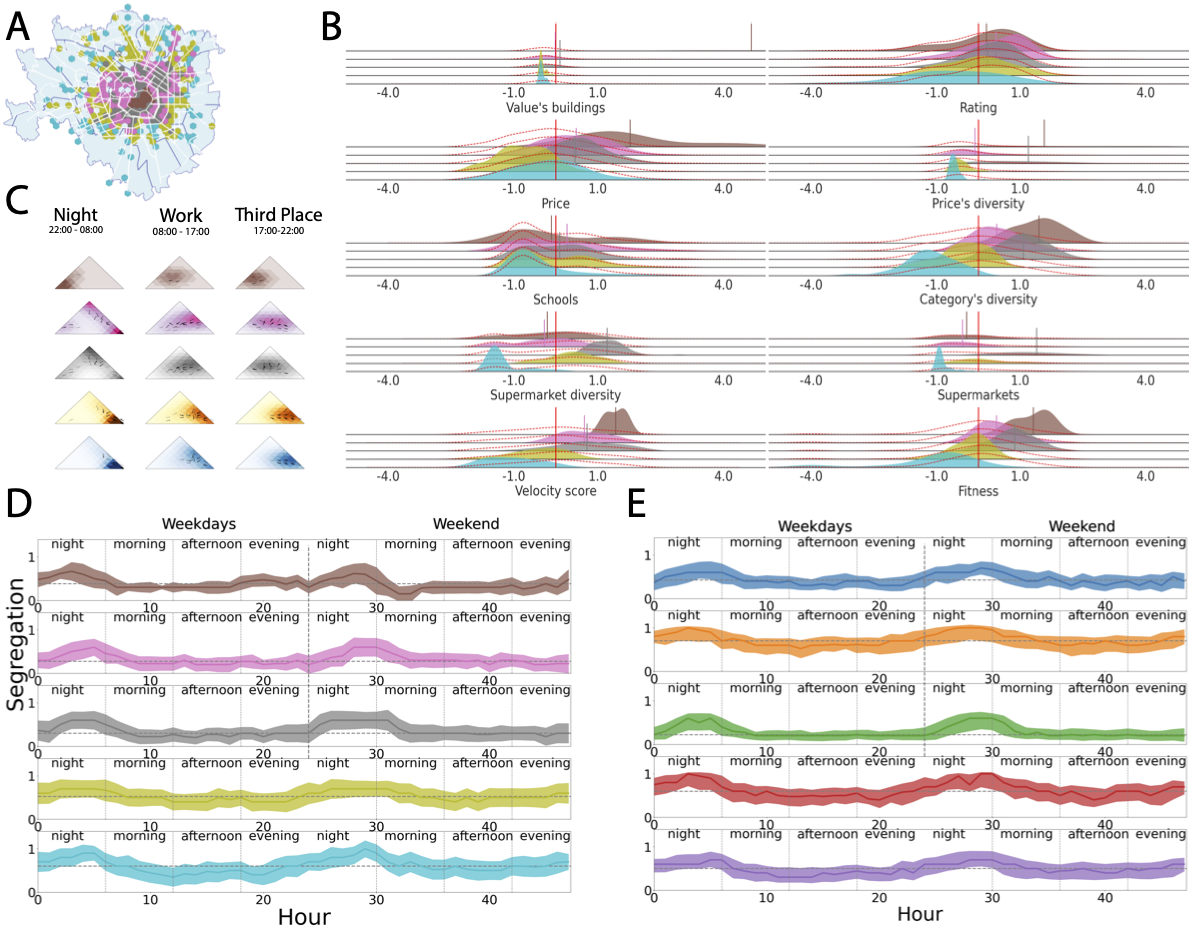}
\caption[Analysis Results for 5 Clusters]{Analysis of the ALA clusters for 5 clusters. \textbf{A}: Map of ALA clusters. \textbf{B}: Distribution of ALA metrics per cluster. \textbf{C}: Ternary plot for the 5 ALA clusters. \textbf{D}: Segregation trends within the ALA clusters over weekdays and weekends. \textbf{E}: Segregation trends for the temporal mixing clusters.}
\label{fig:all_5}
\end{figure}

For city-wide clusterisation, the elbow method, as depicted in Figure S\ref{fig:elbow_city}, indicated three as the optimal number of clusters. To gain a deeper understanding of Milan's spatial dynamics, we explored configurations with higher cluster counts, specifically 4 and 5 clusters, as presented in Figures \ref{fig:all_4} and \ref{fig:all_5}, respectively. While these configurations yielded coherent results, we observed an overlap in the characteristics of the clusters, suggesting that the increase in cluster count did not necessarily lead to a more distinct separation of urban features.

The decision to adopt k=3 was driven by our objective to maintain clarity in our analysis and effectively capture the diverse urban landscape of Milan. The three-cluster model provided a balanced representation, offering meaningful differentiation without overcomplicating the spatial representation.

We employed different clustering methods to ensure the robustness of our findings. Specifically, for the ALA metrics we used k-means and agglomerative clustering~\cite{Murtagh2017}
techniques for clusterization. After clusterization we assessed the similarity of the resulting clusters obtaining the Fowlkes-Mallows score~\cite{Fowlkes1983} of 0.73. For the Temporal Mixing we used k-means, agglomerative
and spectral~\cite{Ng2002} 
clustering techniques for clusterization. After clusterization we assessed the similarity of the resulting clusters obtaining the Fowlkes-Mallows score of 0.84.

\begin{figure}[]
\centering
\includegraphics[scale=0.4]{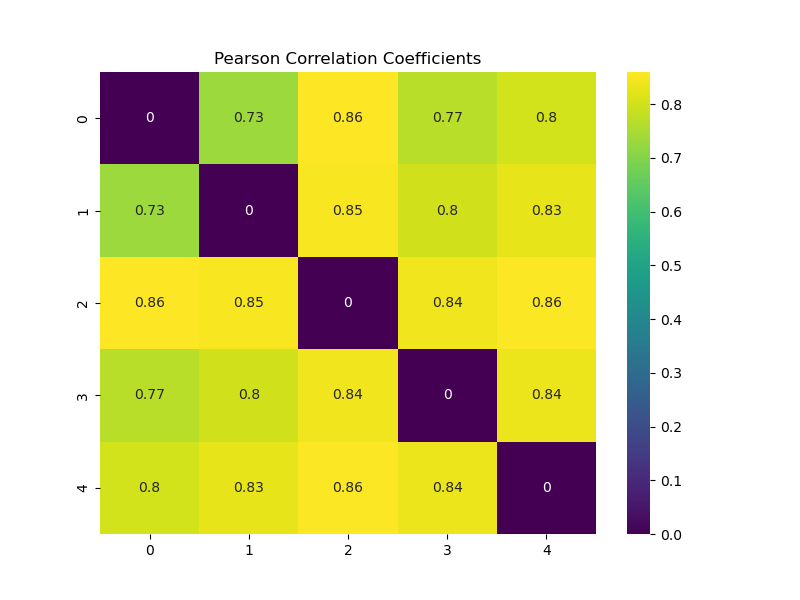}
\caption[Cluster Similarity]{Heatmap of Pearson Correlation Coefficients between Different Clusters. This heatmap visualises the pairwise Pearson correlation coefficients for curves derived from each cluster in the dataset. Each cell represents the correlation coefficient between two clusters, with values closer to 1 indicating a higher degree of similarity in the patterns of the respective curves.}
\label{fig:similitudine}
\end{figure}

\section{Venues}\label{venue}

\begin{figure}[]
\centering
\includegraphics[scale=0.2]{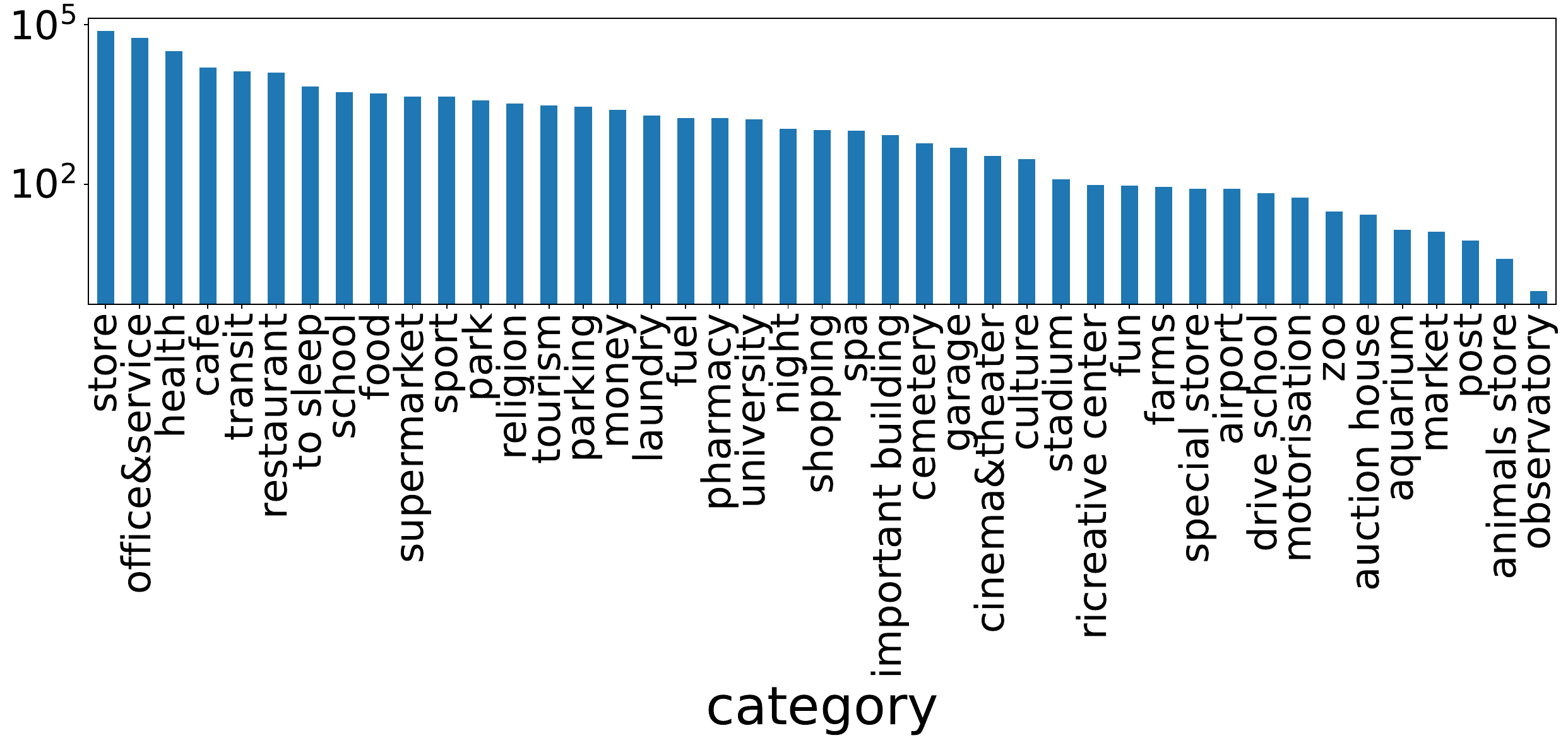}
\caption[Categories distribution]{Number of categories in Google Place dataset}
\label{fig:cat_hist}
\end{figure}

Utilising the Google Place classifications, we categorised the venues frequented by individuals. For generic classifications, such as "point of interest", we referred to Bing for specific categorisation. If unavailable, these generic classifications were discarded. The venues were then grouped systematically according to our Taxonomy as illustrated in (Fig. S\ref{fig:cat_hist}).

\section{Metrics}
\begin{figure}[]
\centering
\includegraphics[scale=0.4]{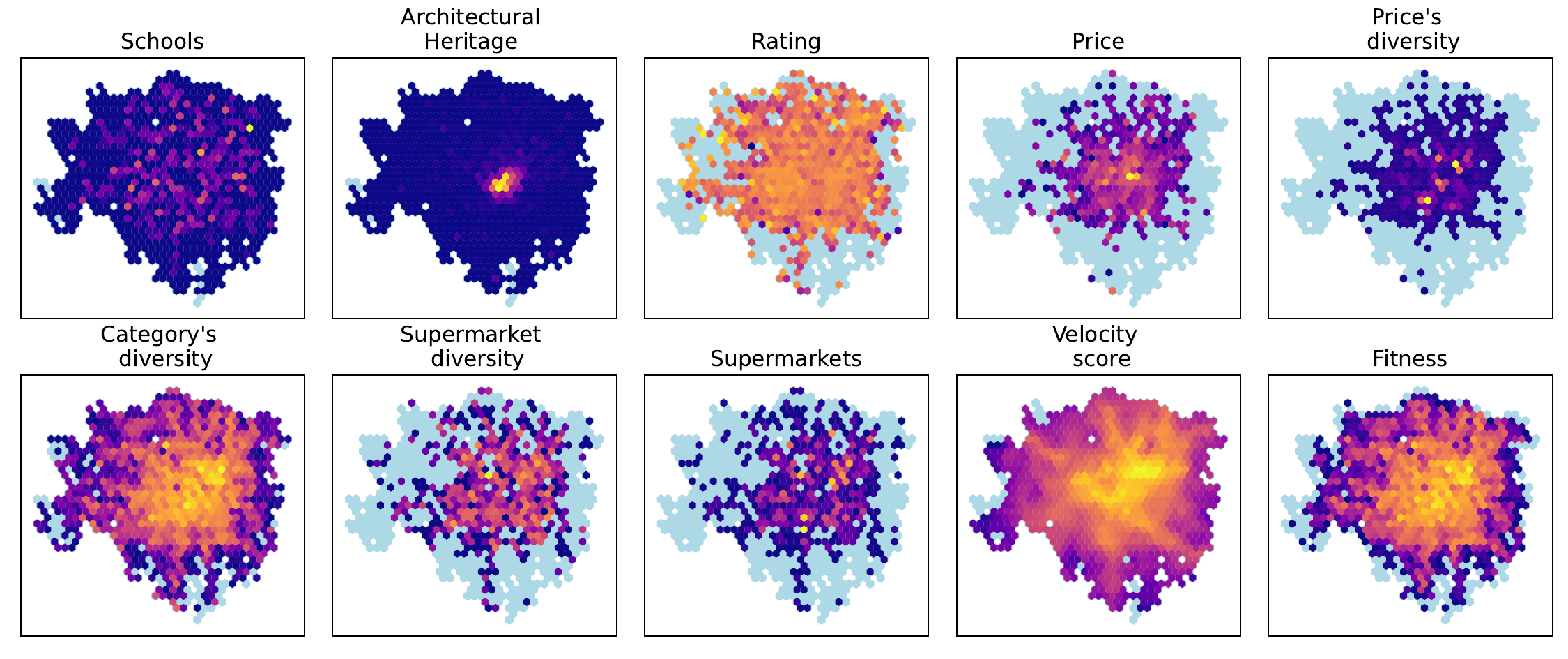}
\caption[Map distributions for all the hexagons]{Map distributions for each hexagon representing the ten ALA metrics across Milan. Each map, computed using Geopandas, illustrates the spatial distribution of a particular feature within the city. The colour corresponds to the value of the feature in the hexagon, with the yellow indicating higher values and the blue having lower values.}
\label{fig:feature_all}
\end{figure}

\begin{figure}[]
\centering
\includegraphics[scale=0.4]{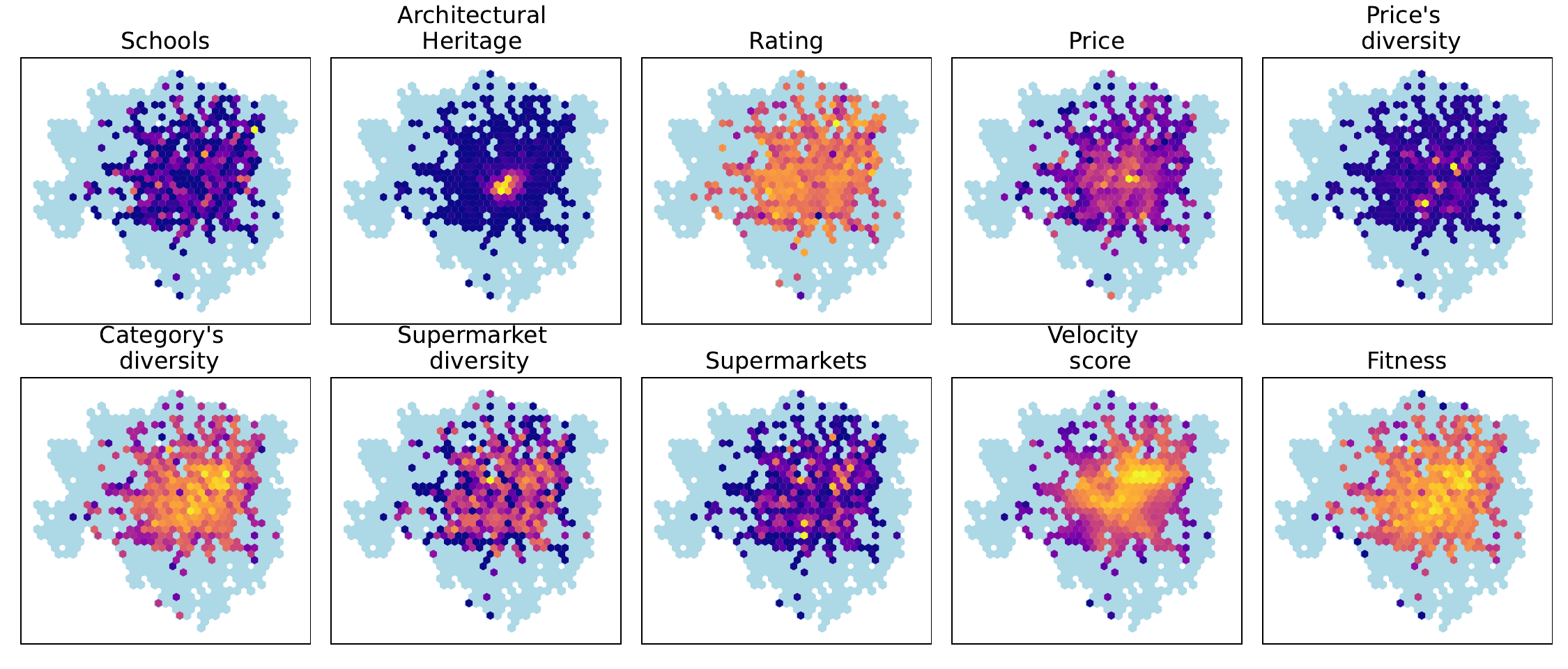}
\caption[Map distributions]{Map distributions for each hexagon representing the ten ALA metrics across Milan, considering only the hexagons where all ALA features are non-null. Each map, computed using Geopandas, illustrates the spatial distribution of a particular feature within the city. The colour corresponds to the value of the feature in the hexagon, with the yellow indicating higher values and the blue having lower values.}
\label{fig:feature_taglio}
\end{figure}

Maps distributions for the ten ALA metrics are shown in (Fig. S\ref{fig:feature_all}, S\ref{fig:feature_taglio}).

\subsection{Liveability}\label{sec:liability}
Neighbourhood liveability is influenced by many factors~\cite{Mouratidis2020}, three essential indicators of a neighbourhood's liveability and potential deprivation or prosperity are its educational accessibility~\cite{Greenberg1999}, shopping options~\cite{cerrato, Gustafson2012}, and architectural heritage~\cite{Cudny2019}. We chose to consider both the number and diversity of schools and supermarkets and the value of a building.
We chose not to incorporate green spaces as an additional criterion for liveability, given their inherently dualistic nature~\cite{Baran2014}: they may provide benefits when well-maintained~\cite{Kruizse2019}, but also evoke concerns and apprehension when in a degraded state~\cite{Mak2018}. Due to the absence of information regarding the condition of parks, we have opted to leave them out of our analysis.
\subsection{Attrattivity: Fitness and Complexity}
The framework of the Economic-Complexity, defined by Tacchella et al.~\cite{Tacchella2012}, is based on the interaction between countries and products, expressed by the application of the Revealed Comparative Advantage (RCA)~\cite{rca} threshold over the entries:
\begin{align}\label{rca}
    RCA^{(y)}_{cp} = \left. \frac{M^{(y)}_{cp}}{\sum_{p'}M^{(y)}_{cp'}}   \middle/ \frac{\sum_{c'}M^{(y)}_{c'p}}{\sum_{c'p'}M^{(y)}_{c'p'}} \right.
\end{align}

\myequations{RCA}
The initial conditions are 1 for both variables. $M_{hp}$ is the bi-adjacency matrix of the undirected bipartite network: its elements are 1 if the neighbourhood has a POI category p and 0 otherwise.
The Fitness $F_c$ for the generic country c and the Quality $Q_p$ for the generic product p at the $n-$th step of the iteration are defined as:

\begin{equation}\label{fitness}
\left\{
\begin{aligned}
\tilde{F}^{(n)}_c &= \sum_p M_{cp}Q^{(n-1)}_p \\
\tilde{Q}^{(n)}_p &= \frac{1}{\sum_c M_{cp}\frac{1}{F^{(n-1)}_c}}
\end{aligned}
\right.
\quad \Longrightarrow \quad
\left\{
\begin{aligned}
F^{n}_c &= \frac{\tilde{F}^{(n)}_c}{<\tilde{F}^{(n)}_c>_c} \\
Q^{n}_p &= \frac{\tilde{Q}^{(n)}_p}{<\tilde{Q}^{(n)}_p>_p}
\end{aligned}
\right.
\end{equation}

\myequations{Fitness and Complexity}
where the symbols $<\cdot>$ indicate the average taken over the proper set. The initial conditions are taken as $F_0=Q_0=1 \forall c \in N_c, \forall p \in N_p $,  where $N_c$ and $N_p$ are the numbers respectively of countries and products.

The concept is applied to different data sources such as urbanisation~\cite{urbanzation}, patents~\cite{stracca}, companies~\cite{giamba}, or scientific publications~\cite{cimini}.

\section{Regression}
Adopting a Lindeman, Merenda, and Gold (LMG)~\cite{lmg}, in our analysis, we opted for the Spatial Lag Model over the Ordinary Least Squares (OLS) and the Spatial Error Model due to its capacity to explicitly incorporate spatial dependency in the regression equation.
Our preference for the Spatial Lag Model over the Spatial Error Model was influenced by a combination of factors. Firstly, the Spatial Lag Model offered a more robust explanation of the data, as evidenced by a higher pseudo R-squared value. This indicated that the Spatial Lag Model accounted for a larger portion of the variance in the dependent variable compared to the Spatial Error Model.

Secondly, the analysis of residuals played a crucial role. The Spatial Lag Model effectively addressed the issue of spatial autocorrelation in the residuals. In spatial analysis, it's essential not only to fit the model well to the observed data but also to ensure that the residuals (the differences between observed and predicted values) do not exhibit spatial autocorrelation. If spatial autocorrelation exists in the residuals, it suggests that some important spatial processes might have been omitted from the model. The Spatial Lag Model, in our case, showed better performance in this aspect, ensuring that the residuals were more randomly distributed without patterns of spatial autocorrelation, thus providing a more reliable and accurate representation of the spatial relationships in the data.

\newenvironment{sidewaystableS}{
    \begin{sidewaystable}
    \centering
}{
    \end{sidewaystable}
}

\begin{sidewaystableS}
\begin{tabular}{llllllll}
\hline
Group & Variable & Only & Only & Only & Only  & All & LASSO \\
 & & Accessibility & Liveability & Attractivity & Population & Together & \\
\hline
Accessibility & Velocity Score & -1.16*** & & & &-0.6*** & -0.65*** \\
\hline
 & Schools & & -0.08 & & & -0.02\\
Liveability & Supermarkets & & 0.01 & & & 0.01 \\
 & Supermarket diversity & & -0.32 & & & -0.08 \\
& Architectural Heritage & & 0.31** & & & 0.36***& 0.37***\\
\hline
 & Fitness & & &-0.38** & & -0.19*&-0.29**\\
 & Category's diversity & & &-0.35* & & -0.11 \\
Attractivity & Price's diversity & & &-0.20* & & -0.14*&-0.18*\\
 & Price & & &-0.50*** & & -0.41*** &-0.38***\\
 & Rating & & & -0.23& & -0.17\\
\hline
Population & Density & & & &-0.152*** & -0.0295 & \\
\hline
\multicolumn{1}{r}{$R^2$} & & 0.428 & 0.116 & 0.412 &0.513&0.165 &0.595  \\
\hline
\end{tabular}
\caption[OSL model Regression ]{Regression results of the OSL model for predicting segregation using combinations of ALA metrics.}
\label{tab:regression_osl}
\end{sidewaystableS}

\begin{sidewaystableS}
\begin{tabular}{llllllll}
\hline
Group & Variable & Only & Only & Only & Only & All & LASSO \\
 & & Accessibility & Liveability & Attractivity & Density & Together & \\
\hline 
Accessibility & Velocity Score & -0.75*** & & & & -0.59*** & -0.58*** \\
 & Lambda & 0.67*** & & & & 0.63*** & 0.63*** \\
\hline
 & Schools & & -0.08 &  & & -0.03\\
Liveability & Supermarkets  & & 0.14 & & & 0.14 & \\
 & Supermarket diversity & & -0.09 & & & -0.06 & \\
& Architectural Heritage  & & 0.10 & & & 0.11 & 0.10\\
 & Lambda & & 0.88*** & & & 0.63*** & 0.63*** \\
\hline
 & Fitness & & & -0.18* & & -0.07 & -0.12*\\
 & Category's diversity & & & -0.05 & & -0.03 & \\
Attractivity & Price's diversity & & &-0.16* &  &  -0.15*&-0.15*\\
 & Price  & & & -0.13* & & -0.11* & -0.11*\\
 & Rating & & & -0.01 & & -0.01 & \\
 & Lambda & & & 0.76*** & & 0.63*** & 0.63*** \\
\hline
Population & Density & & & & -0.01 & -0.01 & \\
 & Lambda & & & & 0.65*** & 0.63*** & 0.63*** \\
\hline
\multicolumn{1}{r}{$R^2$} & & 0.49 & 0.05 & 0.48 & 0.53 & 0.57 & 0.57  \\
\hline
\end{tabular}
\caption[Spatial Error model Regression]{Regression results of the Spatial Error model for predicting segregation using combinations of ALA metrics.}
\label{tab:regression_error}
\end{sidewaystableS}


\begin{thebibliography}{100}
\urlstyle{rm}
\expandafter\ifx\csname url\endcsname\relax
  \def\url#1{\texttt{#1}}\fi
\expandafter\ifx\csname urlprefix\endcsname\relax\def\urlprefix{URL }\fi
\expandafter\ifx\csname doiprefix\endcsname\relax\def\doiprefix{DOI: }\fi
\providecommand{\bibinfo}[2]{#2}
\providecommand{\eprint}[2][]{\url{#2}}

\bibitem{Sharifi2019}
\bibinfo{author}{Sharifi, A.}
\newblock \bibinfo{journal}{\bibinfo{title}{Resilient urban forms: A
  macro-scale analysis}}.
\newblock {\emph{\JournalTitle{Cities}}} \textbf{\bibinfo{volume}{85}},
  \doiprefix\url{10.1016/j.cities.2018.11.023} (\bibinfo{year}{2019}).

\bibitem{He2018}
\bibinfo{author}{He, Q.} \emph{et~al.}
\newblock \bibinfo{journal}{\bibinfo{title}{The impact of urban growth patterns
  on urban vitality in newly built-up areas based on an association rules
  analysis using geographical ‘big data’}}.
\newblock {\emph{\JournalTitle{Land Use Policy}}}
  \textbf{\bibinfo{volume}{78}},
  \doiprefix\url{10.1016/j.landusepol.2018.07.020} (\bibinfo{year}{2018}).

\bibitem{Ravallion2007}
\bibinfo{author}{Ravallion, M.}, \bibinfo{author}{Chen, S.} \&
  \bibinfo{author}{Sangraula, P.}
\newblock \bibinfo{journal}{\bibinfo{title}{New evidence on the urbanization of
  global poverty}}.
\newblock {\emph{\JournalTitle{Population and Development Review}}}
  \textbf{\bibinfo{volume}{33}},
  \doiprefix\url{10.1111/j.1728-4457.2007.00193.x} (\bibinfo{year}{2007}).

\bibitem{Galster2007}
\bibinfo{author}{Galster, G.} \& \bibinfo{author}{Booza, J.}
\newblock \bibinfo{journal}{\bibinfo{title}{The rise of the bipolar
  neighborhood}}.
\newblock {\emph{\JournalTitle{Journal of the American Planning Association}}}
  \textbf{\bibinfo{volume}{73}}, \doiprefix\url{10.1080/01944360708978523}
  (\bibinfo{year}{2007}).

\bibitem{Chetty2016}
\bibinfo{author}{Chetty, R.} \emph{et~al.}
\newblock \bibinfo{journal}{\bibinfo{title}{The association between income and
  life expectancy in the united states, 2001-2014}}.
\newblock {\emph{\JournalTitle{JAMA - Journal of the American Medical
  Association}}} \textbf{\bibinfo{volume}{315}},
  \doiprefix\url{10.1001/jama.2016.4226} (\bibinfo{year}{2016}).

\bibitem{Tammaru2021}
\bibinfo{author}{Tammaru, T.}, \bibinfo{author}{Marcińczak, S.},
  \bibinfo{author}{Aunap, R.} \& \bibinfo{author}{van Ham, M.}
\newblock \bibinfo{journal}{\bibinfo{title}{Inequalities and segregation across
  the long-term economic cycle: An analysis of south and north european
  cities}}.
\newblock {\emph{\JournalTitle{SSRN Electronic Journal}}}
  \doiprefix\url{10.2139/ssrn.3029852} (\bibinfo{year}{2021}).

\bibitem{Thomas2014}
\bibinfo{author}{Thomas, T.~L.}, \bibinfo{author}{Diclemente, R.} \&
  \bibinfo{author}{Snell, S.}
\newblock \bibinfo{journal}{\bibinfo{title}{Overcoming the triad of rural
  health disparities: How local culture, lack of economic opportunity, and
  geographic location instigate health disparities}}.
\newblock {\emph{\JournalTitle{Health Education Journal}}}
  \textbf{\bibinfo{volume}{73}}, \doiprefix\url{10.1177/0017896912471049}
  (\bibinfo{year}{2014}).

\bibitem{Xu2022}
\bibinfo{author}{Xu, W.}
\newblock \bibinfo{journal}{\bibinfo{title}{The contingency of neighbourhood
  diversity: Variation of social context using mobile phone application data}}.
\newblock {\emph{\JournalTitle{Urban Studies}}} \textbf{\bibinfo{volume}{59}},
  \doiprefix\url{10.1177/00420980211019637} (\bibinfo{year}{2022}).

\bibitem{Bor2017}
\bibinfo{author}{Bor, J.}, \bibinfo{author}{Cohen, G.~H.} \&
  \bibinfo{author}{Galea, S.}
\newblock \bibinfo{journal}{\bibinfo{title}{Population health in an era of
  rising income inequality: Usa, 1980–2015}}.
\newblock {\emph{\JournalTitle{The Lancet}}} \textbf{\bibinfo{volume}{389}},
  \doiprefix\url{10.1016/S0140-6736(17)30571-8} (\bibinfo{year}{2017}).

\bibitem{Owens2019}
\bibinfo{author}{Owens, A.} \& \bibinfo{author}{Candipan, J.}
\newblock \bibinfo{journal}{\bibinfo{title}{Social and spatial inequalities of
  educational opportunity: A portrait of schools serving high- and low-income
  neighbourhoods in us metropolitan areas}}.
\newblock {\emph{\JournalTitle{Urban Studies}}} \textbf{\bibinfo{volume}{56}},
  \doiprefix\url{10.1177/0042098018815049} (\bibinfo{year}{2019}).

\bibitem{Bosquet2019}
\bibinfo{author}{Bosquet, C.} \& \bibinfo{author}{Overman, H.~G.}
\newblock \bibinfo{journal}{\bibinfo{title}{Why does birthplace matter so
  much?}}
\newblock {\emph{\JournalTitle{Journal of Urban Economics}}}
  \textbf{\bibinfo{volume}{110}}, \doiprefix\url{10.1016/j.jue.2019.01.003}
  (\bibinfo{year}{2019}).

\bibitem{Echenique2007}
\bibinfo{author}{Echenique, F.} \& \bibinfo{author}{Fryer, R.~G.}
\newblock \bibinfo{journal}{\bibinfo{title}{A measure of segregation based on
  social interactions}}.
\newblock {\emph{\JournalTitle{Quarterly Journal of Economics}}}
  \textbf{\bibinfo{volume}{122}}, \doiprefix\url{10.1162/qjec.122.2.441}
  (\bibinfo{year}{2007}).

\bibitem{Musterd2017}
\bibinfo{author}{Musterd, S.}, \bibinfo{author}{Marcińczak, S.},
  \bibinfo{author}{van Ham, M.} \& \bibinfo{author}{Tammaru, T.}
\newblock \bibinfo{journal}{\bibinfo{title}{Socioeconomic segregation in
  european capital cities. increasing separation between poor and rich}}.
\newblock {\emph{\JournalTitle{Urban Geography}}}
  \textbf{\bibinfo{volume}{38}}, \doiprefix\url{10.1080/02723638.2016.1228371}
  (\bibinfo{year}{2017}).

\bibitem{ric1}
\bibinfo{author}{Tóth, G.} \emph{et~al.}
\newblock \bibinfo{journal}{\bibinfo{title}{Inequality is rising where social
  network segregation interacts with urban topology}}.
\newblock {\emph{\JournalTitle{Nature Communications}}}
  \textbf{\bibinfo{volume}{12}}, \doiprefix\url{10.1038/s41467-021-21465-0}
  (\bibinfo{year}{2021}).

\bibitem{Wissink2016}
\bibinfo{author}{Wissink, B.}, \bibinfo{author}{Schwanen, T.} \&
  \bibinfo{author}{van Kempen, R.}
\newblock \bibinfo{journal}{\bibinfo{title}{Beyond residential segregation:
  Introduction}}.
\newblock {\emph{\JournalTitle{Cities}}} \textbf{\bibinfo{volume}{59}},
  \doiprefix\url{10.1016/j.cities.2016.08.010} (\bibinfo{year}{2016}).

\bibitem{Xu2019}
\bibinfo{author}{Xu, Y.}, \bibinfo{author}{Belyi, A.}, \bibinfo{author}{Santi,
  P.} \& \bibinfo{author}{Ratti, C.}
\newblock \bibinfo{journal}{\bibinfo{title}{Quantifying segregation in an
  integrated urban physical-social space}}.
\newblock {\emph{\JournalTitle{Journal of the Royal Society Interface}}}
  \textbf{\bibinfo{volume}{16}}, \doiprefix\url{10.1098/rsif.2019.0536}
  (\bibinfo{year}{2019}).

\bibitem{Massey1988}
\bibinfo{author}{Massey, D.~S.} \& \bibinfo{author}{Denton, N.~A.}
\newblock \bibinfo{journal}{\bibinfo{title}{The dimensions of residential
  segregation}}.
\newblock {\emph{\JournalTitle{Social Forces}}} \textbf{\bibinfo{volume}{67}},
  \doiprefix\url{10.1093/sf/67.2.281} (\bibinfo{year}{1988}).

\bibitem{Macedo2022}
\bibinfo{author}{Macedo, M.}, \bibinfo{author}{Lotero, L.},
  \bibinfo{author}{Cardillo, A.}, \bibinfo{author}{Menezes, R.} \&
  \bibinfo{author}{Barbosa, H.}
\newblock \bibinfo{journal}{\bibinfo{title}{Differences in the spatial
  landscape of urban mobility: Gender and socioeconomic perspectives}}.
\newblock {\emph{\JournalTitle{PLoS ONE}}} \textbf{\bibinfo{volume}{17}},
  \doiprefix\url{10.1371/journal.pone.0260874} (\bibinfo{year}{2022}).

\bibitem{dong}
\bibinfo{author}{Dong, X.} \emph{et~al.}
\newblock \bibinfo{journal}{\bibinfo{title}{Social bridges in urban purchase
  behavior}}.
\newblock {\emph{\JournalTitle{ACM Trans. Intell. Syst. Technol.}}}
  \textbf{\bibinfo{volume}{9}}, \doiprefix\url{10.1145/3149409}
  (\bibinfo{year}{2017}).

\bibitem{Chetty2022}
\bibinfo{author}{Chetty, R.} \emph{et~al.}
\newblock \bibinfo{journal}{\bibinfo{title}{Social capital ii: determinants of
  economic connectedness}}.
\newblock {\emph{\JournalTitle{Nature}}} \textbf{\bibinfo{volume}{608}},
  \doiprefix\url{10.1038/s41586-022-04997-3} (\bibinfo{year}{2022}).

\bibitem{Wang2018}
\bibinfo{author}{Wang, Q.}, \bibinfo{author}{Phillips, N.~E.},
  \bibinfo{author}{Small, M.~L.} \& \bibinfo{author}{Sampson, R.~J.}
\newblock \bibinfo{journal}{\bibinfo{title}{Urban mobility and neighborhood
  isolation in america‚Äôs 50 largest cities}}.
\newblock {\emph{\JournalTitle{Proceedings of the National Academy of Sciences
  of the United States of America}}} \textbf{\bibinfo{volume}{115}},
  \doiprefix\url{10.1073/pnas.1802537115} (\bibinfo{year}{2018}).

\bibitem{lbs}
\bibinfo{author}{S., J.} \& \bibinfo{author}{V., A.}
\newblock \bibinfo{journal}{\bibinfo{title}{Location-based services}}.
\newblock {\emph{\JournalTitle{Elsevier Inc}}} \textbf{\bibinfo{volume}{24}},
  \doiprefix\url{https://www.elsevier.com/books/location-based-services/schiller/978-1-55860-929-7}
  (\bibinfo{year}{2004}).

\bibitem{Oldenburg1999}
\bibinfo{author}{Oldenburg, R.}
\newblock \bibinfo{journal}{\bibinfo{title}{The great good place}}.
\newblock {\emph{\JournalTitle{The Great Good Place}}}  (\bibinfo{year}{1999}).

\bibitem{Moro2021}
\bibinfo{author}{Moro, E.}, \bibinfo{author}{Calacci, D.},
  \bibinfo{author}{Dong, X.} \& \bibinfo{author}{Pentland, A.}
\newblock \bibinfo{journal}{\bibinfo{title}{Mobility patterns are associated
  with experienced income segregation in large us cities}}.
\newblock {\emph{\JournalTitle{Nature Communications}}}
  \textbf{\bibinfo{volume}{12}}, \doiprefix\url{10.1038/s41467-021-24899-8}
  (\bibinfo{year}{2021}).

\bibitem{Moro2023}
\bibinfo{author}{Fan, Z.} \emph{et~al.}
\newblock \bibinfo{journal}{\bibinfo{title}{Diversity beyond density:
  experienced social mixing of urban streets}}.
\newblock {\emph{\JournalTitle{PNAS Nexus}}}
  \doiprefix\url{10.1093/pnasnexus/pgad077} (\bibinfo{year}{2023}).

\bibitem{Wong2011}
\bibinfo{author}{Wong, D.~W.} \& \bibinfo{author}{Shaw, S.~L.}
\newblock \bibinfo{journal}{\bibinfo{title}{Measuring segregation: An activity
  space approach}}.
\newblock {\emph{\JournalTitle{Journal of Geographical Systems}}}
  \textbf{\bibinfo{volume}{13}}, \doiprefix\url{10.1007/s10109-010-0112-x}
  (\bibinfo{year}{2011}).

\bibitem{nilforoshan2023human}
\bibinfo{author}{Nilforoshan, H.} \emph{et~al.}
\newblock \bibinfo{title}{Human mobility networks reveal increased segregation
  in large cities}, \doiprefix\url{https://doi.org/10.1038/s41586-023-06757-3}
  (\bibinfo{year}{2023}).

\bibitem{Sampson}
\bibinfo{author}{Sampson, R.~J.}
\newblock \emph{\bibinfo{title}{Great American City}} (\bibinfo{publisher}{The
  University of Chicago Press}, \bibinfo{year}{2012}).

\bibitem{sweden}
\bibinfo{author}{Östh, J.}, \bibinfo{author}{Shuttleworth, I.} \&
  \bibinfo{author}{Niedomysl, T.}
\newblock \bibinfo{journal}{\bibinfo{title}{Spatial and temporal patterns of
  economic segregation in sweden’s metropolitan areas: A mobility approach}}.
\newblock {\emph{\JournalTitle{Environment and Planning A}}}
  \textbf{\bibinfo{volume}{50}}, \doiprefix\url{10.1177/0308518X18763167}
  (\bibinfo{year}{2018}).

\bibitem{marta}
\bibinfo{author}{González, M.~C.}, \bibinfo{author}{Hidalgo, C.~A.} \&
  \bibinfo{author}{Barabási, A.~L.}
\newblock \bibinfo{journal}{\bibinfo{title}{Understanding individual human
  mobility patterns}}.
\newblock {\emph{\JournalTitle{Nature}}} \textbf{\bibinfo{volume}{453}},
  \doiprefix\url{10.1038/nature06958} (\bibinfo{year}{2008}).

\bibitem{Yue2017}
\bibinfo{author}{Yue, Y.} \emph{et~al.}
\newblock \bibinfo{journal}{\bibinfo{title}{Measurements of poi-based mixed use
  and their relationships with neighbourhood vibrancy}}.
\newblock {\emph{\JournalTitle{International Journal of Geographical
  Information Science}}} \textbf{\bibinfo{volume}{31}},
  \doiprefix\url{10.1080/13658816.2016.1220561} (\bibinfo{year}{2017}).

\bibitem{Dimaggio2012}
\bibinfo{author}{Dimaggio, P.} \& \bibinfo{author}{Garip, F.}
\newblock \bibinfo{title}{Network effects and social inequality},
  \doiprefix\url{10.1146/annurev.soc.012809.102545} (\bibinfo{year}{2012}).

\bibitem{Mehta2019}
\bibinfo{author}{Mehta, V.}
\newblock \bibinfo{journal}{\bibinfo{title}{Streets and social life in cities:
  a taxonomy of sociability}}.
\newblock {\emph{\JournalTitle{Urban Design International}}}
  \textbf{\bibinfo{volume}{24}}, \doiprefix\url{10.1057/s41289-018-0069-9}
  (\bibinfo{year}{2019}).

\bibitem{Madden2014}
\bibinfo{author}{Madden, D.~J.}
\newblock \bibinfo{journal}{\bibinfo{title}{Neighborhood as spatial project:
  Making the urban order on the downtown brooklyn waterfront}}.
\newblock {\emph{\JournalTitle{International Journal of Urban and Regional
  Research}}} \textbf{\bibinfo{volume}{38}},
  \doiprefix\url{10.1111/1468-2427.12068} (\bibinfo{year}{2014}).

\bibitem{hex}
\bibinfo{author}{Ben-Joseph, E.} \& \bibinfo{author}{Gordon, D.}
\newblock \bibinfo{journal}{\bibinfo{title}{Hexagonal planning in theory and
  practice}}.
\newblock {\emph{\JournalTitle{Journal of Urban Design}}}
  \textbf{\bibinfo{volume}{5}}, \doiprefix\url{10.1080/713683965}
  (\bibinfo{year}{2000}).

\bibitem{Zhong2014}
\bibinfo{author}{Zhong, C.}, \bibinfo{author}{Arisona, S.~M.},
  \bibinfo{author}{Huang, X.}, \bibinfo{author}{Batty, M.} \&
  \bibinfo{author}{Schmitt, G.}
\newblock \bibinfo{journal}{\bibinfo{title}{Detecting the dynamics of urban
  structure through spatial network analysis}}.
\newblock {\emph{\JournalTitle{International Journal of Geographical
  Information Science}}} \textbf{\bibinfo{volume}{28}},
  \doiprefix\url{10.1080/13658816.2014.914521} (\bibinfo{year}{2014}).

\bibitem{amenity}
\bibinfo{author}{Juhász, S.} \emph{et~al.}
\newblock \bibinfo{title}{Amenity complexity and urban locations of
  socio-economic mixing} (\bibinfo{year}{2022}).
\newblock \eprint{2212.07280}.

\bibitem{Salesses2013}
\bibinfo{author}{Salesses, P.}, \bibinfo{author}{Schechtner, K.} \&
  \bibinfo{author}{Hidalgo, C.~A.}
\newblock \bibinfo{journal}{\bibinfo{title}{The collaborative image of the
  city: Mapping the inequality of urban perception}}.
\newblock {\emph{\JournalTitle{PLoS ONE}}} \textbf{\bibinfo{volume}{8}},
  \doiprefix\url{10.1371/journal.pone.0068400} (\bibinfo{year}{2013}).

\bibitem{Nieuwenhuis2020}
\bibinfo{author}{Nieuwenhuis, J.}, \bibinfo{author}{Tammaru, T.},
  \bibinfo{author}{van Ham, M.}, \bibinfo{author}{Hedman, L.} \&
  \bibinfo{author}{Manley, D.}
\newblock \bibinfo{journal}{\bibinfo{title}{Does segregation reduce
  socio-spatial mobility? evidence from four european countries with different
  inequality and segregation contexts}}.
\newblock {\emph{\JournalTitle{Urban Studies}}} \textbf{\bibinfo{volume}{57}},
  \doiprefix\url{10.1177/0042098018807628} (\bibinfo{year}{2020}).

\bibitem{Hidalgo2020}
\bibinfo{author}{Hidalgo, C.~A.}, \bibinfo{author}{Castaner, E.} \&
  \bibinfo{author}{Sevtsuk, A.}
\newblock \bibinfo{journal}{\bibinfo{title}{The amenity mix of urban
  neighborhoods}}.
\newblock {\emph{\JournalTitle{Habitat International}}}
  \textbf{\bibinfo{volume}{106}},
  \doiprefix\url{10.1016/j.habitatint.2020.102205} (\bibinfo{year}{2020}).

\bibitem{Biazzo2019}
\bibinfo{author}{Biazzo, I.}, \bibinfo{author}{Monechi, B.} \&
  \bibinfo{author}{Loreto, V.}
\newblock \bibinfo{journal}{\bibinfo{title}{General scores for accessibility
  and inequality measures in urban areas}}.
\newblock {\emph{\JournalTitle{Royal Society Open Science}}}
  \textbf{\bibinfo{volume}{6}}, \doiprefix\url{10.1098/rsos.190979}
  (\bibinfo{year}{2019}).

\bibitem{Saeidi2012}
\bibinfo{author}{Saeidi, S.} \& \bibinfo{author}{Oktay, D.}
\newblock \bibinfo{journal}{\bibinfo{title}{Diversity for better quality of
  community life: Evaluations in famagusta neighbourhoods}}.
\newblock {\emph{\JournalTitle{Procedia - Social and Behavioral Sciences}}}
  \textbf{\bibinfo{volume}{35}}, \doiprefix\url{10.1016/j.sbspro.2012.02.115}
  (\bibinfo{year}{2012}).

\bibitem{Tacchella2012}
\bibinfo{author}{Tacchella, A.}, \bibinfo{author}{Cristelli, M.},
  \bibinfo{author}{Caldarelli, G.}, \bibinfo{author}{Gabrielli, A.} \&
  \bibinfo{author}{Pietronero, L.}
\newblock \bibinfo{journal}{\bibinfo{title}{A new metrics for countries'
  fitness and products' complexity}}.
\newblock {\emph{\JournalTitle{Scientific Reports}}}
  \textbf{\bibinfo{volume}{2}}, \doiprefix\url{10.1038/srep00723}
  (\bibinfo{year}{2012}).

\bibitem{Belmonte2014}
\bibinfo{author}{Belmonte, A.}, \bibinfo{author}{{Di Clemente}, R.} \&
  \bibinfo{author}{Buldyrev, S.~V.}
\newblock \bibinfo{journal}{\bibinfo{title}{The italian primary school-size
  distribution and the city-size: A complex nexus}}.
\newblock {\emph{\JournalTitle{Scientific Reports}}}
  \textbf{\bibinfo{volume}{4}}, \doiprefix\url{10.1038/srep05301}
  (\bibinfo{year}{2014}).

\bibitem{radial}
\bibinfo{author}{Chakravorty, S.}
\newblock \bibinfo{journal}{\bibinfo{title}{A measurement of spatial disparity:
  The case of income inequality}}.
\newblock {\emph{\JournalTitle{Urban Studies}}} \textbf{\bibinfo{volume}{33}},
  \bibinfo{pages}{1671--1686}, \doiprefix\url{10.1080/0042098966556}
  (\bibinfo{year}{1996}).
\newblock \eprint{https://doi.org/10.1080/0042098966556}.

\bibitem{Ellis2004}
\bibinfo{author}{Ellis, M.}, \bibinfo{author}{Wright, R.} \&
  \bibinfo{author}{Parks, V.}
\newblock \bibinfo{journal}{\bibinfo{title}{Work together, live apart?
  geographies of racial and ethnic segregation at home and at work}}.
\newblock {\emph{\JournalTitle{Annals of the Association of American
  Geographers}}} \textbf{\bibinfo{volume}{94}},
  \doiprefix\url{10.1111/j.1467-8306.2004.00417.x} (\bibinfo{year}{2004}).

\bibitem{Candipan2021}
\bibinfo{author}{Candipan, J.}, \bibinfo{author}{Phillips, N.~E.},
  \bibinfo{author}{Sampson, R.~J.} \& \bibinfo{author}{Small, M.}
\newblock \bibinfo{journal}{\bibinfo{title}{From residence to movement: The
  nature of racial segregation in everyday urban mobility}}.
\newblock {\emph{\JournalTitle{Urban Studies}}} \textbf{\bibinfo{volume}{58}},
  \doiprefix\url{10.1177/0042098020978965} (\bibinfo{year}{2021}).

\bibitem{Blumenberg2019}
\bibinfo{author}{Blumenberg, E.} \& \bibinfo{author}{King, H.}
\newblock \bibinfo{journal}{\bibinfo{title}{Low-income workers, residential
  location, and the changing commute in the united states}}.
\newblock {\emph{\JournalTitle{Built Environment}}}
  \textbf{\bibinfo{volume}{45}}, \bibinfo{pages}{563--581},
  \doiprefix\url{10.2148/benv.45.4.563} (\bibinfo{year}{2019}).

\bibitem{gini}
\bibinfo{author}{Maio, F. G.~D.}
\newblock \bibinfo{journal}{\bibinfo{title}{Income inequality measures}}.
\newblock {\emph{\JournalTitle{Journal of Epidemiology and Community Health}}}
  \textbf{\bibinfo{volume}{61}}, \doiprefix\url{10.1136/jech.2006.052969}
  (\bibinfo{year}{2007}).

\bibitem{florida}
\bibinfo{author}{Perucich, J. F.~V.}
\newblock \bibinfo{journal}{\bibinfo{title}{Florida, richard (2017). the new
  urban crisis: How our cities are increasing inequality, deepening
  segregation, and failing the middle class‚Äîand what we can do about
  it}}.
\newblock {\emph{\JournalTitle{Documents d'analisi Geografica}}}
  \textbf{\bibinfo{volume}{65}}, \doiprefix\url{10.5565/rev/dag.535}
  (\bibinfo{year}{2019}).

\bibitem{Hidalgo2009}
\bibinfo{author}{Hidalgo, C.~A.} \& \bibinfo{author}{Hausmann, R.}
\newblock \bibinfo{journal}{\bibinfo{title}{The building blocks of economic
  complexity}}.
\newblock {\emph{\JournalTitle{Proceedings of the National Academy of Sciences
  of the United States of America}}} \textbf{\bibinfo{volume}{106}},
  \doiprefix\url{10.1073/pnas.0900943106} (\bibinfo{year}{2009}).
\bibitem{mobilkit}
Enrico Ubaldi, Takahiro Yabe, Nicholas K.~W. Jones, Maham~Faisal Khan,
  Satish~V. Ukkusuri, and Riccardo Di Clemente~Emanuele Strano.
\newblock Mobilkit: {A} python toolkit for urban resilience and disaster risk
  management analytics using high frequency human mobility data.
\newblock {\em CoRR}, abs/2107.14297, 2021.

\bibitem{Hariharan2004}
Ramaswamy Hariharan and Kentaro Toyama.
\newblock Project lachesis: Parsing and modeling location histories.
\newblock {\em Lecture Notes in Computer Science (including subseries Lecture
  Notes in Artificial Intelligence and Lecture Notes in Bioinformatics)}, 3234,
  2004.

\bibitem{oecd}
Delineating functional areas in all territories.
\newblock \url{https://data.oecd.org/}.

\bibitem{Phithakkitnukoon2012}
Santi Phithakkitnukoon, Zbigniew Smoreda, and Patrick Olivier.
\newblock Socio-geography of human mobility: A study using longitudinal mobile
  phone data.
\newblock {\em PLoS ONE}, 7, 2012.

\bibitem{Alexander2015}
Lauren Alexander, Shan Jiang, Mikel Murga, and Marta~C. Gonzalez.
\newblock Origin-destination trips by purpose and time of day inferred from
  mobile phone data.
\newblock {\em Transportation Research Part C: Emerging Technologies}, 58,
  2015.

\bibitem{Yabe2023}
Takahiro Yabe, Bernardo Garcia~Bulle Bueno, Xiaowen Dong, Alex Pentland, and
  Esteban Moro.
\newblock Behavioral changes during the covid-19 pandemic decreased income
  diversity of urban encounters.
\newblock {\em Nature Communications}, 14, 2023.

\bibitem{caasa}
Caasa.
\newblock \url{https://www.caasa.it/m}.

\bibitem{hex}
E.~Ben-Joseph and D.~Gordon.
\newblock Hexagonal planning in theory and practice.
\newblock {\em Journal of Urban Design}, 5, 2000.

\bibitem{Murtagh2017}
Fionn Murtagh and Pedro Contreras.
\newblock Algorithms for hierarchical clustering: an overview, ii, 2017.

\bibitem{Fowlkes1983}
E.~B. Fowlkes and C.~L. Mallows.
\newblock A method for comparing two hierarchical clusterings.
\newblock {\em Journal of the American Statistical Association}, 78, 1983.

\bibitem{Ng2002}
Andrew~Y. Ng, Michael~I. Jordan, and Yair Weiss.
\newblock On spectral clustering: Analysis and an algorithm.
\newblock 2002.

\bibitem{Mouratidis2020}
Kostas Mouratidis.
\newblock Neighborhood characteristics, neighborhood satisfaction, and
  well-being: The links with neighborhood deprivation.
\newblock {\em Land Use Policy}, 99, 2020.

\bibitem{Greenberg1999}
Michael~R. Greenberg.
\newblock Improving neighborhood quality: A hierarchy of needs.
\newblock {\em Housing Policy Debate}, 10, 1999.

\bibitem{cerrato}
Belkis~Cerrato Caceres and Jacqueline Geoghegan.
\newblock Effects of new grocery store development on inner-city neighborhood
  residential prices, 2017.

\bibitem{Gustafson2012}
Alison~A. Gustafson, Sarah Lewis, Corey Wilson, and Stephanie Jilcott-Pitts.
\newblock Validation of food store environment secondary data source and the
  role of neighborhood deprivation in appalachia, kentucky, 2012.

\bibitem{Cudny2019}
Waldemar Cudny and Hakan Appelblad.
\newblock Monuments and their functions in urban public space.
\newblock {\em Norsk Geografisk Tidsskrift}, 73, 2019.

\bibitem{Baran2014}
Perver~K. Baran, William~R. Smith, Robin~C. Moore, Myron~F. Floyd, Jason~N.
  Bocarro, Nilda~G. Cosco, and Thomas~M. Danninger.
\newblock Park use among youth and adults: Examination of individual, social,
  and urban form factors.
\newblock {\em Environment and Behavior}, 46, 2014.

\bibitem{Kruizse2019}
Hanneke Kruizse, Nina van~der Vliet, Brigit Staatsen, Ruth Bell, Aline Chiabai,
  Gabriel Muinos, Sahran Higgins, Sonia Quiroga, Pablo Martinez-Juarez,
  Monica~Aberg Yngwe, Fotis Tsichlas, Pania Karnaki, Maria~Luisa Lima,
  Silvestre~Garcia de~Jalon, Matluba Khan, George Morris, and Ingrid Stegeman.
\newblock Urban green space: creating a triple win for environmental
  sustainability, health, and health equity through behavior change, 2019.

\bibitem{Mak2018}
Bonnie~K.L. Mak and C.~Y. Jim.
\newblock Examining fear-evoking factors in urban parks in hong kong.
\newblock {\em Landscape and Urban Planning}, 171, 2018.


\bibitem{rca}
Bela Balassa.
\newblock Trade liberalisation and “revealed” comparative advantage.
\newblock {\em The Manchester School}, 33, 1965.

\bibitem{urbanzation}
Riccardo {Di Clemente}, Emanuele Strano, and Michael Batty.
\newblock Urbanization and economic complexity.
\newblock {\em Scientific Reports}, 11, 2021.

\bibitem{stracca}
Straccamore Matteo, Bruno Matteo, Monechi Bernardo, and Vittorio Loreto.
\newblock Urban economic fitness and complexity from patent data.
\newblock {\em Sci Rep}, 13, 2023.

\bibitem{giamba}
Giambattista Albora and Andrea Zaccaria.
\newblock Machine learning to assess relatedness: The advantage of using
  firm-level data.
\newblock {\em Complexity}, 2022.

\bibitem{cimini}
Giulio Cimini, Andrea Gabrielli, and Francesco~Sylos Labini.
\newblock The scientific competitiveness of nations.
\newblock {\em PLoS ONE}, 9, 2014.

\bibitem{lmg}
P.~K. Sen, Richard~H. Lindeman, Peter~F. Merenda, and Ruth~Z. Gold.
\newblock Introduction to bivariate and multivariate analysis.
\newblock {\em Journal of the American Statistical Association}, 76, 1981.


\end{thebibliography}
\end{document}